\documentclass[preprint,10pt]{elsarticle}

\usepackage{lineno}
\usepackage{subfig}
\usepackage{mathtools}
\usepackage{comment}
\usepackage{booktabs}
\usepackage{amssymb}
\usepackage{url}
\usepackage{tikz} 
\usetikzlibrary{shapes,arrows,positioning,chains,fit,arrows.meta}

\usepackage{float}
\usepackage{tabularx}
\usepackage{enumerate}
\usepackage{array}
\usepackage{xspace}

\usepackage{amsthm}
\usepackage{amsmath}
\usepackage{algorithm}
\usepackage[noend]{algpseudocode}
\usepackage{booktabs}

\usepackage{ulem}
\usepackage[a4paper,left=3cm,right=3cm,top=2.5cm,bottom=2.5cm]{geometry}

\newcommand{\fig}{\text{Fig.~}}
\newcommand{\tab}{\text{Tab.~}}
\newcommand{\eq}{\text{Eq.~}}
\newcommand{\pb}{\text{problem~}}
\newcommand{\sez}{\text{Sec.~}}

\DeclareFontEncoding{LS1}{}{}
\DeclareFontSubstitution{LS1}{stix}{m}{n}
\DeclareSymbolFont{symbolsstix}{LS1}{stixscr}{m}{n}
\SetSymbolFont{symbolsstix}{bold}{LS1}{stixscr}{b}{n}
\DeclareMathSymbol{\y}{0}{symbolsstix}{'171}
\DeclareMathSymbol{\I}{0}{symbolsstix}{'111}
\DeclareMathSymbol{\jj}{0}{symbolsstix}{'152}
\DeclareMathSymbol{\kk}{0}{symbolsstix}{'153}
\DeclareMathSymbol{\EE}{0}{symbolsstix}{'105}

\makeatletter
\def\ps@pprintTitle{%
  \let\@oddhead\@empty
  \let\@evenhead\@empty
  \def\@oddfoot{\reset@font\hfil\thepage\hfil}
  \let\@evenfoot\@oddfoot
}
\makeatother

\begin{document}

\begin{frontmatter}
\title{Enhancing Bayesian model updating in structural health monitoring via learnable mappings}

\author[1]{Matteo~Torzoni\corref{cor1}}
\ead{matteo.torzoni@polimi.it}

\author[2]{Andrea~Manzoni}
\ead{andrea1.manzoni@polimi.it}

\author[1]{Stefano~Mariani}
\ead{stefano.mariani@polimi.it}


\affiliation[1]{organization={Dipartimento di Ingegneria Civile e Ambientale, Politecnico di Milano},
city={Milan},
postcode={20133},
country={Italy}}
\affiliation[2]{organization={MOX, Dipartimento di Matematica, Politecnico di Milano},
city={Milan},
postcode={20133},
country={Italy}}

\cortext[cor1]{Corresponding author}

\begin{abstract}
In the context of structural health monitoring (SHM), the selection and extraction of damage-sensitive features from raw sensor recordings represent a critical step towards solving the inverse problem underlying the identification of structural health conditions. This work introduces a novel approach that employs deep neural networks to enhance stochastic SHM methods. A learnable feature extractor and a feature-oriented surrogate model are synergistically exploited to evaluate a likelihood function within a Markov chain Monte Carlo sampling algorithm. The feature extractor undergoes pairwise supervised training to map sensor recordings onto a low-dimensional metric space, which encapsulates the sensitivity to structural health parameters. The surrogate model maps structural health parameters to their feature representation. The procedure enables the updating of beliefs about structural health parameters, eliminating the need for computationally expensive numerical models. A preliminary offline phase involves the generation of a labeled dataset to train both the feature extractor and the surrogate model. Within a simulation-based SHM framework, training vibration responses are efficiently generated using a multi-fidelity surrogate modeling strategy to approximate sensor recordings under varying damage and operational conditions. The multi-fidelity surrogate exploits model order reduction and artificial neural networks to speed up the data generation phase while ensuring the damage-sensitivity of the approximated signals. The proposed strategy is assessed through three synthetic case studies, demonstrating high accuracy in the estimated parameters and strong computational efficiency.
\end{abstract}

\begin{keyword} Bayesian model updating \sep Deep learning \sep Markov chain Monte Carlo \sep Structural health monitoring \sep Multi-fidelity methods \sep Reduced-order modeling \sep Contrastive learning.
\end{keyword}

\end{frontmatter}

\section{Introduction}

Ensuring the safety of civil structural systems is a key societal challenge. This is daily threatened by material deterioration, cyclic and extraordinary loading conditions, and more and more by effects triggered by climate change, such as anomalous heat waves and destructive storms~\cite{committee2018climate}. Since the lifecycle (economic, social and safety) costs entailed by such structural systems may be extremely high, it is now critical to enable a condition-based maintenance approach instead of time-based ones~\cite{art:Glaser,art:Achenbach}. To this aim, non-destructive tests and in situ inspections are not suitable to implement a continuous and automated global monitoring; on the other hand, by assimilating vibration response data acquired with permanently installed data collection systems~\cite{proc:Rosafalco2022,proc:Springer_Ubertini}, vibration-based structural health monitoring (SHM) techniques allow for damage identification and evolution tracking.

Data-driven approaches to SHM~\cite{art:Torzoni_DML,art:Worden_detection} rely on a pattern recognition paradigm~\cite{art:Farrar01} involving: ($i$) operational evaluation; ($ii$) data acquisition; ($iii$) feature selection and extraction; ($iv$) statistical modeling to unveil the relationship between the selected features and sought damage patterns~\cite{book:Bishop,art:Avci_review}. In this process, the selection of compact and informative features is the most critical step, as it requires problem-specific knowledge subject to the available expertise. To this aim, deep learning (DL) represents a promising solution to automatize the selection and extraction of features optimized for the task at hand.

Within a different strategy, Bayesian model-based approaches to SHM~\cite{art:Ierimonti,art:Demetrio_2,art:kamariotis_voi,art:Azam_Mariani} assess damage via parameter estimation and model updating. Such a probabilistic framework has the advantage of naturally dealing with the ill-posedness of the SHM problem, and allows to account for and quantify uncertainty.

In this paper, we propose a hybrid approach to SHM leveraging the strengths of both data-driven and model-based approaches. Learnable features, optimized for the structure to be monitored, are automatically selected and extracted by a DL-based feature extractor that relies on an autoencoder architecture to map the input vibration recordings onto their feature representation in a low-dimensional space. During training, the autoencoder is equipped with a Siamese appendix~\cite{art:Siamese} of the encoder, optimized through pairwise contrastive learning~\cite{proc:LeCun_Contrastive_2005,proc:LeCun_Contrastive}. Such a deep metric learning~\cite{art:Deep_metric_survey,art:Metric_survey} strategy enables the learning of a distance function that conforms to a task-specific definition of similarity, so that the neighbors of a data point are mapped closer than non-neighbors in the learned metric space~\cite{proc:Deep_Metric_Learning_Rank}. The resulting mapping encodes the sensitivity to the sought parameters according to the chosen metric, thereby enabling a manifold to describe suitably the parametric space underlying the processed measurements. The extracted features are exploited within a Markov chain Monte Carlo (MCMC) algorithm~\cite{art:VBI,art:AM_Green,art:Lam2018} to estimate parameters describing the variability of the structural system. The likelihood function underlying the MCMC sampler is evaluated by means of a feature-oriented surrogate model, which maps the parameters to be inferred onto the corresponding feature representation.

The proposed strategy takes advantage of a preliminary offline learning phase. The training of the feature extractor and the feature-oriented surrogate model is carried out in a supervised fashion. Labeled data corresponding to specific damage conditions are generated in an inexpensive way through a multi-fidelity (MF) surrogate modeling strategy. In this work, such MF surrogate modeling is chosen as an effective strategy to reduce the computational cost, while ensuring the accuracy of the approximated signals in terms of damage sensitivity. The vibration response data required to fit the MF surrogate are generated through physics-based numerical simulations, so that the effect of damage on the structural response is systematically reproduced.

\begin{figure}[t]
\begin{center}
\includegraphics[width=1\textwidth]{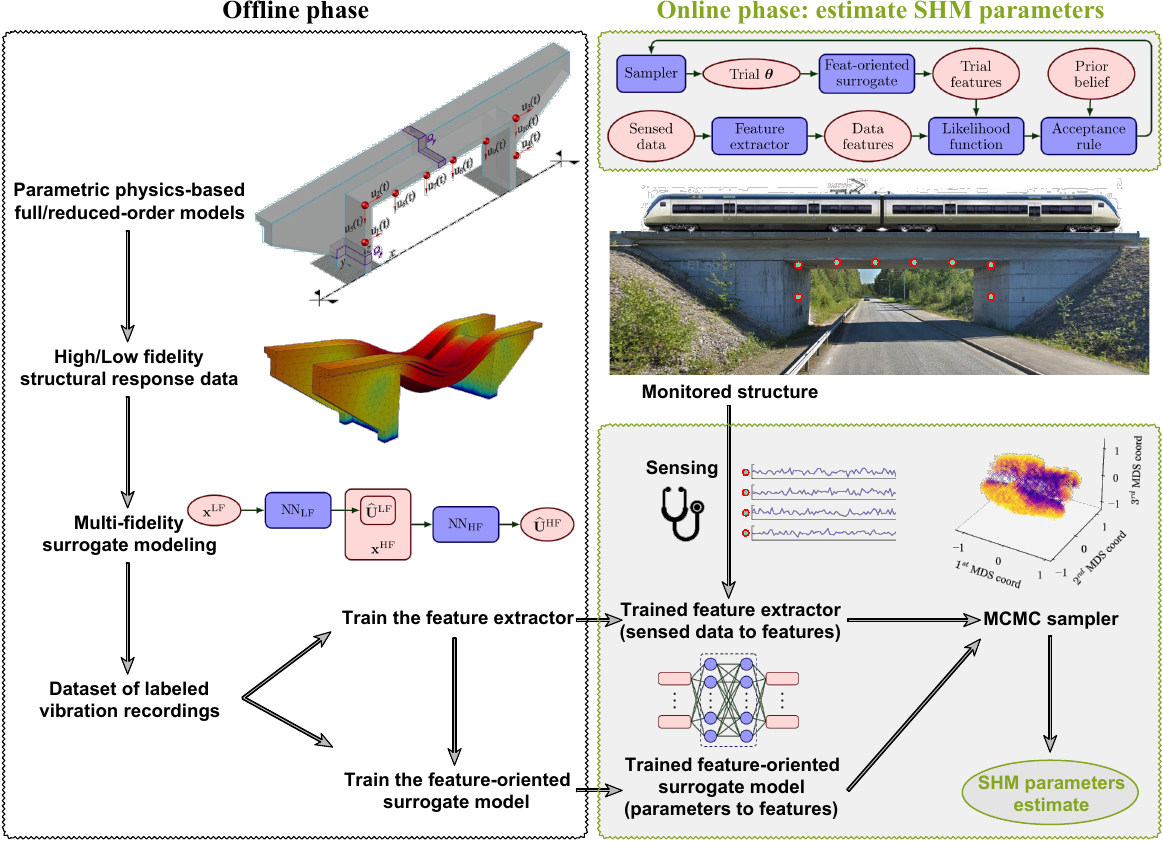}
\caption{{Graphical abstraction of the proposed methodology.}\label{fig:map}}
\end{center}
\end{figure}

A graphical abstraction of the proposed framework is reported in \fig\ref{fig:map}. Vibration responses of different fidelity levels are simulated offline using physics-based full/reduced-order numerical models, similarly to~\cite{art:Metodologico,art:Torzoni_temperature}. These data are then exploited to train a MF surrogate model, following the strategy proposed in~\cite{art:Torzoni_MF}. Once trained, the MF surrogate model is employed to provide an arbitrarily large training dataset. This dataset is used to train the deep-metric-learning-based feature extractor, following a strategy similar to that proposed in~\cite{art:Torzoni_DML}, and the feature-oriented surrogate model, which is employed to approximate the functional link between the parameters to be updated and the low-dimensional feature space. During the online monitoring phase, the feature extractor and the surrogate model are eventually exploited by an MCMC sampling algorithm to update the prior belief about the structural state. 

The elements of novelty that characterize this work are the following. First, the assimilation of data related to vibration responses is carried out by exploiting DL models, which allow the automatic selection and extraction of optimized features from raw vibration recordings. Second, the employed low-dimensional feature space benefits from a geometrical structure, which encodes the sensitivity to the parameters to be updated. The resulting MCMC framework enjoys: a competitive computational cost due to the low dimensionality of the involved features; fast convergence due to the geometrical structure characterizing the feature space; accurate estimates due to the informativeness of the extracted features.

The remainder of the paper is organized as follows. Section~\ref{sec:data} reviews the MF surrogate modeling strategy employed for dataset population purposes. Section~\ref{sec:methodology} describes the proposed parameter estimation framework. Section~\ref{sec:Results} assesses the computational procedure on three test cases, respectively related to a cantilever beam, a portal frame, and a railway bridge. Conclusions and future developments are finally drawn in \sez\ref{sec:conclusions}.

\section{Population of training datasets}
\label{sec:data}

In this section, we describe the population of training datasets within the simulation-based paradigm of SHM. Section~\ref{subsec:data_specification} specifies the composition of the handled vibration responses. Section~\ref{subsec:num_models} describes the numerical models underlying the generation of labeled data corresponding to specific damage conditions. Section~\ref{subsec:surrogate} reviews the MF surrogate modeling strategy employed to populate large training datasets.

\subsection{Data specification}
\label{subsec:data_specification}

The SHM strategy relies on the assimilation of vibration recordings shaped as multivariate time series $\mathbf{U}^\text{EXP}(\boldsymbol{\theta})=[\mathbf{u}^\text{EXP}_1(\boldsymbol{\theta}),\ldots,\mathbf{u}^\text{EXP}_{N_u}(\boldsymbol{\theta})]\in\mathbb{R}^{L\times N_u}$, comprising $N_u$ series, each consisting of $L$ measurements equally spaced in time. For instance, measurements can be provided as accelerations or displacements at structural nodes. The vector $\boldsymbol{\theta} \in\mathbb{R}^{N_\text{par}}$ comprises $N_\text{par}$ parameters, representing variability in structural health and operational conditions, for which we aim to update the belief. Each recording corresponds to a time interval $(0,T)$, during which measurements are recorded with a sampling rate $f_\text{s}$. For the problem setting we consider herewith, the time interval $(0,T)$ is assumed short enough for the operational, environmental, and damage conditions to be considered time-invariant, yet long enough not to compromise the identification of the structural behavior.

\subsection{Low/high fidelity physics-based models}
\label{subsec:num_models}

The labeled dataset required to train the feature extractor and the feature-oriented surrogate model is populated by exploiting the MF surrogate modeling strategy proposed in~\cite{art:Torzoni_MF}. The resulting surrogate model relies on a composition of deep neural network (DNN) models and is therefore termed MF-DNN. The MF surrogate is trained on synthetic data, generated by means of physics-based models. In this section, we describe the models employed to systematically reproduce the effect of damage on the structural response, while \sez\ref{subsec:surrogate} reviews the MF-DNN surrogate.

The chosen physics-based numerical models are: a low-fidelity (LF) reduced-order model (ROM), obtained by using a proper orthogonal decomposition (POD)-Galerkin reduced basis method for parametrized finite element models~\cite{book:RB,art:Metodologico,art:Torzoni_temperature}; and a high-fidelity (HF) finite element model. The two models are employed to simulate the structural responses under varying operational conditions, respectively in the absence or in the presence of a structural damage. In particular, LF data are generated always referring to a baseline condition, while HF data must account for potential degradation processes. Thanks to this modeling choice, the LF component never needs to be updated, and whenever a deterioration of the structural health is detected, the MF surrogate can be updated by adjusting only its HF component. Without loss of generality, we will refer to the initial monitoring phase of an undamaged reference condition.

The HF model describes the dynamic response of the monitored structure to the applied loadings, under the assumption of linearized kinematics. By modeling the structure as a linear-elastic continuum, and by discretizing it in space through finite elements, the HF model consists of the following semi-discretized form:
\begin{linenomath*}
\begin{equation}
\left\{
\begin{array}{ll}
\mathbf{M}_\text{HF}\ddot{\mathbf{d}}^\text{HF}(t) + \mathbf{C}_\text{HF}(\mathbf{x}^\text{HF})\dot{\mathbf{d}}^\text{HF}(t) + \mathbf{K}_\text{HF}(\mathbf{x}^\text{HF})\mathbf{d}^\text{HF}(t)=\mathbf{f}_\text{HF}(t,\mathbf{x}^\text{HF})~, & t\in(0,T)\\ \mathbf{d}^\text{HF}(0)=\mathbf{d}^\text{HF}_{0}~, & \\ \dot{\mathbf{d}}^\text{HF}(0)=\dot{\mathbf{d}}^\text{HF}_{0}~, &
\end{array}
\right.
\label{eq:HF_model}
\end{equation}
\end{linenomath*}
which is referred to as the HF full-order model (FOM). In \pb\eqref{eq:HF_model}: $t\in(0,T)$ denotes time; $\mathbf{d}^\text{HF}(t)$, $\dot{\mathbf{d}}^\text{HF}(t)$, $\ddot{\mathbf{d}}^\text{HF}(t)\in\mathbb{R}^{N_\text{FE}}$ are the vectors of nodal displacements, velocities, and accelerations, respectively; $N_\text{FE}$ is the number of degrees of freedom (dofs); $\mathbf{M}_\text{HF}\in\mathbb{R}^{N_\text{FE}\times N_\text{FE}}$ is the mass matrix; $\mathbf{C}_\text{HF}(\mathbf{x}^\text{HF})\in\mathbb{R}^{N_\text{FE}\times N_\text{FE}}$ is the damping matrix, assembled according to the Rayleigh's model; $\mathbf{K}_\text{HF}(\mathbf{x}^\text{HF})\in\mathbb{R}^{N_\text{FE}\times N_\text{FE}}$ is the stiffness matrix; $\mathbf{f}_\text{HF}(t,\mathbf{x}^\text{HF})\in\mathbb{R}^{N_\text{FE}}$ is the vector of nodal forces induced by the external loadings; $\mathbf{x}^\text{HF}\in\mathbb{R}^{N_\text{par}^\text{HF}}$ is a vector of $N_\text{par}^\text{HF}$ input parameters ruling the operational, damage and (possibly) environmental conditions, such that $\boldsymbol{\theta}\subseteq\mathbf{x}^\text{HF}$; $\mathbf{d}^\text{HF}_{0}$ and $\dot{\mathbf{d}}^\text{HF}_{0}$ are the initial conditions at $t=0$, respectively in terms of nodal displacements and velocities. The solution of \pb\eqref{eq:HF_model} is advanced in time using an implicit Newmark integration scheme (constant average acceleration method).

With reference to civil structures, we focus on the early detection of damage patterns characterized by a small evolution rate, whose prompt identification can reduce lifecycle costs and increase the safety and availability of the structure. In this context, structural damage is often modeled as a localized reduction of the material stiffness~\cite{book:ML_perspective,art:Kapteyn_nature,art:TEUGHELS2002}, which is obtained here by means of a suitable parametrization of the stiffness matrix. In practical terms, we parametrize a damage condition through its position $\boldsymbol{\y}\in\mathbb{R}^{3}$ and magnitude $\delta\in\mathbb{R}$, both included in the parameter vector $\mathbf{x}^\text{HF}$.

The POD-based LF model approximates the solution to \pb\eqref{eq:HF_model} by providing $\mathbf{d}^\text{LF}(t,\mathbf{x}^\text{LF})\approx\mathbf{W}\mathbf{r}(t,\mathbf{x}^\text{LF})$, where $\mathbf{W}=[\mathbf{w}_1,\ldots,\mathbf{w}_{N_\text{RB}}] \in \mathbb{R}^{N_\text{FE} \times N_\text{RB}}$ is a basis matrix featuring $N_\text{RB}\ll N_\text{FE}$ POD basis functions as columns, and $\mathbf{r}(t,\mathbf{x}^\text{LF})\in \mathbb{R}^{N_\text{RB}}$ is the vector of unknown POD coefficients. The approximation is provided for a given vector of LF parameters $\mathbf{x}^\text{LF}\in\mathbb{R}^{N_\text{par}^\text{LF}}$, collecting $N_\text{par}^\text{LF}$ parameters that rule the operational conditions undergone by the structure, with $N_\text{par}^\text{LF}<N_\text{par}^\text{HF}$. By enforcing the orthogonality between the residual and the subspace spanned by the first $N_\text{RB}$ POD modes through a Galerkin projection, the following $N_\text{RB}$-dimensional semi-discretized form is obtained:
\begin{linenomath*}
\begin{equation}
\left\{
\begin{array}{ll}
\mathbf{M}_{r}\ddot{\mathbf{r}}(t) + \mathbf{C}_{r}\dot{\mathbf{r}}(t) + \mathbf{K}_{r}\mathbf{r}(t)=\mathbf{f}_{r}(t,\mathbf{x}^\text{LF})~, &  t\in(0,T)\\ \mathbf{r}(0)=\mathbf{W}^\top\mathbf{d}^\text{LF}_{0}~, & \\ \dot{\mathbf{r}}(0)=\mathbf{W}^\top\dot{\mathbf{d}}^\text{LF}_{0}~. &
\end{array}
\right.
\label{eq:LF_model}
\end{equation}
\end{linenomath*}
The solution of this low-dimensional dynamical system is advanced in time using the same strategy employed for the HF model and is then projected onto the original LF-FOM space as $\mathbf{d}^\text{LF}(t,\mathbf{x}^\text{LF})\approx\mathbf{W}\mathbf{r}(t,\mathbf{x}^\text{LF})$. Here, the reduced-order matrices $\mathbf{M}_{r}$, $\mathbf{C}_{r}$, and $\mathbf{K}_{r}$, and vector $\mathbf{f}_{r}$ play the same role as their HF counterparts, yet with dimension $N_\text{RB} \times N_\text{RB}$ instead of $N_\text{FE} \times N_\text{FE}$, and read:
\begin{linenomath*}
\begin{equation}
\begin{array}{lll}
\mathbf{M}_{r} \equiv \mathbf{W}^\top \mathbf{M}_{\text{HF}} \mathbf{W}~, &\qquad &\mathbf{C}_{r} \equiv \mathbf{W}^\top \mathbf{C}_{\text{HF}} \mathbf{W}~,\\ 
\mathbf{K}_{r} \equiv \mathbf{W}^\top \mathbf{K}_{\text{LF}}\mathbf{W}~, &\qquad&
\mathbf{f}_{r}(t,\mathbf{x}^\text{LF}) \equiv \mathbf{W}^\top \mathbf{f}_{\text{HF}}(t,\mathbf{x}^\text{LF})~.
\end{array}
\end{equation}
\end{linenomath*}

The matrix $\mathbf{W}$ is obtained by exploiting the so-called method of snapshots as follows~\cite{art:sirovich,art:Kerschen_1,art:Kerschen_2}. First, a LF-FOM, similar to that defined by \pb\eqref{eq:HF_model}, but excluding the presence of damage, is employed to assemble a snapshot matrix $\mathbf{S}=[\mathbf{d}^\text{LF}_1,\ldots,\mathbf{d}^\text{LF}_{N_\text{S}}]\in\mathbb{R}^{N_\text{FE}\times N_\text{S}}$ from $N_\text{S}$ solution snapshots, computed by time integration of the LF-FOM for different values of parameters $\mathbf{x}^\text{LF}$. The computation of an optimal reduced basis is then carried out by factorizing $\mathbf{S}$ through a singular value decomposition. We use a standard energy-based criterion to set the order $N_\text{RB}$ of the approximation; for further details, see~\cite{art:Metodologico,art:Torzoni_temperature,art:Torzoni_MF,art:Torzoni_DML}.

To populate the LF and HF datasets, $\mathbf{D}_\text{LF}$ and $\mathbf{D}_\text{HF}$, the parameter vectors $\mathbf{x}^\text{LF}$ and $\mathbf{x}^\text{HF}$ are assumed to follow uniform distributions, and then sampled via the latin hypercube rule. Although this choice is not restrictive, the number of samples is set equal to the number of instances $I_\text{LF}$ and $I_\text{HF}$, with $I_\text{LF}>I_\text{HF}$, collected in $\mathbf{D}_\text{LF}$ and $\mathbf{D}_\text{HF}$, respectively, as:
\begin{linenomath*}
\begin{equation}
\mathbf{D}_\text{LF}=\lbrace(\mathbf{x}^\text{LF}_i,\mathbf{U}^\text{LF}_i)\rbrace_{i=1}^{I_\text{LF}}~,\quad\mathbf{D}_\text{HF}=\lbrace(\mathbf{x}^\text{HF}_j,\mathbf{U}^\text{HF}_j)\rbrace_{j=1}^{I_\text{HF}}~,
\label{eq:Dataset}
\end{equation}
\end{linenomath*}
where the LF and HF vibration recordings $\mathbf{U}^\text{LF}_i(\mathbf{x}^\text{LF}_i)=[\mathbf{u}^\text{LF}_1(\mathbf{x}^\text{LF}),\ldots,\mathbf{u}^\text{LF}_{N_u}(\mathbf{x}^\text{LF})]_i\in\mathbb{R}^{L\times N_u}$ and $\mathbf{U}^\text{HF}_j(\mathbf{x}^\text{HF}_j)=[\mathbf{u}^\text{HF}_1(\mathbf{x}^\text{HF}),\ldots,\mathbf{u}^\text{HF}_{N_u}(\mathbf{x}^\text{HF})]_j\in\mathbb{R}^{L\times N_u}$, are labeled by the corresponding $i$-th and $j$-th samples of $\mathbf{x}^\text{LF}$ and $\mathbf{x}^\text{HF}$, respectively, and are computed as detailed below.

By dropping indices $i$ and $j$ for ease of notation and with reference to displacement recordings, nodal values in $(0,T)$ are first collected as $\mathbf{V}_\text{LF}=[\mathbf{W}\mathbf{r}_1,\ldots,\mathbf{W}\mathbf{r}_{L}] \in \mathbb{R}^{N_\text{FE} \times L}$ and $\mathbf{V}_\text{HF}=[\mathbf{d}^\text{HF}_1,\ldots,\mathbf{d}^\text{HF}_{L}] \in \mathbb{R}^{N_\text{FE} \times L}$, by solving \pb\eqref{eq:LF_model} and \pb\eqref{eq:HF_model}, respectively. The relevant vibration recordings $\mathbf{U}^\text{LF}$ and $\mathbf{U}^\text{HF}$ are then obtained as:
\begin{linenomath*}
\begin{equation}
\mathbf{U}^\text{LF}=(\mathbf{T}\mathbf{V}_\text{LF})^\top~, \qquad \mathbf{U}^\text{HF}=(\mathbf{T}\mathbf{V}_\text{HF})^\top~,
\label{eq:bool_sensors}
\end{equation}
\end{linenomath*}
where $\mathbf{T}\in\mathbb{B}^{N_u \times N_\text{FE}}$ is a Boolean matrix whose $(n,m)$-th entry is equal to $1$ only if the $n$-th sensor coincides with the $m$-th dof. For the problem setting we consider, the sampling frequency $f_\text{s}$, and the number and location of monitored dofs are assumed to be the same for both fidelity levels. However, there are no restrictions in this regard, and LF and HF data with different dimensions can also be considered. Moreover, we note that the matrix product $\mathbf{T}\mathbf{W}\in\mathbb{R}^{N_u \times N_\text{RB}}$ can be computed once for all simulations, to extract $\mathbf{U}^\text{LF}$ for any given set of LF input parameters $\mathbf{x}^\text{LF}$.

\subsection{Multi-fidelity surrogate modeling for structural health monitoring}
\label{subsec:surrogate}

We now review the MF-DNN surrogate modeling strategy proposed in~\cite{art:Torzoni_MF}, which is employed here to generate data corresponding to specific damage conditions in an inexpensive way. The generated data will serve to carry out the training of the feature extractor and of the feature-oriented surrogate. The employed surrogate modeling strategy falls into the wider framework of MF methods, see for instance~\cite{art:MF_survey,art:Meng2020a,art:conti2022MF}. These methods are characterized by the use of multiple models with varying accuracy and computational cost. By blending LF and HF models, MF methods allow for improved approximation accuracy compared to the LF solution, while carrying a lower computational burden than the HF solver. Indeed, LF samples often supply useful information on the major trends of the problem, allowing the MF setting to outperform single-fidelity methods in terms of prediction accuracy and computational efficiency. In addition, MF surrogate models based on DNNs enjoy several appealing features: they are suitable for high-dimensional problems and benefit from large LF training datasets, provide real-time predictions, can deal with linear and nonlinear correlations in an adaptive fashion without requiring prior information, and can handle the approximation of strongly discontinuous trajectories.

Our MF-DNN surrogate model is devised to map damage and operational parameters onto sensor recordings. It leverages an LF part and an HF part, sequentially trained, and respectively denoted by $\text{N\hspace{-1px}N}_\text{LF}$ and $\text{N\hspace{-1px}N}_\text{HF}$. The resulting surrogate model reads as:
\begin{linenomath*}
\begin{equation}
\text{N\hspace{-1px}N}_\text{MF}(\mathbf{x}^\text{HF}, \mathbf{x}^\text{LF}) = \text{N\hspace{-1px}N}_\text{HF}(\mathbf{x}^\text{HF})\circ\text{N\hspace{-1px}N}_\text{LF}(\mathbf{x}^\text{LF})~,
\label{eq:Surrogate_eval}
\end{equation}
\end{linenomath*}
where $\circ$ stands for function composition, see \fig\ref{fig:MF_flow}. 

\tikzstyle{block} = [rectangle, draw=blue!40!black, thick, fill=blue!40, text width=4em,  node distance=2.1cm, text centered, rounded corners, minimum height=2.5em]
\tikzstyle{block2} = [rectangle, draw=blue!40!black, thick, fill=blue!40, text width=4em,  node distance=2.2cm, text centered, rounded corners, minimum height=2.5em]
\tikzstyle{cloud} = [draw=red!40!black, ellipse, fill=red!15, thick, node distance=2.04cm, text badly centered, minimum height=2em, text width=2em]
\tikzstyle{cloud_2} = [node distance=1cm, text badly centered, minimum height=2em, text width=1em]
\tikzstyle{cloud_3} = [rectangle, draw=red!40!black, fill=red!15, thick, node distance=2cm, text badly centered, minimum height=2em, text width=1.8em, rounded corners]
\tikzstyle{cloud_null} = [rectangle, draw=red!40!black, fill=red!15, thick, node distance=1cm, rounded corners, text width=4em, minimum height=5em]
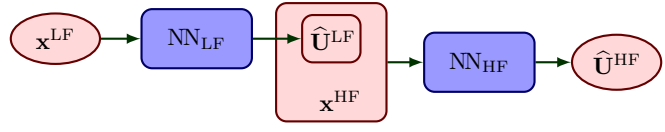
\begin{figure}[t]
\center
\begin{tikzpicture}[node distance = 2cm, scale=0.89, every node/.style={scale=0.89}]
\node [cloud] (LF_param) {$\mathbf{x}^\text{LF}$};
\node [block, right of=LF_param] (NN_LF) {$\text{N\hspace{-1px}N}_\text{LF}$};
\node [cloud_3, right of=NN_LF] (null_0) {$\widehat{\mathbf{U}}^\text{LF}$};
\node [cloud_null, below=-2.3em of null_0] (null) {};
\node [cloud_3, right of=NN_LF] (U_LF) {$\widehat{\mathbf{U}}^\text{LF}$};
\node [cloud_2, below=0.5em of null_0] (HF_param) {$\mathbf{x}^\text{HF}$};
\node [block2, right of=null] (NN_HF) {$\text{N\hspace{-1px}N}_\text{HF}$};
\node [cloud, right of=NN_HF] (U_HF) {$\widehat{\mathbf{U}}^\text{HF}$};
\draw[-latex,thick,green!20!black] (LF_param) to (NN_LF);
\draw[-latex,thick,green!20!black] (NN_LF) to (U_LF);
\draw[-latex,thick,green!20!black] (null) to (NN_HF);
\draw[-latex,thick,green!20!black] (NN_HF) to (U_HF);
\end{tikzpicture}
\caption{Scheme of the MF-DNN surrogate model: red nodes denote the input/output quantities; blue nodes refer to the learnable components of the surrogate model; hat variables denote quantities obtained from neural network approximations. Figure adapted from~\cite{art:Torzoni_MF}.\label{fig:MF_flow}}
\end{figure}

$\text{N\hspace{-1px}N}_\text{LF}$ is set as a fully-connected DL model, exploited to approximate the LF vibration recordings for any given set of LF input parameters $\mathbf{x}^\text{LF}$. In particular, $\text{N\hspace{-1px}N}_\text{LF}$ provides an approximation to a set of POD coefficients encoding $\mathbf{U}^\text{LF}$, allowing the number of trainable parameters of $\text{N\hspace{-1px}N}_\text{LF}$ to be largely reduced.

$\text{N\hspace{-1px}N}_\text{HF}$ is a DNN built upon the long short-term memory (LSTM) model, which exploits the time correlation between the two fidelity levels. Indeed, an LSTM model for $\text{N\hspace{-1px}N}_\text{HF}$ can capitalize on the temporal structure of the LF signals $\widehat{\mathbf{U}}^\text{LF}$ provided through $\text{N\hspace{-1px}N}_\text{LF}$. At each time step, $\text{N\hspace{-1px}N}_\text{HF}$ takes as inputs the HF parameters $\mathbf{x}^\text{HF}$, the current time instant $t$, and the corresponding LF approximation $\widehat{\mathbf{U}}_t^\text{LF}$, to enrich the latter with the effects of damage and provide the HF approximation $\widehat{\mathbf{U}}^\text{HF}(t)$.

The main steps involved in our MF-DNN surrogate modeling strategy are outlined in \fig\ref{fig:meth_flow} and consist of:($1^\text{st}$) the definition of a parametric LF-FOM; the construction of a parametric LF-ROM by means of POD; the population of $\mathbf{D}_\text{LF}$ with LF vibration recordings at sensor locations through LF-ROM simulations; the training and validation of the LF component $\text{N\hspace{-1px}N}_\text{LF}$, employed to approximate $\mathbf{U}^\text{LF}$ for any given $\mathbf{x}^\text{LF}$; the testing of the generalization capabilities of $\text{N\hspace{-1px}N}_\text{LF}$ on LF-FOM data; ($2^\text{nd}$) the definition of a parametric HF structural model accounting for the effects of damage; the population of $\mathbf{D}_\text{HF}$ through HF-FOM simulations; the training and validation of the HF component $\text{N\hspace{-1px}N}_\text{HF}$, employed to enrich the $\widehat{\mathbf{U}}^\text{LF}$ approximation with the effects of damage for any given $\mathbf{x}^\text{HF}$; the testing of the generalization capabilities of $\text{N\hspace{-1px}N}_\text{MF}$. For the interested reader, the detailed steps of our MF-DNN surrogate modeling strategy are reported in~\cite{art:Torzoni_MF}.

\tikzstyle{flo1} = [rectangle, draw=green!20!black, thick, text width=8.2em,  node distance=2.1cm, text centered, rounded corners, minimum height=2.5em]
\tikzstyle{flo2} = [rectangle, draw=green!20!black, thick, text width=5em,  node distance=2.1cm, text centered, rounded corners, minimum height=2.5em]
\tikzstyle{flo3} = [rectangle, draw=green!20!black, thick, text width=6.7em,  node distance=2.1cm, text centered, rounded corners, minimum height=1.5em]
\tikzstyle{flo4} = [rectangle, draw=green!20!black, thick, text width=7em,  node distance=2.1cm, text centered, rounded corners, minimum height=1.5em, dashed]

\begin{figure}[t]
\center
\begin{tikzpicture}[node distance = 2cm, scale=0.82, every node/.style={scale=0.82}]
\linespread{1}
\node [flo1] (param) at (-1.5,0) {Parametrize operational and damage conditions};
\node [flo2,] (LFom_mod) at (2,1.2) {Build LF-FOM};
\node [flo2] (LRom_mod) at (4.8,1.2) {Derive LF-ROM};
\node [flo4] (LF_test) at (4.8,0.15) {LF testing data};
\node [flo3] (LF_data) at (7.8,1.2) {Generate LF dataset $\mathbf{D}_\text{LF}$};
\node [flo3] (LF_train) at (11,1.55) {Train $\text{N\hspace{-1px}N}_\text{LF}$};
\node [flo3] (LF_vali) at (11,0.85) {Validate $\text{N\hspace{-1px}N}_\text{LF}$};
\node [flo3,dashed] (LF_testing) at (11,0.15) {Test $\text{N\hspace{-1px}N}_\text{LF}$};
\node [flo2] (HF_mod) at (2,-1.2) {Build HF-FOM};
\node [flo3] (HF_data) at (7.8,-1.2) {Generate HF dataset $\mathbf{D}_\text{HF}$};
\node [flo4] (HF_test) at (4.8,-2.25) {HF testing data};
\node [flo3] (HF_train) at (11,-0.85) {Train $\text{N\hspace{-1px}N}_\text{HF}$};
\node [flo3] (HF_vali) at (11,-1.55) {Validate $\text{N\hspace{-1px}N}_\text{HF}$};
\node [flo3,dashed] (HF_testing) at (11,-2.25) {Test $\text{N\hspace{-1px}N}_\text{HF}$};

\draw[thick,-latex,rounded corners,green!20!black] (param.east) [xshift=2mm] |-  (LFom_mod.west);
\draw[thick,-latex,rounded corners,green!20!black] (param.east) [xshift=2mm] |-  (HF_mod.west);
\draw[thick,-latex,rounded corners,green!20!black] (LFom_mod) to (LRom_mod);
\draw[thick,-latex,rounded corners,green!20!black] (LRom_mod) to  (LF_data);
\draw[thick,-latex,rounded corners,green!20!black,dashed] (LFom_mod.south) [xshift=2mm] |-  (LF_test.west);
\draw[thick,-latex,rounded corners,green!20!black,dashed] (LF_test) to (LF_testing);
\draw[thick,-latex,rounded corners,green!20!black] (HF_mod) to (HF_data);
\draw[thick,-latex,rounded corners,green!20!black] (LF_data.east) [xshift=2mm] |-  (LF_train.west);
\draw[thick,-latex,rounded corners,green!20!black] (LF_data.east) [xshift=2mm] |-  (LF_vali.west);
\draw[thick,-latex,rounded corners,green!20!black,dashed] (HF_mod.south) [xshift=2mm] |-  (HF_test.west);
\draw[thick,-latex,rounded corners,green!20!black,dashed] (HF_test) to  (HF_testing);
\draw[thick,-latex,rounded corners,green!20!black] (HF_data.east) [xshift=2mm] |-  (HF_train.west);
\draw[thick,-latex,rounded corners,green!20!black] (HF_data.east) [xshift=2mm] |-  (HF_vali.west);
\node [green!20!black] at (0.55,1.45) {$1^{\text{st}}$};
\node [green!20!black] at (0.55,-1.45) {$2^{\text{nd}}$};
\end{tikzpicture}
\caption{Flowchart of the MF-DNN surrogate modeling strategy.}\label{fig:meth_flow}
\end{figure}
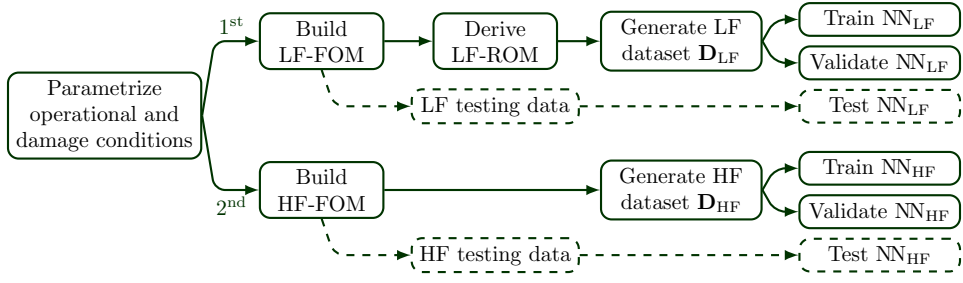

The key feature of $\text{N\hspace{-1px}N}_\text{MF}$ is that the effect of damage on the training data is reproduced with the HF model only, which is considered to be the most accurate description enabling to account for unexperienced damage scenarios. The $\text{N\hspace{-1px}N}_\text{MF}$ training is carried out offline, and is characterized by a limited number of evaluations of the HF finite element solver. At the same time, the computational time required to evaluate $\text{N\hspace{-1px}N}_\text{MF}$ for new input parameters is negligible. This latter aspect greatly speeds up the generation of a large number of training instances, compared to what would be required if relying solely on the HF finite element solver. Finally, it is worth noting that the MF-DNN surrogate modeling paradigm can be easily adapted to application domains other than SHM, even when the number of fidelity levels differs, and potentially extended to handle full-field or feature-based data.

The trained MF-DNN surrogate is eventually exploited to populate a large labeled dataset $\mathbf{D}_\text{train}$, according to:
\begin{linenomath*}
\begin{equation}
\mathbf{D}_\text{train}=\lbrace(\mathbf{x}^\text{HF}_k,\widehat{\mathbf{U}}^\text{HF}_k=\text{N\hspace{-1px}N}_\text{MF}(\mathbf{x}^\text{HF}_k, \mathbf{x}^\text{LF}_k))\rbrace_{k=1}^{I_\text{train}}~,
\label{eq:Dataset_training}
\end{equation}
\end{linenomath*}
where $I_\text{train}$ is the number of instances collected in $\mathbf{D}_\text{train}$. These instances are provided through $\text{N\hspace{-1px}N}_\text{MF}$ for varying input parameters $\mathbf{x}^\text{HF}$ (with $\mathbf{x}^\text{LF}$ being a subset of them) sampled via the latin hypercube rule. In order to mimic measurement noise, each vibration recording in $\mathbf{D}_\text{train}$ is then corrupted by adding independent and identically distributed Gaussian noise, whose statistical properties depend on the target accuracy of the sensors.

\section{Deep learning-enhanced Bayesian model updating}
\label{sec:methodology}

In this section, we describe the proposed methodology to enhance an MCMC algorithm for model updating purposes through learnable mappings. The key components are a learnable feature extractor, which extracts informative features from the sensed structural response, and a feature-oriented surrogate model, which maps the $\boldsymbol{\theta}$ parameters to be updated into the low-dimensional feature space. Both the feature extractor and the feature-oriented surrogate rely on DL models. These models are trained using the $\mathbf{D}_\text{train}$ dataset, populated through the MF-DNN surrogate model described above. Section~\ref{subsec:DL_models} discusses the architectures of the two models and the technical aspects related to their training and evaluation. Section~\ref{subsec:MCMC} explains how the feature extractor and the feature-oriented surrogate are employed to sample the posterior distribution of $\boldsymbol{\theta}$ conditioned on observational data.

\subsection{Feature extractor and feature-oriented surrogate: models specification and training}
\label{subsec:DL_models}

Before training, the synthetic data generated through the MF-DNN surrogate model and collected in $\mathbf{D}_\text{train}$ are preprocessed to be transformed into images as described below. We highlight that this is not a restrictive choice; indeed, the proposed methodology is general and can be easily adapted to deal with data of a different nature.

The recent developments in computer vision suggest the possibility of transforming time series into images for SHM purposes, see for instance~\cite{proc:Giglioni_EWSHM,art:MTF_Fink,art:MTF_rocking}. Imaging time series is reported to help highlighting local patterns that might otherwise be spread over or lying outside the time domain. In particular, the Markov transition field (MTF) technique~\cite{art:MTF} is here employed to preprocess the multivariate time histories collected in $\mathbf{D}_\text{train}$. The MTF technique is chosen over other conversion methods, such as Gramian angular fields~\cite{art:MTF}, recurrence plots~\cite{art:rec_plots} and grayscale encoding~\cite{art:gray_scale}, as it has been reported to offer better performance for SHM applications~\cite{art:MTF_Fink,art:MTF_rocking}. However, MTF is a signal processing algorithm not yet employed in practice as frequently as those based on spectral analysis, such as the spectrogram or scalogram representations. The MTF technique is reviewed in \ref{sec:MTF}.

Each instance $\widehat{\mathbf{U}}^\text{HF}_k$, with $k=1,\ldots,I_\text{train}$, is transformed into a grayscale mosaic $\I_k\in\mathbb{R}^{h_\I\times w_\I}$, with $h_\I$ and $w_\I$ respectively being the height and the width of the mosaic. Each mosaic is composed of the juxtaposition of $N_u$ MTF representations, or tesserae, obtained via MTF encoding of the $N_u$ time series collected in $\widehat{\mathbf{U}}^\text{HF}_k$. Accordingly, $\mathbf{D}_\text{train}$ is reassembled as: 
\begin{linenomath*}
\begin{equation}
\mathbf{D}^\I_\text{train}=\lbrace(\mathbf{x}^\text{HF}_k,\I_k)\rbrace_{k=1}^{I_\text{train}}~.
\label{eq:mosaic_data}
\end{equation}
\end{linenomath*}

The feature extractor and the feature-oriented surrogate model are learned through a sequential training process involving two learning steps (see \fig\ref{fig:training_flow}). The first learning step involves training the feature extractor to map structural response data onto their feature representation in a low-dimensional space. The second learning step involves training the feature-oriented surrogate to map the parameters to be inferred onto the low-dimensional feature space. Once trained, the two components are exploited within an MCMC algorithm to sample the posterior distribution of $\boldsymbol{\theta}$ conditioned on observational data, as detailed next.

\tikzstyle{timeseries} = [draw=red!40!black, thick, rectangle, fill=red!15, text width=1em, text centered, minimum height=2em]
\tikzstyle{encoder} = [draw=blue!40!black, thick, fill=blue!40, trapezium, trapezium angle=50, rotate=270, minimum width=8em, minimum height=1.4em, text centered]
\tikzstyle{decoder} = [draw=blue!40!black, thick, fill=blue!40, trapezium, trapezium angle=-50, rotate=270, minimum width=8em, minimum height=1.6em, text centered]
 \tikzstyle{big_block} = [rectangle, draw=red!40!black, thick, text width=3.2em, rounded corners, minimum height=5em, dashed]
  \tikzstyle{big_block_blue} = [rectangle, draw=blue!40!black, thick, text width=5em, rounded corners, minimum height=1.75em, dashed, , text centered]
 \tikzstyle{add_node} = [circle,draw=blue!40!black,thick,align=center,minimum size=0.5cm,fill=blue!40]
 \tikzstyle{input_node} = [draw=red!40!black, thick, rectangle, fill=red!15, text width=3em, text centered, rounded corners]

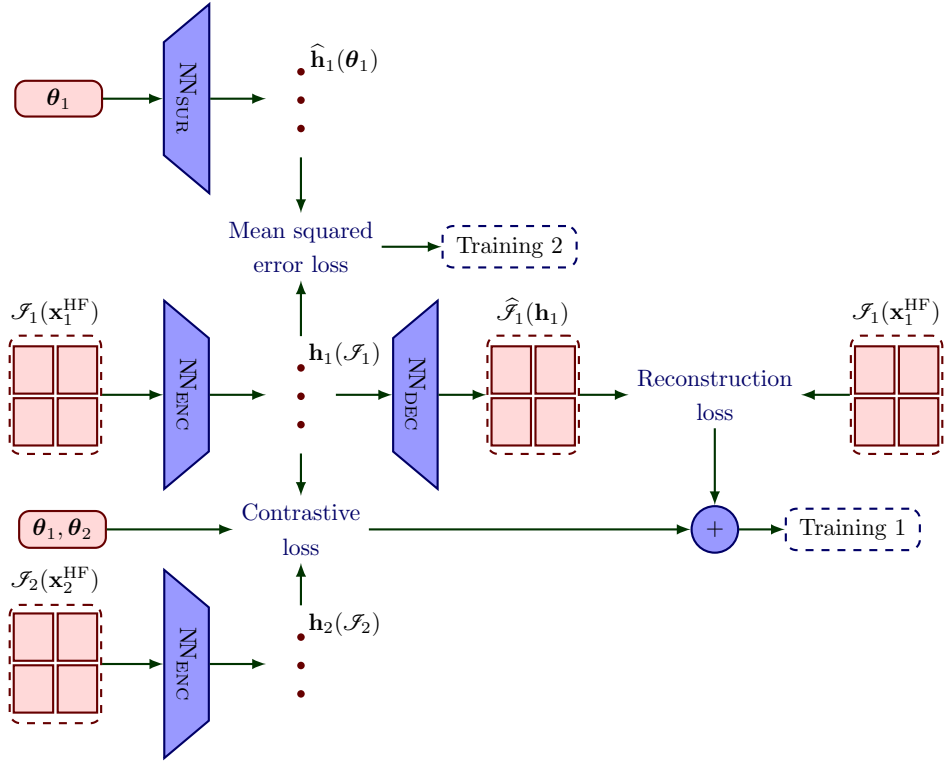
\begin{figure}[t]
\center
\begin{tikzpicture}[scale=0.89, every node/.style={scale=0.89}]

\node [timeseries] (t_1_1) at (0.025,1.05em) {};
\node [below=0.1em of t_1_1,timeseries] (t_1_1b) {};
\node [right=0.1em of t_1_1,timeseries] (t_1_2) {};
\node [below=0.1em of t_1_2,timeseries] (t_1_2b) {};
\node [big_block, centered] (frame_1) at (1em,0) {};
\node [encoder] (encoder_1) at (2.3,0){$\text{N\hspace{-1px}N}_\text{ENC}$};
\node [scale=3,text height=0.333cm, red!40!black] (feat_1) at (4,0) {$\vdots$};
\node [decoder] (decoder) at (5.7,0){$\text{N\hspace{-1px}N}_\text{DEC}$};
\node [timeseries] (tr_1_1) at (7.125,1.05em) {};
\node [below=0.1em of tr_1_1,timeseries] (tr_1_1b) {};
\node [right=0.1em of tr_1_1,timeseries] (tr_1_2) {};
\node [below=0.1em of tr_1_2,timeseries] (tr_1_2b) {};
\node [big_block, centered] (frame_1r) at (7.45,0) {};

\node [timeseries] (t0_1_1) at (12.525,1.05em) {};
\node [below=0.1em of t0_1_1,timeseries] (t0_1_1b) {};
\node [right=0.1em of t0_1_1,timeseries] (t0_1_2) {};
\node [below=0.1em of t0_1_2,timeseries] (t0_1_2b) {};
\node [big_block, centered] (frame_10) at (12.85,0) {};

\node [timeseries] (t_2_1) at (0.025,-3.6) {};
\node [below=0.1em of t_2_1,timeseries] (t_2_1b) {};
\node [right=0.1em of t_2_1,timeseries] (t_2_2) {};
\node [below=0.1em of t_2_2,timeseries] (t_2_2b) {};
\node [encoder] (encoder_2) at (2.3,-4){$\text{N\hspace{-1px}N}_\text{ENC}$};
\node [scale=3,text height=0.333cm, red!40!black] (feat_2) at (4,-4) {$\vdots$};
\node [big_block, centered] (frame_2) at (1em,-4) {};

\node [input_node] (inputs) at (1.3em,-2) {$\boldsymbol{\theta}_1,\boldsymbol{\theta}_2$};

\node [align=center,blue!40!black] (contrastive) at (4,-2) {Contrastive\\ loss};
\node [align=center,blue!40!black] (L2) at (10.15,0) {Reconstruction\\ loss};
\node [add_node] (add) at (10.15,-2) {$+$};

\node [input_node] (inputs_1) at (0.4,4.4) {$\boldsymbol{\theta}_1$};
\node [decoder] (surrogate) at (2.3,4.4){$\text{N\hspace{-1px}N}_\text{SUR}$};
\node [scale=3,text height=0.333cm, red!40!black] (feat_1r) at (4,4.4) {$\vdots$};
\node [align=center,blue!40!black] (L2_2) at (4,2.2) {Mean squared\\ error loss};

\node [big_block_blue] (t_1) at (12.2,-2) {Training 1};
\node [big_block_blue] (t_2) at (7.1,2.2) {Training 2};

\node [] at (1em,3.5em) {$\I_1(\mathbf{x}_1^\text{HF})$};
\node [] at (7.45,3.5em) {$\widehat{\I}_1(\mathbf{h}_1)$};
\node [] at (12.85,3.5em) {$\I_1(\mathbf{x}_1^\text{HF})$};
\node [] at (1em,-2.75) {$\I_2(\mathbf{x}_2^\text{HF})$};
\node [] at (4.65,0.6) {$\mathbf{h}_1(\I_1)$};
\node [] at (4.65,-3.4) {$\mathbf{h}_2(\I_2)$};
\node [] at (4.65,5) {$\widehat{\mathbf{h}}_1(\boldsymbol{\theta}_1)$};

\draw[-latex,thick,green!20!black] (frame_1) to (encoder_1);
\draw[-latex,thick,green!20!black] (encoder_1) to (feat_1);
\draw[-latex,thick,green!20!black] (feat_1) to (decoder);
\draw[-latex,thick,green!20!black] (decoder) to (frame_1r);
\draw[-latex,thick,green!20!black] (frame_1r) to (L2);
\draw[-latex,thick,green!20!black] (frame_10) to (L2);
\draw[-latex,thick,green!20!black] (frame_2) to (encoder_2);
\draw[-latex,thick,green!20!black] (encoder_2) to (feat_2);
\draw[-latex,thick,green!20!black] (feat_1) to (contrastive);
\draw[-latex,thick,green!20!black] (feat_2) to (contrastive);
\draw[-latex,thick,green!20!black] (inputs) to (contrastive);
\draw[-latex,thick,green!20!black] (L2) to (add);
\draw[-latex,thick,green!20!black] (contrastive) to (add);
\draw[-latex,thick,green!20!black] (add) to (t_1);
\draw[-latex,thick,green!20!black] (inputs_1) to (surrogate);
\draw[-latex,thick,green!20!black] (surrogate) to (feat_1r);
\draw[-latex,thick,green!20!black] (feat_1r) to (L2_2);
\draw[-latex,thick,green!20!black] (feat_1) to (L2_2);
\draw[-latex,thick,green!20!black] (L2_2) to (t_2);
\end{tikzpicture}
\caption{Learnable feature extractor and feature-oriented surrogate: flowchart of the sequential training process. Red nodes refer to the input/output quantities and blue nodes denote the corresponding computational blocks. $\text{N\hspace{-1px}N}_\text{ENC}$ is the feature extractor, $\text{N\hspace{-1px}N}_\text{DEC}$ is the decoder branch, and $\text{N\hspace{-1px}N}_\text{SUR}$ is the feature-oriented surrogate model. $\I(\mathbf{x}^\text{HF})$ denotes the input mosaic, $\widehat{\I}(\mathbf{h})$ denotes the reconstructed mosaic, and $\boldsymbol{\theta}$ is the vector of parameters for which we aim to update the belief. $\mathbf{h}(\I)$ is the low-dimensional feature representation of  $\I(\mathbf{x}^\text{HF})$ provided by $\text{N\hspace{-1px}N}_\text{ENC}$, and $\widehat{\mathbf{h}}(\boldsymbol{\theta})$ is the corresponding approximation provided by $\text{N\hspace{-1px}N}_\text{SUR}$.
\label{fig:training_flow}}
\end{figure}

The feature extractor is built upon an autoencoder equipped with a Siamese appendix~\cite{art:Siamese} of the encoder branch (refer to \mbox{``Training 1''} in \fig\ref{fig:training_flow}). This model enhances the dimensionality reduction capabilities achieved through the unsupervised training of an autoencoder by introducing a distance function in the latent space via pairwise contrastive learning~\cite{proc:LeCun_Contrastive}. Within the resulting latent space, features extracted from similar data points are pushed to be as close as possible, while those provided for dissimilar data points are kept away. The concept of similarity refers to a task-specific distance measure, with respect to the $\boldsymbol{\theta}$ parameters describing the variability of the monitored system.

The learnable components of the feature extractor are the encoder $\text{N\hspace{-1px}N}_\text{ENC}$ and decoder $\text{N\hspace{-1px}N}_\text{DEC}$ branches of an autoencoder. $\text{N\hspace{-1px}N}_\text{ENC}$ provides the feature representation $\mathbf{h}\in\mathbb{R}^{D_h}$ of the input mosaic $\I$ in a low-dimensional space of size $D_h$, while $\text{N\hspace{-1px}N}_\text{DEC}$ takes $\mathbf{h}$ and provides the reconstructed mosaic $\widehat{\I}$, as follows:
\begin{linenomath*}
\begin{align}
&\mathbf{h}(\I)=\text{N\hspace{-1px}N}_\text{ENC}(\I(\mathbf{x}^\text{HF}))~,\\
&\widehat{\I}(\mathbf{h})=\text{N\hspace{-1px}N}_\text{DEC}(\mathbf{h}(\I))~.
\label{eq:encoder}
\end{align}
\end{linenomath*}
The key component that links $\text{N\hspace{-1px}N}_\text{ENC}$ and $\text{N\hspace{-1px}N}_\text{DEC}$ is the bottleneck layer characterized by the low-dimensional feature size $D_h$. $D_h$ is much smaller than the dimension of the input and output layers of the autoencoder, thus forcing the data through a compressed representation while attempting to recreate the input as closely as possible in the output. The unsupervised training of an autoencoder is a well-known procedure in the literature, see for instance~\cite{book:DL_book}. On the other hand, the Siamese appendix of the encoder branch affects the training process through a contrastive loss function linking two $\text{N\hspace{-1px}N}_\text{ENC}$ twins. Data points are thus processed in pairs, yielding two outputs $\mathbf{h}_1=\text{N\hspace{-1px}N}_\text{ENC}(\I_1(\mathbf{x}^\text{HF}_1))$ and $\mathbf{h}_2=\text{N\hspace{-1px}N}_\text{ENC}(\I_2(\mathbf{x}^\text{HF}_2))$. The data pairing process is carried out as follows. First, a threshold distance $\overline{\mathcal{E}_{\theta}}$ is fixed to characterize the similarity for the parametric space of $\boldsymbol{\theta}$. The mosaics dataset $\mathbf{D}^\I_\text{train}$ is then augmented by assembling $\zeta_{+}$ positive pairs for each instance, characterized by $\lVert \boldsymbol{\theta}_1 -  \boldsymbol{\theta}_2 \rVert_{2}\leq\overline{\mathcal{E}_{\theta}}$, and $\zeta_{-}$ negative pairs, characterized by $\lVert \boldsymbol{\theta}_1 - \boldsymbol{\theta}_2 \rVert_{2}>\overline{\mathcal{E}_{\theta}}$, according to:
\begin{linenomath*}
\begin{equation}
\mathbf{D}^\I_\text{P}=\lbrace(\mathbf{x}^\text{HF}_1,\I_1,\mathbf{x}^\text{HF}_2,\I_2)_\iota\rbrace_{\iota=1}^{I_\text{train}^\text{P}}~,
\end{equation}
\end{linenomath*}
with $I_\text{train}^\text{P}=I_\text{train}(\zeta_{+}+\zeta_{-})$ being the total number of pairs.

The set of weights and biases parametrizing the autoencoder is denoted as $\boldsymbol{\Omega}_\text{AE}$. During \mbox{``Training 1''}, this is optimized by minimizing the following loss function over $\mathbf{D}^\I_\text{P}$:
\begin{linenomath*}
\begin{equation}
\begin{split}
\mathcal{L}_\text{AE}(\boldsymbol{\Omega}_\text{AE},\mathbf{D}^\I_\text{P}) = &\displaystyle\frac{1}{I_\text{train}^{\text{P}}}\sum^{I_\text{train}^{\text{P}}}_{\iota=1}\biggl\{\lVert \I_1(\mathbf{x}_1^\text{HF}) - \text{N\hspace{-1px}N}_\text{DEC}(\text{N\hspace{-1px}N}_\text{ENC}(\I_1(\mathbf{x}_1^\text{HF})))\rVert_{2}^2+\smallskip
\\
& \qquad\qquad\quad \Bigl[\frac{1-\gamma}{2} (\mathcal{E}_h)^2 +\frac{\gamma}{2}\left[\max{(0,\psi-\mathcal{E}_h)}\right]^2\Bigr]\biggr\}_\iota +\lambda_\text{AE}\lVert \boldsymbol{\Omega_\text{AE}} \rVert_{2}^2~,
\label{eq:AE_loss}
\end{split}
\end{equation}
\end{linenomath*}
where: the first term is the reconstruction loss function, typically employed to train autoencoders; the second term is the pairwise contrastive loss function, useful to induce a geometrical structure in the feature space; and the last term is an $L^2$ regularization of rate $\lambda_\text{AE}$ applied over the model parameters $\boldsymbol{\Omega}_\text{AE}$. In \eq\eqref{eq:AE_loss}: $\gamma=\{0,1\}$, if $\boldsymbol{\theta}_1$ and $\boldsymbol{\theta}_2$ identify either a positive or a negative pair, respectively; $\psi > 0$ is a margin beyond which the loss ignores negative pairs; $\mathcal{E}_h = \lVert\mathbf{h}_1-\mathbf{h}_2\rVert_{2}$ is the Euclidean distance between any pair of mappings $\mathbf{h}_1=\text{N\hspace{-1px}N}_\text{ENC}(\I_1(\mathbf{x}^\text{HF}_1))$ and $\mathbf{h}_2=\text{N\hspace{-1px}N}_\text{ENC}(\I_2(\mathbf{x}^\text{HF}_2))$. Minimizing the contrastive loss function is equivalent to learning a distance function $\mathcal{E}_h$ that approximates the Euclidean distance $\lVert \boldsymbol{\theta}_1 -  \boldsymbol{\theta}_2 \rVert_{2}$ between the target labels $\boldsymbol{\theta}_1$ and $\boldsymbol{\theta}_2$ of the processed pair of data points. The label information is thus exploited to guide the dimensionality reduction, so that the sensitivity to damage and (possibly) operational conditions described by $\boldsymbol{\theta}$ is encoded in the low-dimensional feature space.

After the first learning step, $\text{N\hspace{-1px}N}_\text{DEC}$, the Siamese appendix, and $\mathbf{D}^\I_\text{train}$ are discarded, and only $\text{N\hspace{-1px}N}_\text{ENC}$ and $\mathbf{D}^\I_\text{train}$ are retained to train the feature-oriented surrogate $\text{N\hspace{-1px}N}_\text{SUR}$ (refer to \mbox{``Training 2''} in \fig\ref{fig:training_flow}). $\text{N\hspace{-1px}N}_\text{SUR}$ is set as a fully-connected DL model, which approximates the functional link between the parametric space of $\boldsymbol{\theta}$ and the low-dimensional feature space described by $\text{N\hspace{-1px}N}_\text{ENC}$, as follows:
\begin{linenomath*}
\begin{equation}
\widehat{\mathbf{h}}=\text{N\hspace{-1px}N}_\text{SUR}(\boldsymbol{\theta})~,
\label{eq:surrogate}
\end{equation}
\end{linenomath*}
where $\widehat{\mathbf{h}}$ denotes the $\text{N\hspace{-1px}N}_\text{SUR}$ approximation to the low-dimensional features provided through $\text{N\hspace{-1px}N}_\text{ENC}$.

The dataset dedicated to the training of $\text{N\hspace{-1px}N}_\text{SUR}$ is derived from the mosaics dataset $\mathbf{D}^\I_\text{train}$ in \eq\eqref{eq:mosaic_data} by mapping the mosaics in $\mathbf{D}^\I_\text{train}$ onto the feature space, once and for all, to provide:
\begin{linenomath*}
\begin{equation}
\mathbf{D}^h_\text{train}=\lbrace(\boldsymbol{\theta}_k,\mathbf{h}_k)\rbrace_{k=1}^{I_\text{train}}~,
\end{equation}
\end{linenomath*}
collecting the feature representations $\mathbf{h}$ of the training data and their corresponding labels, in terms of the sought parameters $\boldsymbol{\theta}$. The set of weights and biases $\boldsymbol{\Omega}_\text{SUR}$ parametrizing $\text{N\hspace{-1px}N}_\text{SUR}$ is then learned through the minimization of the following loss function:
\begin{linenomath*}
\begin{equation}
\mathcal{L}_\text{SUR}(\boldsymbol{\Omega}_\text{SUR},\mathbf{D}^h_\text{train}) =\frac{1}{I_\text{train}}\sum^{I_\text{train}}_{k=1}\lVert \mathbf{h}_k(\I_k) - \text{N\hspace{-1px}N}_\text{SUR}(\boldsymbol{\theta}_k)\rVert_{2}^2+\lambda_\text{SUR}\lVert \boldsymbol{\Omega_\text{SUR}} \rVert_{2}^2~.
\label{eq:SUR_loss}
\end{equation}
\end{linenomath*}
Equation~\eqref{eq:SUR_loss} provides a measure of the distance between the target low-dimensional feature vector $\mathbf{h}(\I)$, obtained through the feature extractor $\text{N\hspace{-1px}N}_\text{ENC}$, and its approximated counterpart $\widehat{\mathbf{h}}=\text{N\hspace{-1px}N}_\text{SUR}(\boldsymbol{\theta})$, obtained via the feature-oriented surrogate model. 

The implementation details of the DL models are reported in \ref{sec:implementation}. It is worth noting that our modeling choices are tailored to the specific characteristics of the observational data considered in this paper. However, the overall framework is fairly general and can accommodate alternative modeling choices adapted to the specific data and features of the problem at hand. In our case, vibration recordings are encoded into images via MTF preprocessing to highlight structures and patterns in the data. While we thus show how to extract informative features in a low-dimensional metric space in the case of images, data of a different nature can be addressed in a similar way through an appropriate choice of the DL architectures. For instance, one-dimensional convolutional layers could be exploited in place of two-dimensional ones to deal with time series data. Moreover, there may be cases where the decoding branch $\text{N\hspace{-1px}N}_\text{DEC}$ should be avoided. Retaining $\text{N\hspace{-1px}N}_\text{DEC}$ is motivated by the fact that the reconstruction term in \eq\eqref{eq:AE_loss} acts as a regularizer during training. As a by-product, the trained $\text{N\hspace{-1px}N}_\text{DEC}$ and $\text{N\hspace{-1px}N}_\text{SUR}$ models can also serve as a surrogate of the form $\text{N\hspace{-1px}N}_\text{DEC}(\text{N\hspace{-1px}N}_\text{SUR}(\boldsymbol{\theta}))$, following an approach similar to~\cite{art:Fresca2020}, to approximate observational data for any given parameter vector $\boldsymbol{\theta}$. In our case, the contrastive term in \eq\eqref{eq:AE_loss} is minimized by exploiting label information that completely characterize the parametrization of the physics-based model. However, when the number $N_\text{par}^\text{HF}$ of parameters in $\mathbf{x}^\text{HF}$ becomes large, the paring process underlying the contrastive term can become computationally demanding. Although this issue does not arise in the present work, it could be addressed by including in $\boldsymbol{\theta}$ only a subset of $\mathbf{x}^\text{HF}$, limited to the parameters that actually need to be inferred. In this eventuality, the decoding branch $\text{N\hspace{-1px}N}_\text{DEC}$ should be omitted to avoid the latent space exhibiting dependence on parameters not included in $\boldsymbol{\theta}$, which would prevent a one-to-one correspondence between $\boldsymbol{\theta}$ and $\mathbf{h}$ that may not be captured by $\text{N\hspace{-1px}N}_\text{SUR}$. The same consideration applies to systems subject to unknown stochastic inputs, such as seismic or wind loads acting on civil structures.

\subsection{Feature-based MCMC sampling algorithm}
\label{subsec:MCMC}

The feature extractor  $\text{N\hspace{-1px}N}_\text{ENC}$ and the feature-oriented surrogate model $\text{N\hspace{-1px}N}_\text{SUR}$, trained as described in the previous section, are exploited in the online monitoring phase to enhance an MCMC sampler for model updating purposes. The MCMC algorithm is here employed to update the prior probability density function (pdf) $p(\boldsymbol{\theta})$ of the parameter vector $\boldsymbol{\theta}$, to compute the posterior pdf $p(\boldsymbol{\theta}|\mathbf{U}^{\text{EXP}}_{1,\dots,N_\text{obs}})$ conditioned on a batch of gathered sensor recordings $\mathbf{U}^{\text{EXP}}_{1,\dots,N_\text{obs}}$. Here, $N_\text{obs}$ represents the batch size of processed observations, each consisting of a series of $N_u$ signals with $L$ measurements over time.

By exploiting the Metropolis-Hastings sampler~\cite{art:MH}, the updating procedure is carried out by iteratively generating a chain of samples $\lbrace\boldsymbol{\theta}_{1},\dots,\boldsymbol{\theta}_{L_\text{chain}}\rbrace$ from a proposal distribution, and then deciding whether to accept or reject each sample, based on how likely a $\boldsymbol{\theta}$ sample is to explain $\mathbf{U}^{\text{EXP}}_{1,\dots,N_\text{obs}}$. To this aim, $\text{N\hspace{-1px}N}_\text{ENC}$ and $\text{N\hspace{-1px}N}_\text{SUR}$ are synergistically exploited to provide informative features $\mathbf{h}^{\text{EXP}}_{1,\dots,N_\text{obs}}$ via assimilation of the observational data, and to surrogate the functional link between the parameters to be inferred and the feature space, respectively, as sketched in \fig\ref{fig:MCMC_flow}. The resulting parameter estimation framework benefits from significantly reduced computational cost due to the low dimensionality of the features involved, an improved convergence rate stemming from the geometrical structure of the feature space, and more accurate estimates thanks to the informativeness of the extracted features.

\tikzstyle{b_cloud} = [draw=red!40!black, thick, ellipse, fill=red!15, text badly centered, minimum height=2em, text width=3.5em]
\tikzstyle{b_block} = [rectangle, draw=blue!40!black, thick, fill=blue!40, text width=4em, text centered, rounded corners, minimum height=2.5em]
\tikzstyle{b_cloud2} = [draw=red!40!black, thick, ellipse, fill=red!15, text badly centered, minimum height=2em, text width=4em]
\tikzstyle{b_block2} = [rectangle, draw=blue!40!black, thick, fill=blue!40, text width=5em, text centered, rounded corners, minimum height=2.5em]
\tikzstyle{b_cloud3} = [draw=red!40!black, thick, ellipse, fill=red!15, text badly centered, minimum height=2em, text width=4em]

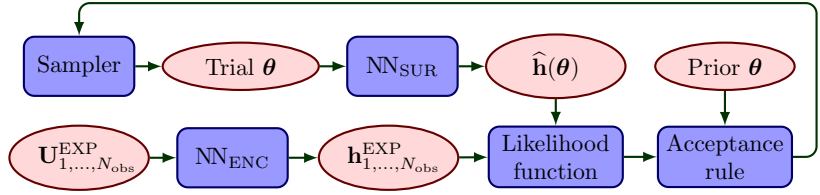
\begin{figure}[t]
\center
\begin{tikzpicture}[node distance = 2cm, scale=0.89, every node/.style={scale=0.89}]
\linespread{1}
\node [b_cloud] (prior) {Prior $\boldsymbol{\theta}$};
\node [b_cloud, left=1.05em of prior] (feat) {${\widehat{\mathbf{h}}}(\boldsymbol{\theta})$};
\node [b_block, left=1em of feat] (surrogate) {$\text{N\hspace{-1px}N}_\text{SUR}$};
\node [b_cloud3, left=1em of surrogate] (parameters) {Trial $\boldsymbol{\theta}$};
\node [b_block, left=1em of parameters] (sampler) {Sampler};
\node [b_cloud, below=1.05em of sampler] (obs) {$\mathbf{U}^{\text{EXP}}_{1,\ldots,N_\text{obs}}$};
\node [b_block, right=1em of obs] (feat_extractor) {$\text{N\hspace{-1px}N}_\text{ENC}$};
\node [b_cloud, right=1.1em of feat_extractor] (exp_feat) {$\mathbf{h}^{\text{EXP}}_{1,\ldots,N_\text{obs}}$};
\node [b_block2, below=1em of feat] (like) {Likelihood function};
\node [b_block2, right=1.25em of like] (rule) {Acceptance rule};
\draw[-latex,thick,green!20!black] (sampler) to (parameters);
\draw[-latex,thick,green!20!black] (parameters) to (surrogate);
\draw[-latex,thick,green!20!black] (surrogate) to (feat);
\draw[-latex,thick,green!20!black] (feat) to (like);
\draw[-latex,thick,green!20!black] (obs) to (feat_extractor);
\draw[-latex,thick,green!20!black] (feat_extractor) to (exp_feat);
\draw[-latex,thick,green!20!black] (exp_feat) to (like);
\draw[-latex,thick,green!20!black] (like) to (rule);
\draw[-latex,thick,green!20!black] (prior) to (rule);
\draw[thick,-latex,rounded corners,green!20!black] (rule.east) -| ++ (1em,6.5em) -| (sampler.north);
\end{tikzpicture}
\caption{Scheme of the MCMC procedure to update the posterior probability distribution of the structural state. Red nodes refer to the input/output quantities and blue nodes denote the corresponding computational blocks. \label{fig:MCMC_flow}}\end{figure}

According to the Bayes’ rule, the posterior pdf $p(\boldsymbol{\theta}|\mathbf{U}^{\text{EXP}}_{1,\dots,N_\text{obs}})$ is given as:
\begin{linenomath*}
\begin{equation}
p(\boldsymbol{\theta}|\mathbf{U}^{\text{EXP}}_{1,\dots,N_\text{obs}}) = \frac{p(\mathbf{U}^{\text{EXP}}_{1,\dots,N_\text{obs}}|\boldsymbol{\theta})p(\boldsymbol{\theta})}{\int p(\mathbf{U}^{\text{EXP}}_{1,\dots,N_\text{obs}}|\boldsymbol{\theta})p(\boldsymbol{\theta})\,d\boldsymbol{\theta}}~,
\label{eq:Bayes_rule}
\end{equation}
\end{linenomath*}
where: $p(\mathbf{U}^{\text{EXP}}_{1,\dots,N_\text{obs}}|\boldsymbol{\theta})$ is the likelihood function that provides the mechanism informing the posterior about the observations; the denominator is a normalizing factor, that is typically analytically intractable. To address this challenge, $p(\boldsymbol{\theta}|\mathbf{U}^{\text{EXP}}_{1,\dots,N_\text{obs}})$ is approximated through an MCMC sampling algorithm. By assuming additive Gaussian noise to represent the uncertainty due to modeling inaccuracies and measurement noise, the likelihood function is assumed to be Gaussian too and to read:
\begin{linenomath*}
\begin{equation}
p(\mathbf{U}^{\text{EXP}}_{1,\dots,N_\text{obs}}|\boldsymbol{\theta})
=\prod_{n=1}^{N_\text{obs}} c^{-1} \textup{exp}\bigg(-\frac{(\mathbf{h}^{\text{EXP}}_{n} - \widehat{\mathbf{h}}(\boldsymbol{\theta}))^\top(\mathbf{h}^{\text{EXP}}_{n} - \widehat{\mathbf{h}}(\boldsymbol{\theta}))}{2\sigma^2}\bigg)~.
\label{eq:Likelihood_1} 
\end{equation}
\end{linenomath*}
In \eq\eqref{eq:Likelihood_1}, the term $c=\sqrt{2\pi\sigma^2}$ is a normalization constant, with $\sigma\in\mathbb{R}$ being the root mean square of the prediction error at each MCMC iteration, which serves as the standard deviation of the uncertainty under the zero-mean assumption. Since $\sigma$ depends on $\boldsymbol{\theta}$, it must be recomputed at each MCMC iteration; however, this has no impact on the overall computational cost of the methodology, thanks to the low dimensionality of the feature vectors.

The proposal pdf is taken as Gaussian. The covariance matrix is initialized as diagonal, with sufficiently small entries to allow the sampler to start moving, and is subsequently adapted during sampling using the adaptive Metropolis algorithm~\cite{art:MHadaptive1}. It is worth noting that the proposed procedure is general and allows for different choices of sampling algorithms. For example, it can be similarly employed with more advanced samplers, such as transitional MCMC or Hybrid Monte Carlo algorithms, along with their recent extensions~\cite{art:TMCMC_Straub,art:No_U_Turn}. Moreover, the entire methodology can be adapted to solve inverse problems in application domains beyond SHM, even when working with data types other than vibration recordings.

In order to check the quality of the estimates and stop the MCMC simulation, the estimated potential scale reduction (EPSR) metric~\cite{art:Gelman-Rubin} is employed to monitor the convergence to a steady distribution. Since it is not possible to monitor the convergence of an MCMC simulation from a single chain, the EPSR test exploits multiple chains from parallel runs. The EPSR metric $\widehat{\EE}$ measures the ratio between the estimate of the between-chain variance of samples to the average within-chain variance of samples. The convergence criterion is considered satisfied, only when all the chains converge to the same stationary distribution. In this work, each MCMC simulation is carried out by randomly initializing five Markov chains that are simultaneously evolved to meet the EPSR tolerance set to $\widehat{\EE}\leq1.01$, with $1.1$ being a recommended tolerance value~\cite{art:Gelman-Rubin}. The first half of each chain is then discarded to eliminate the initialization effect, and 3 out of 4 samples are further discarded to reduce the within-chain autocorrelation of samples.

\section{Numerical results}
\label{sec:Results}

This section aims to demonstrate the performance of the proposed strategy through simulated monitoring of three structural systems with increasing structural complexity: an L-shaped cantilever beam, a portal frame, and a railway bridge.

The FOMs and ROMs have been solved in the \texttt{Matlab} environment, using the \texttt{redbKIT} library~\cite{Redbkit}. All computations have been carried out on a PC featuring an \texttt{AMD Ryzen\textsuperscript{TM} 9 5950X} CPU @ 3.4 GHz and 128 GB RAM. The DL models have been implemented through the \texttt{Tensorflow}-based \texttt{Keras} API~\cite{chollet2015keras}, and trained on a single \texttt{Nvidia GeForce RTX\textsuperscript{TM} 3080} GPU card.

\subsection{L-shaped cantilever beam}
\label{sec:L-shaped}

\begin{figure}[t]
\begin{centering}
\includegraphics[width=.7\textwidth]{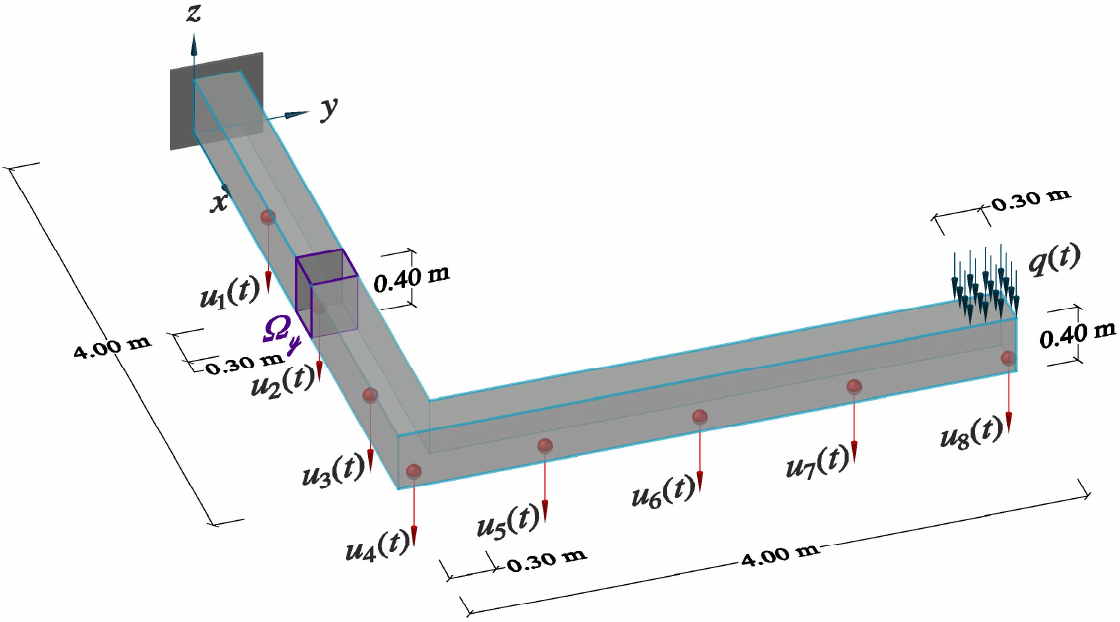}
\caption{{L-shaped cantilever beam: details of the synthetic recordings of displacements $u_1(t),\ldots,u_8(t)$, the loading condition, and the damageable region $\Omega_\y$.}\label{fig:maniglia_model}}
\end{centering}
\end{figure}

The first test case involves the L-shaped cantilever beam depicted in \fig\ref{fig:maniglia_model}. The structure is made of two arms, each with a length of $4~\textup{m}$, a width of $0.3~\textup{m}$ and a height of $0.4~\textup{m}$. The assumed mechanical properties are those of concrete: Young's modulus $E = 30~\textup{GPa}$, Poisson's ratio $\nu = 0.2$, and density $\rho = 2500~\textup{kg/m}^3$. The structure is excited by a distributed vertical load $q(t)$, acting on an area of $(0.3\times0.3)~\textup{m}^2$ close to its tip. The load varies over time according to $q(t) = Q \sin{(2\pi f t)}$, where $Q\in[1, 5]~\textup{kPa}$ and $f\in[10,60]~\textup{Hz}$ represent the load amplitude and frequency, respectively. Following the setup described in \sez\ref{sec:data}, $Q$ and $f$ have a uniform distribution within their ranges.

Synthetic displacement time histories are gathered in relation to $N_u=8$ dofs along the bottom surface of the structure, to mimic a monitoring system arranged as depicted in \fig\ref{fig:maniglia_model}. Each recording is provided for a time interval $(0,T)$, with $T=1~\textup{s}$, and an acquisition frequency $f_\text{s}=200~\textup{Hz}$. Recordings are corrupted with additive Gaussian noise yielding a signal-to-noise ratio of $100$.

The HF numerical model is obtained with a finite element discretization using linear tetrahedral elements and resulting in $N_\text{FE}=4659$ dofs. The structural dissipation is modeled by means of a Rayleigh damping matrix, assembled to account for a $5\%$ damping ratio on the first four structural modes. Damage is simulated by reducing the material stiffness within a subdomain $\Omega_{\y}$ of size $0.3\times0.3\times0.4~\textup{m}^3$. The position of $\Omega_{\y}$ is parametrized by the coordinates of its center of mass $\boldsymbol{\y}= ( x_\Omega, y_\Omega )^\top$, with either $x_\Omega$ or $y_\Omega$ varying in the range $[0.15, 3.85]~\textup{m}$. The magnitude of the stiffness reduction is set to $\delta=25\%$ and held constant within the considered time interval. Accordingly, the vector of HF input parameters is $\mathbf{x}^\text{HF} = (Q,f,x_\Omega,y_\Omega)^\top$.

The basis matrix $\mathbf{W}$ ruling the LF-ROM is obtained from a snapshot matrix $\mathbf{S}$, assembled through $200$ evaluations of the LF-FOM, at varying values of the LF parameters $\mathbf{x}^\text{LF} = (Q,f)^\top$ sampled via the latin hypercube rule. By prescribing a tolerance $\epsilon=10^{-3}$ on the fraction of energy content to be disregarded in the approximation, the order of the LF-ROM approximation turns out to be $N_\text{RB}=14$.

The dataset $\mathbf{D}_\text{LF}$ consists of $I_\text{LF}=10,000$ LF data instances collected using the LF-ROM, while the dataset $\mathbf{D}_\text{HF}$ contains only $I_\text{HF}=1000$ additional HF data instances. These two datasets are used to train the MF-DNN surrogate model: $\mathbf{D}_\text{LF}$ is used to train $\text{N\hspace{-1px}N}_\text{LF}$, and $\mathbf{D}_\text{HF}$ to train $\text{N\hspace{-1px}N}_\text{HF}$. The trained MF-DNN surrogate is then employed to populate the dataset $\mathbf{D}_\text{train}$ with $I_\text{train}=20,000$ instances generated by varying the HF input parameters $\mathbf{x}^\text{HF}$.

\begin{figure}[t]
\begin{centering}
\includegraphics[width=.6\textwidth]{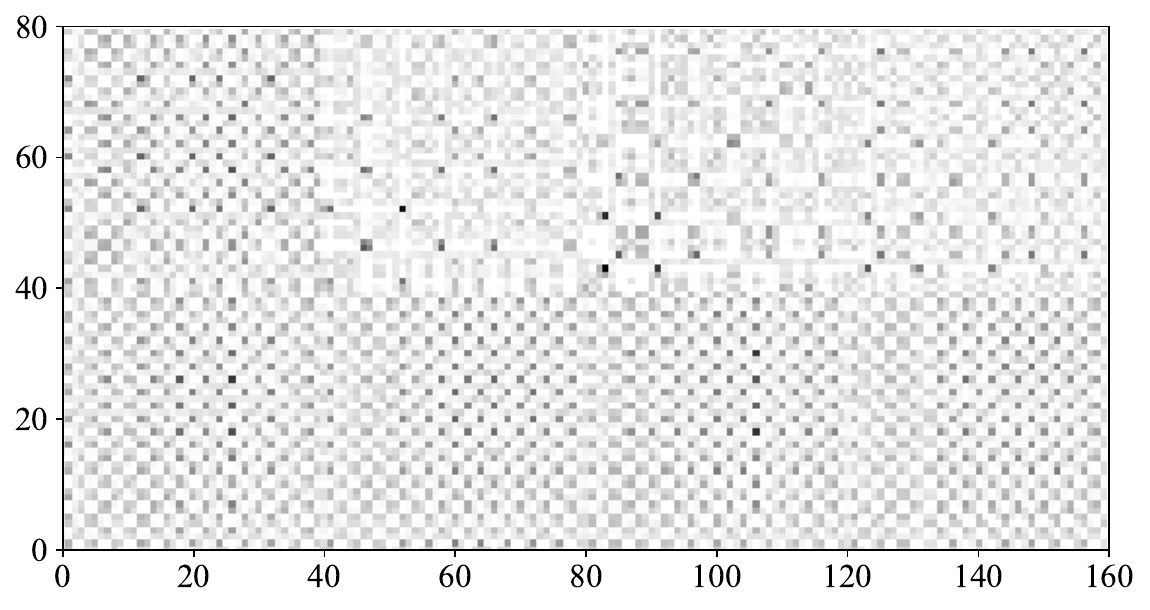}
\caption{{L-shaped cantilever beam - Exemplary MTF mosaic.}\label{fig:amtf}}
\end{centering}
\end{figure}

To train the feature extractor and the feature-oriented surrogate, the vibration recordings in $\mathbf{D}_\text{train}$ are transformed into images via MTF encoding. Each mosaic $\I_k$ in $\mathbf{D}^\I_\text{train}$, with $k=1,\ldots,I_\text{train}$, is obtained by placing the $N_u=8$ MTF tesserae into a $2\times4$ grid, with each MTF tessera being a $40\times40$ pixel image (see \fig\ref{fig:amtf}). The size of the MTF tesserae depends on the length of the time series and on the width of the blurring kernel. For the detailed steps of the mosaics generation via MTF encoding, see \ref{sec:MTF}. In this case, the length $L$ of the vibration recordings in $\mathbf{D}_\text{train}$ is reduced by removing the initial $20\%$ of each time history to eliminate potential inaccuracies induced by the $\text{N\hspace{-1px}N}_\text{HF}$ LSTM model. The chosen width of the blurring kernel is equal to $4$. Moreover, each vibration recording in $\mathbf{D}_\text{train}$ is normalized to follow a standard Gaussian distribution, thus allowing the updating procedure to neglect dependence on the load amplitude $Q$ thanks to the linear-elastic modeling behind $\mathbf{D}_\text{HF}$. The mosaics dataset $\mathbf{D}^\I_\text{train}$ is then exploited to minimize the loss functions in \eq\eqref{eq:AE_loss} and in \eq\eqref{eq:SUR_loss}, as described in \sez\ref{subsec:DL_models}.

\begin{figure}[t]
\captionsetup[subfigure]{justification=centering}\subfloat{\includegraphics[width=.33\textwidth]{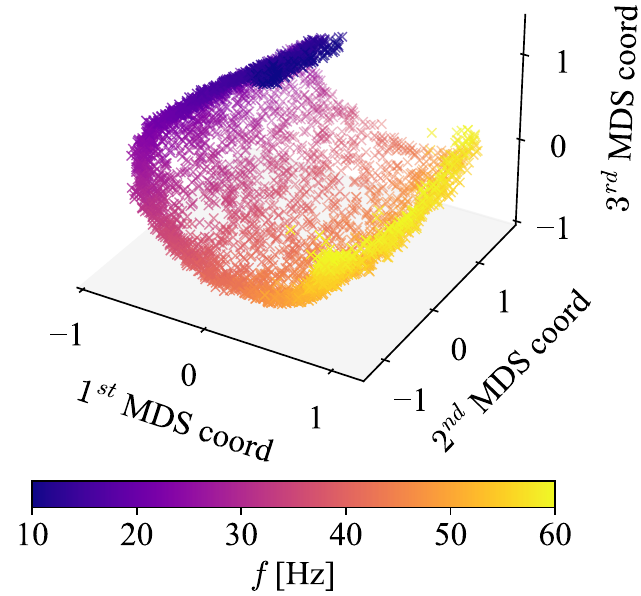}}\subfloat{\includegraphics[width=.33\textwidth]{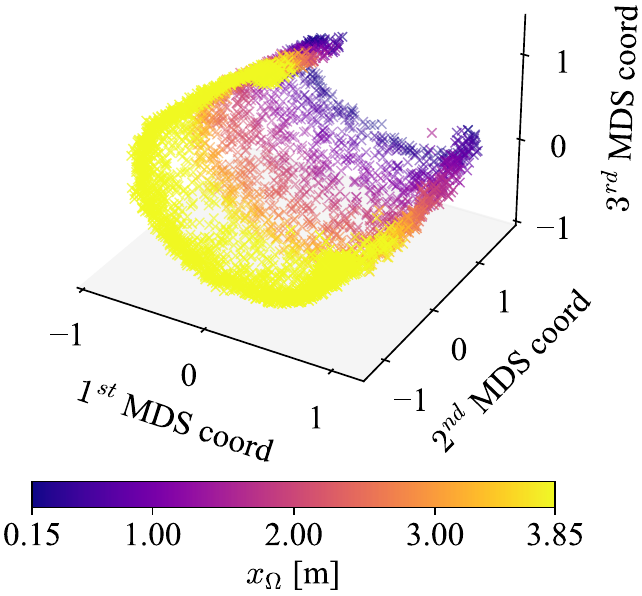}}
\subfloat{\includegraphics[width=.33\textwidth]{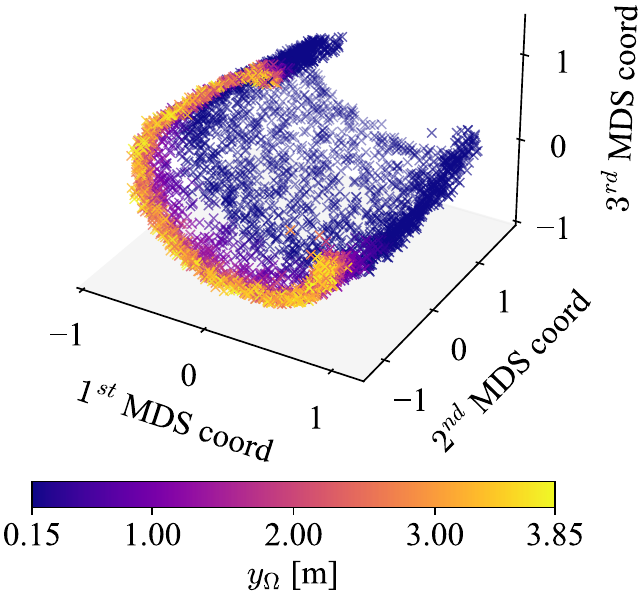}}
\caption{L-shaped cantilever beam - 3D multidimensional scaling representations of the low-dimensional features obtained for the validation data, against the target values of (left) load frequency, (center) damage position along the $x$-direction, and (right) damage position along the $y$-direction.\label{fig:Beam_MDS}}
\end{figure}

A compact representation of the low-dimensional features extracted through $\text{N\hspace{-1px}N}_\text{ENC}$ for the validation set of $\mathbf{D}^\I_\text{train}$ is reported in \fig\ref{fig:Beam_MDS}. The scatter plots report a downsized version of the extracted features, obtained by means of the metric multidimensional scaling (MDS) implemented in \texttt{scikit-learn}~\cite{scikit-learn}. The three-dimensional (3D) MDS representations are reported with a color channel referring to the target values of the load frequency and of the damage position along the $x$ and $y$ directions. Note how the resulting manifold suitably encodes the sensitivity of the processed measurements on the parameters employed to describe the variability of the system. This visual check provides a first qualitative indication about the positive impact of adopting the feature space described by $\text{N\hspace{-1px}N}_\text{ENC}$ for addressing the Bayesian model updating task.

In the absence of experimental data, the MCMC simulations are carried out considering batches of $N_\text{obs}=8$ HF noisy observations. Each observation batch is associated with the same $\theta_\Omega$, where $\theta_\Omega\in[0.15,7.55]~\textup{m}$ is an abscissa running along the axis of the structure and encoding the position of $\Omega_\y$ in place of $x_\Omega$ and $y_\Omega$. Each data instance in the observation batch is generated by sampling parameters $Q$ and $f$ from a Gaussian pdf centered at the ground truth value of the parameters, and featuring a standard deviation equal to $0.25\%$ of their respective ranges.

In the following, results are reported for six MCMC analyses, carried out under different operational conditions while moving the damage position from the clamp to the free-end. Table~\ref{tab:beam_outocomes} reports the outcome of the identification of the damage position, in terms of target value, posterior mean, posterior mode, standard deviation, and chain length. The quality of the estimates is highlighted by the small discrepancy between the target and the posterior mean values, which is only a few centimeters (less than $3\%$ of the admissible support length). Also note the relatively low values of standard deviation, which however increases as the damage position gets far from the clamped side of the structure. This is quite an expected outcome, and it is due to a smaller sensitivity of sensor recordings to damage when the damage is located near the free-end of the beam. The only case characterized by a large discrepancy between the target and the posterior mean values, as well as by a larger uncertainty, is in fact the last one, which features a damage position close to the free-end of the beam. For instance, in Case 4, the discrepancy between the target and the posterior mean values is only $0.044~\textup{m}$ over an admissible support of $7.4~\textup{m}$, while it reaches $0.845~\textup{m}$ in Case 6. Despite the larger discrepancy between the target and the posterior mean, the target value falls within the $95\%$ confidence interval, as in the other cases, thereby demonstrating the reliability of the estimate.

 \begin{table}[t]
\caption{L-shaped cantilever beam - Damage localization results for different operational and damage conditions, in terms of target value, posterior mean, posterior mode, standard deviation, and chain length.}\label{tab:beam_outocomes}
  \centering
   \scriptsize
\begin{tabular}{p{0.5cm} p{1.7cm} p{1.7cm} p{1.7cm} p{1.7cm} p{0.6cm}}
    \toprule
    \mbox{Case} &  \mbox{Target$(\theta_\Omega)$} & $\text{Mean}(\theta_\Omega)$ & \mbox{$\text{Mode}(\theta_\Omega)$} & \mbox{$\text{Stdv}(\theta_\Omega)$} & $L_\text{chain}$\\
    \toprule
     \mbox{1} &\mbox{$0.564~\textup{m}$} &\mbox{$0.580~\textup{m}$} &\mbox{$0.600~\textup{m}$} &\mbox{$0.043~\textup{m}$} &\mbox{$2200$}\\
     \mbox{2} &\mbox{$2.200~\textup{m}$} &\mbox{$2.195~\textup{m}$} &\mbox{$2.225~\textup{m}$} &\mbox{$0.110~\textup{m}$} &\mbox{$2000$}\\ 
     \mbox{3} &\mbox{$2.888~\textup{m}$} &\mbox{$2.830~\textup{m}$} &\mbox{$2.887~\textup{m}$} &\mbox{$0.137~\textup{m}$} &\mbox{$2000$}\\ 
     \mbox{4} &\mbox{$4.435~\textup{m}$} &\mbox{$4.391~\textup{m}$} &\mbox{$4.362~\textup{m}$} &\mbox{$0.077~\textup{m}$} &\mbox{$2000$}\\ 
     \mbox{5} &\mbox{$5.204~\textup{m}$} &\mbox{$5.403~\textup{m}$} &\mbox{$5.412~\textup{m}$} &\mbox{$0.315~\textup{m}$} &\mbox{$2150$}\\ 
     \mbox{6} &\mbox{$7.380~\textup{m}$} &\mbox{$6.535~\textup{m}$} &\mbox{$6.200~\textup{m}$} &\mbox{$0.511~\textup{m}$} &\mbox{$2250$}\\ 
    \bottomrule
\end{tabular}
\end{table}

\begin{figure}[!t]
\center
\captionsetup[subfigure]{justification=centering}\subfloat[]{\includegraphics[width=.49\textwidth]{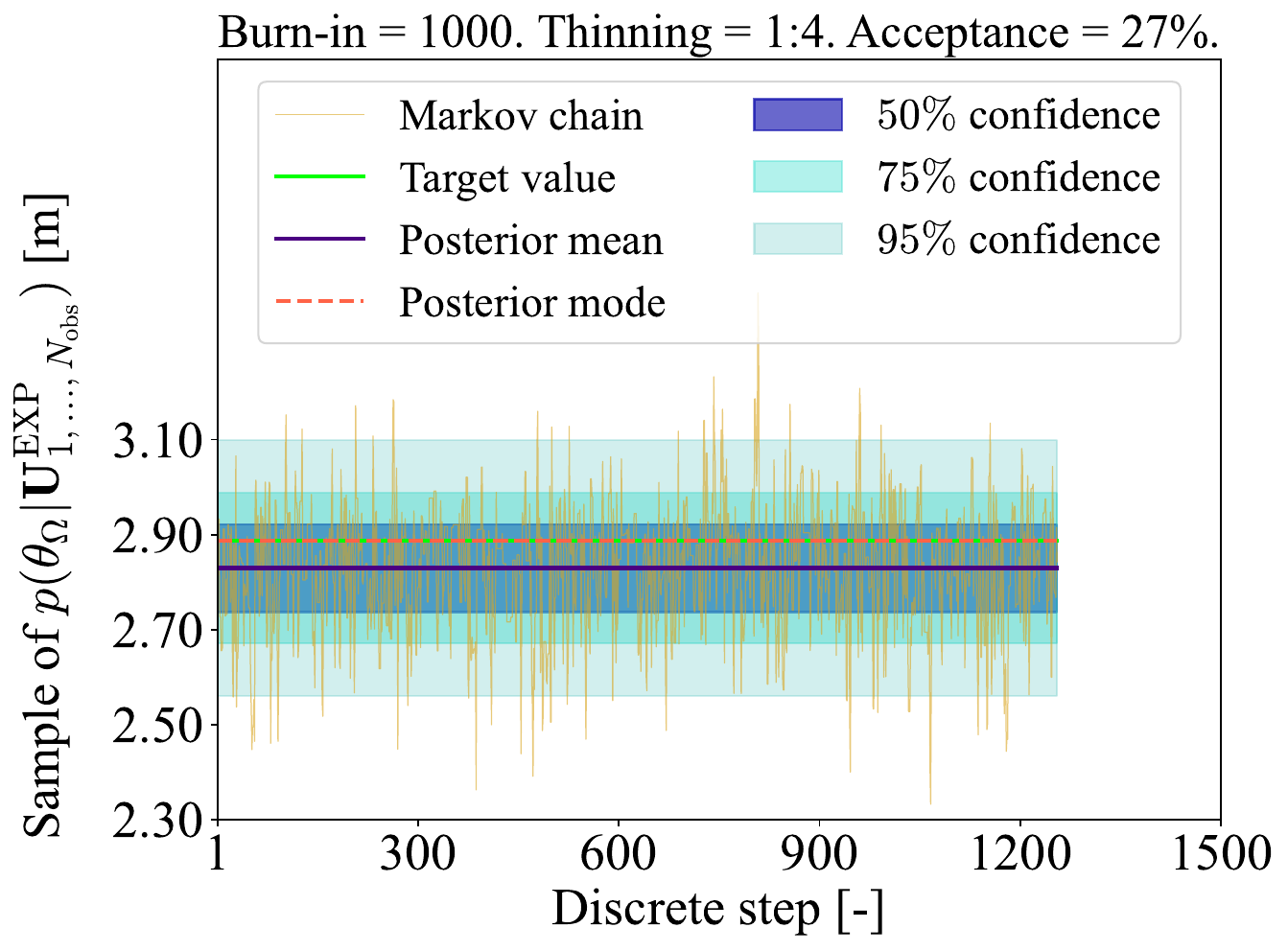}}\hspace{0.25cm}\subfloat[]{\includegraphics[width=.49\textwidth]{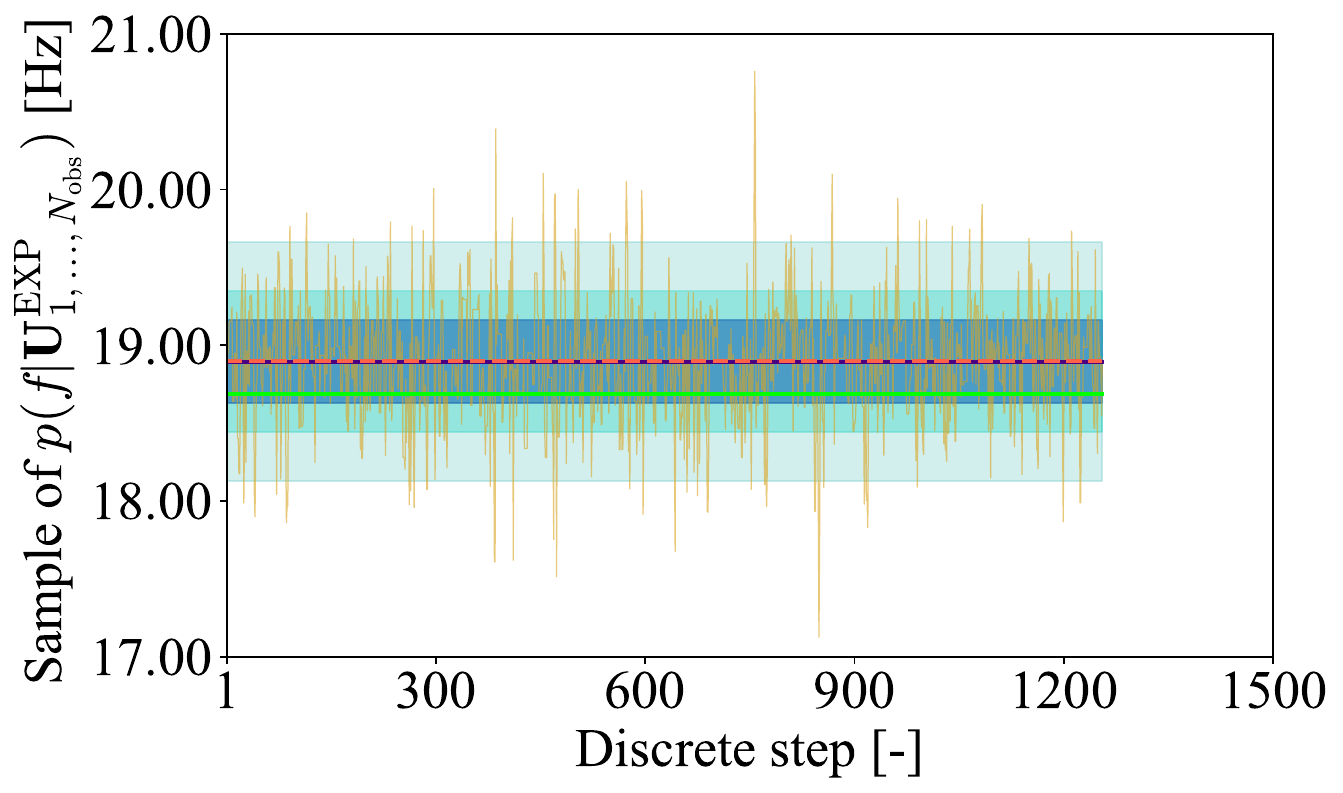}}\vspace{-0.2cm}
\\
\subfloat[]{\includegraphics[width=.42\textwidth]{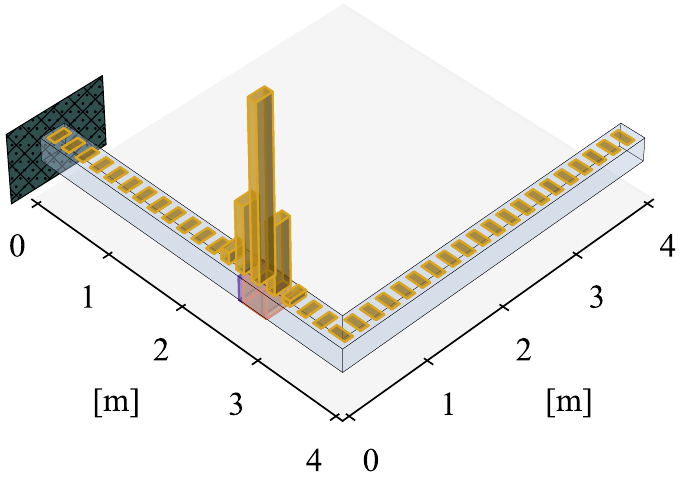}}
\caption{L-shaped cantilever beam - Exemplary MCMC result (Case 3): Markov chain, target value, posterior mean, posterior mode, and credibility intervals related to the estimation of (a) damage position $\theta_\Omega$ and (b) load frequency $f$; (c) histogram of the approximated, unnormalized posterior pdf $p(\theta_\Omega|\mathbf{U}^{\text{EXP}}_{1,\dots,N_\text{obs}})$ over the admissible support.\label{fig:parameter_identification}}
\end{figure}

Figure~\ref{fig:parameter_identification} presents an exemplary outcome of the MCMC simulation for Case 3. The graphs show the sampled Markov chain along with the estimated posterior mean and credibility intervals, for both $\theta_\Omega$ and $f$. Note that the chains are plotted over a relatively small range of values for the sake of visualization. Thanks to the low dimensionality of the involved features, the procedure also enjoys a considerable computational efficiency. The runtime for the parameter estimation is only about $5~\textup{s}$; this is a remarkable result, highlighting the real-time damage identification capabilities of the proposed strategy, all with quantified uncertainty.

 \begin{table}[!t]
\caption{L-shaped cantilever beam - Damage localization results for different operational and damage conditions, without using the feature extractor and the feature-oriented surrogate model. Table adapted from~\cite{art:Torzoni_MF}.}\label{tab:ref_outocomes}
  \centering
   \scriptsize
\begin{tabular}{p{0.5cm} p{1.7cm} p{1.7cm} p{1.7cm} p{1.7cm} p{0.6cm}}
    \toprule
    \mbox{Case} &  \mbox{Target$(\theta_\Omega)$} & $\text{Mean}(\theta_\Omega)$ & \mbox{$\text{Mode}(\theta_\Omega)$} & \mbox{$\text{Stdv}(\theta_\Omega)$} & $L_\text{chain}$\\
    \toprule
     \mbox{1} &\mbox{$0.564~\textup{m}$} &\mbox{$0.631~\textup{m}$} &\mbox{$0.587~\textup{m}$} &\mbox{$0.170~\textup{m}$} &\mbox{$2000$}\\
     \mbox{2} &\mbox{$2.200~\textup{m}$} &\mbox{$2.474~\textup{m}$} &\mbox{$2.414~\textup{m}$} &\mbox{$0.511~\textup{m}$} &\mbox{$2000$}\\
     \mbox{3} &\mbox{$2.888~\textup{m}$} &\mbox{$3.088~\textup{m}$} &\mbox{$2.844~\textup{m}$} &\mbox{$0.710~\textup{m}$} &\mbox{$3400$}\\
     \mbox{4} &\mbox{$4.435~\textup{m}$} &\mbox{$4.834~\textup{m}$} &\mbox{$4.198~\textup{m}$} &\mbox{$0.969~\textup{m}$} &\mbox{$2000$}\\
     \mbox{5} &\mbox{$5.204~\textup{m}$} &\mbox{$5.759~\textup{m}$} &\mbox{$5.397~\textup{m}$} &\mbox{$0.962~\textup{m}$} &\mbox{$3000$}\\
     \mbox{6} &\mbox{$7.380~\textup{m}$} &\mbox{$6.080~\textup{m}$} &\mbox{$7.136~\textup{m}$} &\mbox{$0.866~\textup{m}$} &\mbox{$4000$}\\
    \bottomrule
\end{tabular}
\end{table}

To quantify the impact of using learnable features, additional results related to the identification of the damage location are reported in \tab\ref{tab:ref_outocomes}, as presented in~\cite{art:Torzoni_MF}. In that work, the posterior distribution $p(\boldsymbol{\theta}|\mathbf{U}^{\text{EXP}}_{1,\dots,N_\text{obs}})$ was sampled without employing the feature extractor and the feature-oriented surrogate, relying instead directly on the MF-DNN surrogate model. A comparison with \tab\ref{tab:beam_outocomes} shows that incorporating $\text{N\hspace{-1px}N}_\text{ENC}$ and $\text{N\hspace{-1px}N}_\text{SUR}$ improves parameter identification across all performance indicators. In Case 4, for instance, the discrepancy from the target and the standard deviation are reduced by 7 and 10 times, respectively.

The superiority of the proposed hybrid approach stems from the optimized selection and extraction of informative features performed by $\text{N\hspace{-1px}N}_\text{ENC}$. This preliminary preprocessing of raw vibration recordings enables more accurate and less uncertain estimates. Moreover, the low dimensionality of the extracted features and the computational efficiency of $\text{N\hspace{-1px}N}_\text{SUR}$ enable thousands of likelihood function evaluations to be performed almost instantaneously. While the MF-DNN surrogate $\text{N\hspace{-1px}N}_\text{MF}$ can also be evaluated efficiently, a likelihood mechanism that leverages high-dimensional vibration recordings can not be evaluated with the same efficiency and is less informative due to the low damage sensitivity of raw signals. Another factor contributing to the higher efficiency of the proposed approach is the contrastive loss term, which promotes a geometric structure in the latent space that captures the parametric dependence of the vibration recordings, thereby enabling faster MCMC convergence. For instance, with the minimum allowed chain length of $L_\text{chain}=2000$, the proposed strategy achieves a 25-fold speed-up compared to relying directly on the MF-DNN surrogate model.

\subsection{Portal frame}
\label{sec:Frame}

\begin{figure}[t]
\captionsetup[subfigure]{justification=centering}\subfloat[\label{fig:telaio_load_damage}]{\includegraphics[width=.46\textwidth]{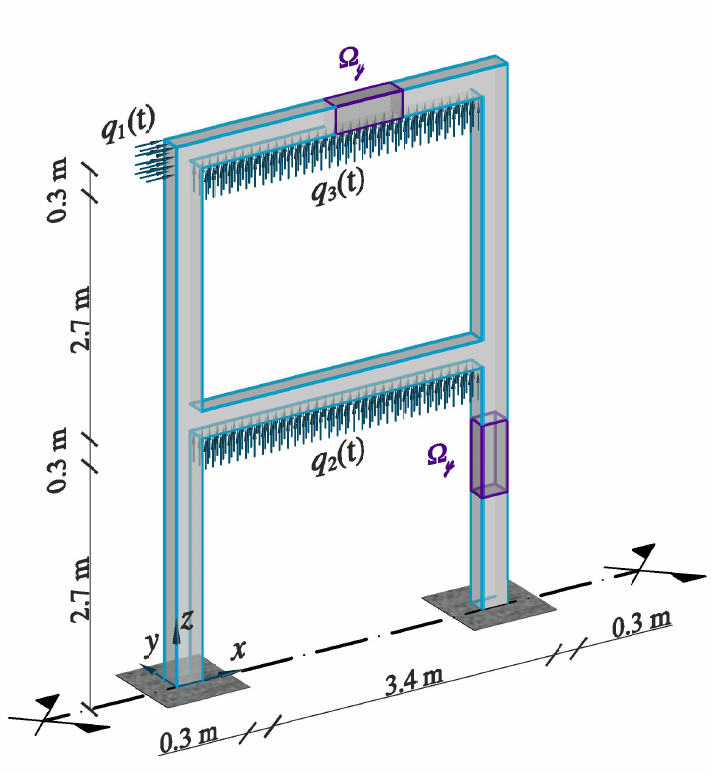}}\hspace{0.2cm}
\subfloat[\label{fig:telaio_sensors}]{\includegraphics[width=.46\textwidth]{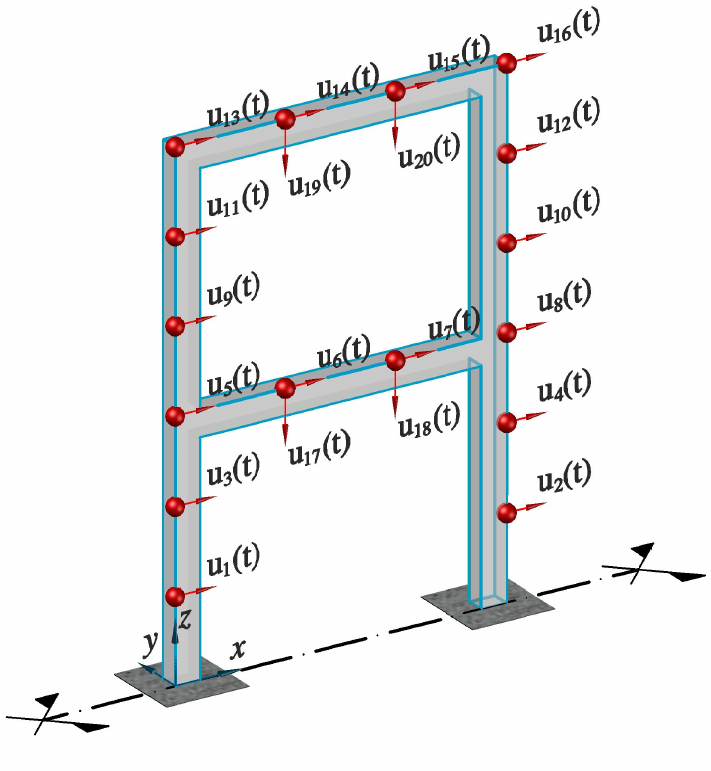}}
\caption{Portal frame: (a) details of the loading condition and the damageable region $\Omega_\y$; (b) synthetic recordings of displacements $u_1(t),\dots,u_{20}(t)$.\label{fig:telaio_model}}
\end{figure}

The second test case involves the two-story portal frame depicted in \fig\ref{fig:telaio_model}. The columns are $0.3~\textup{m}$ wide, the beams are $0.3~\textup{m}$ high, the inter-story height is $2.7~\textup{m}$, the span of the beams is $3.4~\textup{m}$, and the out-of-plane thickness is $0.45~\textup{m}$. The assumed mechanical properties are: Young's modulus $E = 34~\textup{GPa}$, Poisson's ratio $\nu = 0.2$, and density $\rho = 2500~\textup{kg/m}^3$. 

The structure is excited by three distributed loads $q_1(t)$, $q_2(t)$, and $q_3(t)$, respectively applied on top of the left column and on the bottom surface of the two horizontal beams, as shown in \fig\ref{fig:telaio_load_damage}. The three loads vary over time according to:
\begin{linenomath*}
\begin{equation}
q_{1,2,3}(t)=\left\{
\begin{array}{ll}
Q\frac{t}{T_q},&\text{if $t\leq T_q$},\\
0,&\text{if $t>T_q$},
\end{array}\right.
\end{equation}
\end{linenomath*}
with $Q=10~\textup{kPa}$ and $T_q=0.08~\textup{s}$. This fast-linear-ramp actuation may be connected to smart structures, equipped with actuators for forced vibration tests.

Displacement time histories are obtained at $N_u=20$ dofs, mimicking a monitoring system deployed as depicted in \fig\ref{fig:telaio_sensors}. The recordings are provided for a time interval $(0,T)$, with $T=1.12~\textup{s}$, and an acquisition frequency $f_\text{s}=125~\textup{Hz}$. Recordings are corrupted with additive Gaussian noise yielding a signal-to-noise ratio of $150$.

The HF numerical model features $N_\text{FE}=4827$ dofs. The Rayleigh damping matrix is assembled to account for a $2.5\%$ damping ratio on the first two structural modes. In this case, damage is simulated by means of a localized stiffness reduction that can take place anywhere in the frame, within subdomains $\Omega_{\y}$ featuring a different layout for the columns and the beams (see \fig\ref{fig:telaio_load_damage}). The position of $\Omega_{\y}$ is parametrized by the coordinates of its center of mass $\boldsymbol{\y}= (x_\Omega, z_\Omega)^\top$, with $x_\Omega$ and $z_\Omega$ varying in the ranges $[0.15, 3.85]~\textup{m}$ and $[0.4, 5.85]~\textup{m}$, respectively. The magnitude of the stiffness reduction ranges as $\delta\in[40\%,80\%]$, and it remains constant during an excitation event.

In the present case, the LF structural response is not parametrized. The LF dataset $\mathbf{D}_\text{LF}$ consists of a single instance underlying the structural response in the absence of damage. This recording thus replaces $\text{N\hspace{-1px}N}_\text{LF}$ in the MF-DNN surrogate. The HF component $\text{N\hspace{-1px}N}_\text{HF}$ is trained on $I_\text{HF}=1000$ HF data instances, to enrich the LF instance with the effects of damage for any given $\mathbf{x}^\text{HF}=(x_\Omega,z_\Omega,\delta)^\top$. The trained MF-DNN surrogate is then employed to populate $\mathbf{D}_\text{train}$ with $I_\text{train}=20,000$ instances, generated for varying values of the $\mathbf{x}^\text{HF}$ input parameters.

The mosaics dataset $\mathbf{D}^\I_\text{train}$ is obtained by encoding each training instance in $\mathbf{D}_\text{train}$ into a $4\times5$ MTF mosaic, with each MTF tessera being a $32\times32$ pixel image. Before undergoing the MTF encoding, the vibration recordings in $\mathbf{D}_\text{train}$ are normalized to follow a Gaussian distribution with zero mean and unit standard deviation. Additionally, the initial $8\%$ of each time history is removed to eliminate potential inaccuracies induced by $\text{N\hspace{-1px}N}_\text{HF}$. In the present case, the width of the blurring kernel is set equal to $4$.

\begin{figure}[t]
\captionsetup[subfigure]{justification=centering}\subfloat{\includegraphics[width=.33\textwidth]{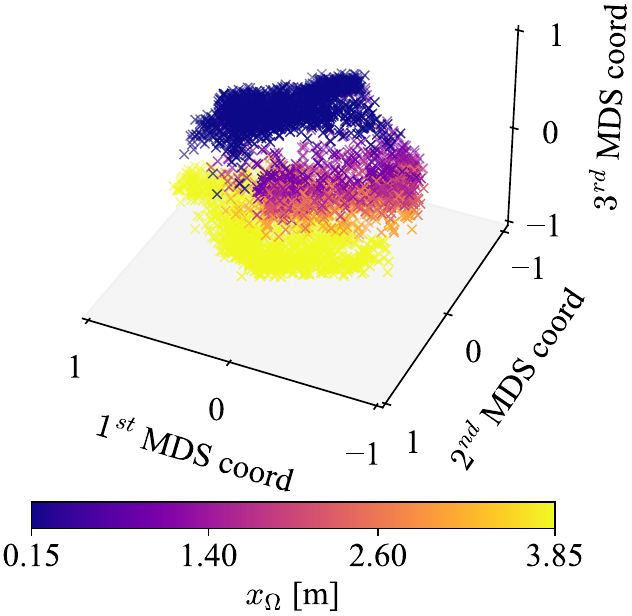}}\subfloat{\includegraphics[width=.33\textwidth]{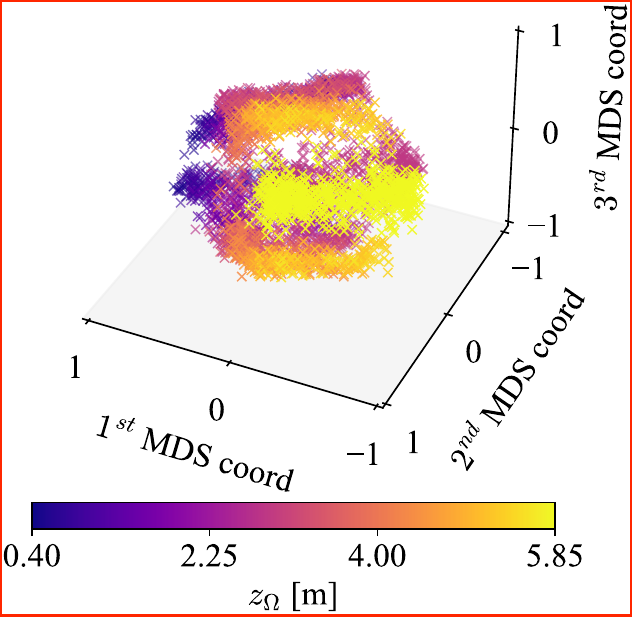}}
\subfloat{\includegraphics[width=.33\textwidth]{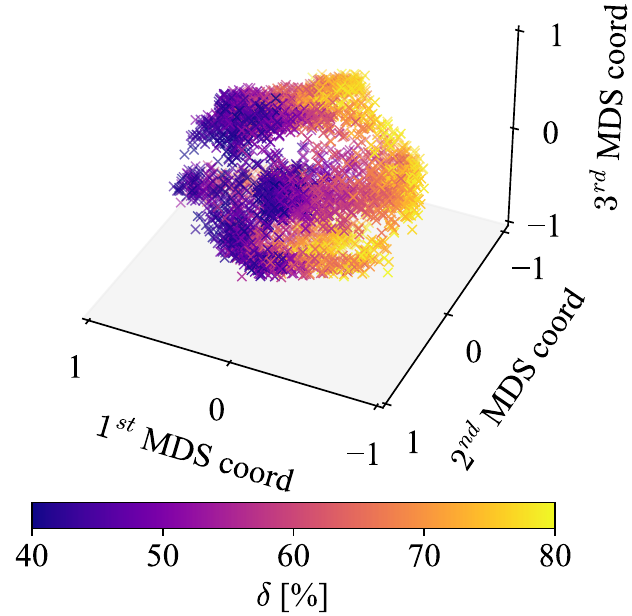}}
\caption{Portal frame - 3D multidimensional scaling representations of the low-dimensional features obtained for the validation data, against the target values of (left) damage position along the $x$-direction, (center) damage position along the $z$-direction, and (right) damage magnitude.\label{fig:Frame_MDS}}
\end{figure}

An MDS representation of the features extracted through $\text{N\hspace{-1px}N}_\text{ENC}$ for the validation set of $\mathbf{D}^\I_\text{train}$ is reported in \fig\ref{fig:Frame_MDS}. In this case, the color channels correspond to $x_\Omega$, $z_\Omega$, and $\delta$. The three plots qualitatively demonstrate the presence of an underlying manifold, which encodes the sensitivity of the recordings to the structural health parameters.

 \begin{table}[t]
\caption{Portal frame - Damage localization and quantification results for different operational and damage conditions, in terms of target value, posterior mean, posterior mode, standard deviation, and chain length.}\label{tab:frame_outocomes}
  \centering
   \scriptsize
\begin{tabular}{p{0.5cm} p{2.8cm} p{2.8cm} p{2.8cm} p{2.8cm} p{0.6cm}}
    \toprule
    \mbox{Case} &  \mbox{Target$(x_\Omega;z_\Omega;\delta)$} & $\text{Mean}(x_\Omega;z_\Omega;\delta)$ & \mbox{$\text{Mode}(x_\Omega;z_\Omega;\delta)$} & \mbox{$\text{Stdv}(x_\Omega;z_\Omega;\delta)$} & $L_\text{chain}$\\
    \toprule
    \mbox{1} &\mbox{$0.15~\textup{m};0.58~\textup{m};74.32\%$} &\mbox{$0.21~\textup{m};0.66~\textup{m};74.54\%$} &\mbox{$0.15~\textup{m};0.40~\textup{m};76.00\%$} &\mbox{$0.06~\textup{m};0.22~\textup{m};2.21\%$} &\mbox{$4550$}\\
    \mbox{2} &\mbox{$0.15~\textup{m};3.68~\textup{m};77.69\%$} &\mbox{$0.21~\textup{m};3.47~\textup{m};74.92\%$} &\mbox{$0.15~\textup{m};3.67~\textup{m};76.00\%$} &\mbox{$0.06~\textup{m};0.16~\textup{m};3.29\%$} &\mbox{$3350$}\\
    \mbox{3} &\mbox{$3.85~\textup{m};2.65~\textup{m};67.58\%$} &\mbox{$3.76~\textup{m};2.61~\textup{m};68.63\%$} &\mbox{$3.85~\textup{m};2.58~\textup{m};68.00\%$} &\mbox{$0.07~\textup{m};0.19~\textup{m};2.65\%$} &\mbox{$4050$}\\ 
    \mbox{4} &\mbox{$3.85~\textup{m};4.94~\textup{m};53.30\%$} &\mbox{$3.76~\textup{m};5.04~\textup{m};53.13\%$} &\mbox{$3.85~\textup{m};5.30~\textup{m};52.00\%$} &\mbox{$0.07~\textup{m};0.19~\textup{m};4.02\%$} &\mbox{$5850$}\\ 
    \mbox{5} &\mbox{$1.94~\textup{m};2.85~\textup{m};56.64\%$} &\mbox{$1.85~\textup{m};2.84~\textup{m};56.37\%$} &\mbox{$1.63~\textup{m};3.58~\textup{m};56.00\%$} &\mbox{$0.26~\textup{m};0.23~\textup{m};2.99\%$} &\mbox{$3850$}\\ 
    \mbox{6} &\mbox{$1.70~\textup{m};5.85~\textup{m};63.70\%$} &\mbox{$1.90~\textup{m};5.70~\textup{m};69.06\%$} &\mbox{$2.00~\textup{m};5.85~\textup{m};68.00\%$} &\mbox{$0.29~\textup{m};0.16~\textup{m};3.35\%$} &\mbox{$2950$}\\ 
    \bottomrule
\end{tabular}
\end{table}

\begin{figure}[t]
\center
\captionsetup[subfigure]{justification=centering}\subfloat[]{\includegraphics[width=.49\textwidth]{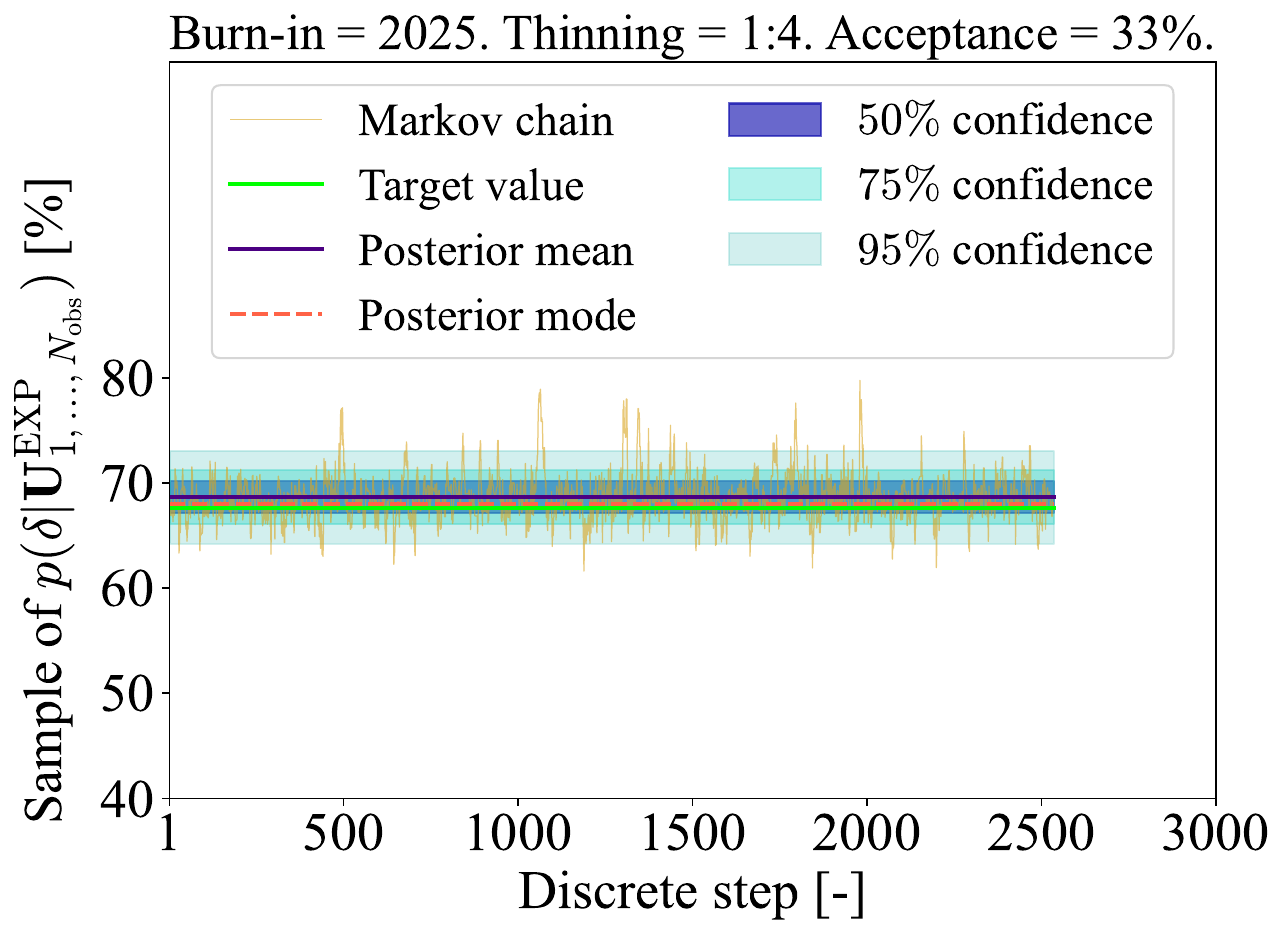}}\\\subfloat[]{\includegraphics[width=.49\textwidth]{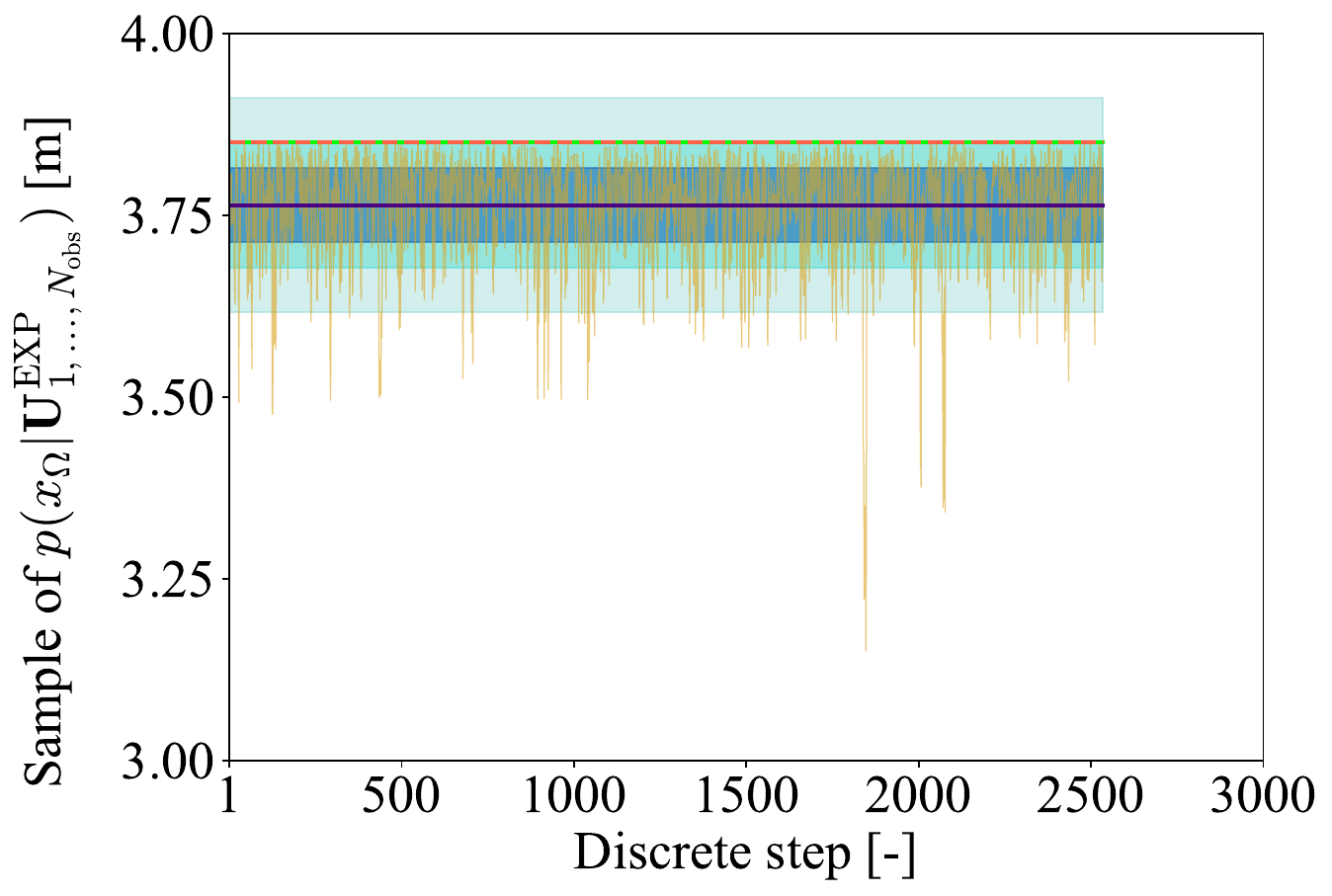}}\hspace{0.25cm}\subfloat[]{\includegraphics[width=.49\textwidth]{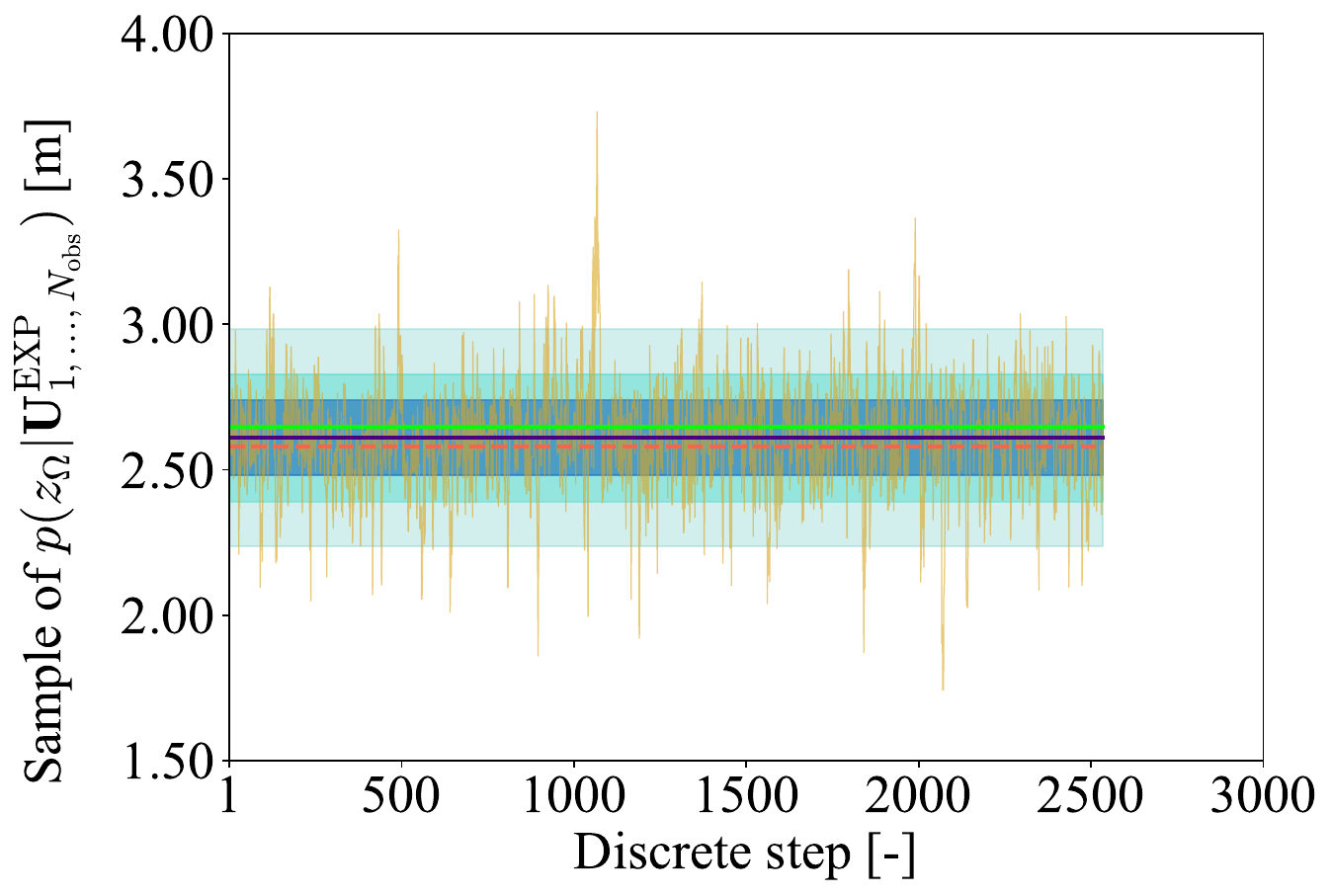}}
\caption{Portal frame - Exemplary MCMC result (Case 3): Markov chain, target value, posterior mean, posterior mode, and credibility intervals related to the estimation of (a) damage magnitude $\delta$, (b) damage position along the $x$-direction, and (c) damage position along the $z$-direction.\label{fig:parameter_identification_frame}}
\end{figure}

The learned feature space is employed to update the prior belief about $\boldsymbol{\theta}=(x_\Omega,z_\Omega,\delta)^\top$ via MCMC sampling. The algorithm is fed with batches of $N_\text{obs}=8$ noisy observations, all related to the same damage location and magnitude. The results obtained from six MCMC simulations are summarized in \tab\ref{tab:frame_outocomes}. Overall, both the damage location and the damage magnitude are identified with very high accuracy and relatively low uncertainty. There are no cases characterized by a significant discrepancy between the target and the posterior mean values. As expected, the standard deviation of either $x_\Omega$ or $z_\Omega$ is larger along the axis in which $\Omega_{\y}$ is free to move. Additionally, the uncertainty in $\delta$ increases as $\Omega_{\y}$ moves farther from the clamped sides, due to a smaller sensitivity of sensor recordings to damage in such cases. For visualization purposes, \fig\ref{fig:parameter_identification_frame} presents an exemplary MCMC-recovered posterior for Case 3.

\subsection{H{\"o}rnefors railway bridge}
\label{sec:Bridge}

\begin{figure}[t]
\begin{center}
\includegraphics[width=1\textwidth]{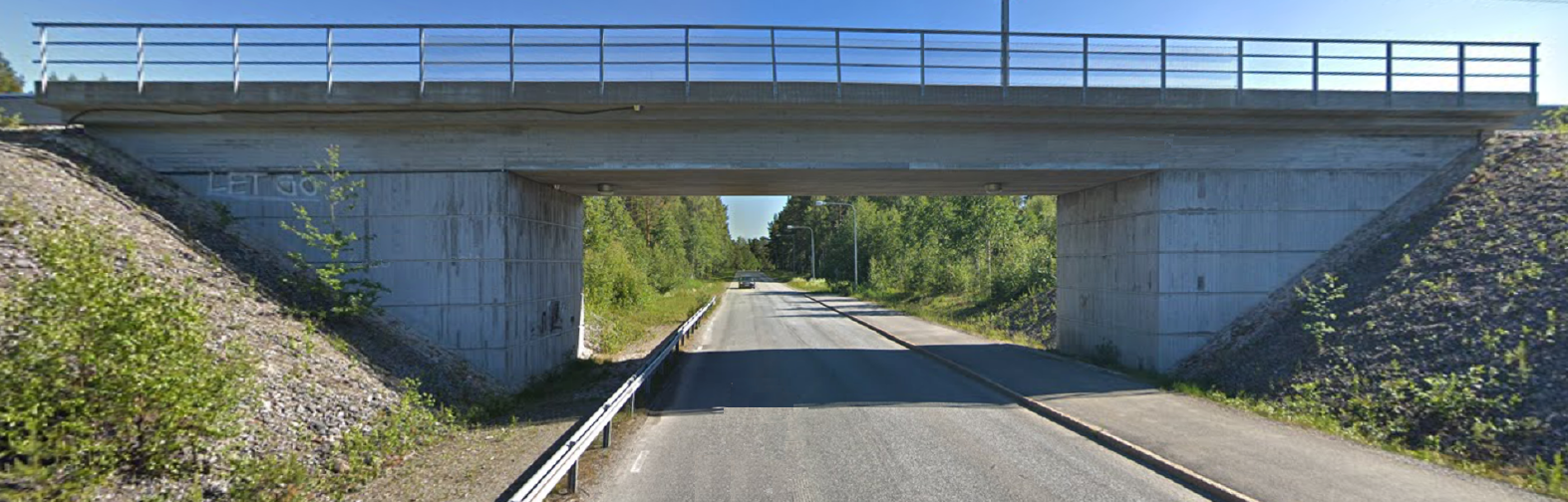}
\caption{{H{\"o}rnefors railway bridge.}\label{fig:bridge_photo}}
\end{center}
\end{figure}

This third test case aims to assess the performance of the proposed strategy in a more complex situation, involving the railway bridge depicted in \fig\ref{fig:bridge_photo}. It is an integral concrete bridge located along the Bothnia line in H{\"o}rnefors (Sweden). It features a span of $15.7~\textup{m}$, a clear height of $4.7~\textup{m}$, and a width of $5.9~\textup{m}$ (edge beams excluded). The thickness of the structural elements is $0.5~\textup{m}$ for the deck, $0.7~\textup{m}$ for the frame walls, and $0.8~\textup{m}$ for the wing walls. The bridge is founded on two plates connected by stay beams and supported by pile groups. The concrete is of class C35/45, whose mechanical properties are: $E=34~\textup{GPa}$, $\nu= 0.2$, and $\rho=2500~\textup{kg/m}^3$. The superstructure consists of a single track with sleepers spaced $0.65~\textup{m}$ apart, resting on a $0.6~\textup{m}$ deep, $4.3~\textup{m}$ wide ballast layer with a density of $\rho_B=1800~\textup{kg/m}^3$. The geometrical and mechanical modeling data have been adapted from former research activities~\cite{thesis:kth3,thesis:kth2}.

The bridge is subjected to the transit of Gr{\"o}na T{\r a}get trains, traveling at a speed $\upsilon\in[160,215]~\textup{km/h}$. We specifically consider trains composed of two wagons, totaling 8 axles. Each axle carries a mass  $\phi\in[16,22]~\textup{ton}$. The corresponding load model is described in~\cite{art:Metodologico}, and consists of 25 equivalent distributed forces transmitted from the sleepers to the ballast layer, which are then propagated to the deck with a $4:1$ slope, according to Eurocode 1~\cite{code:EC1}.

\begin{figure}[t]
\begin{center}
\includegraphics[width=0.65\textwidth]{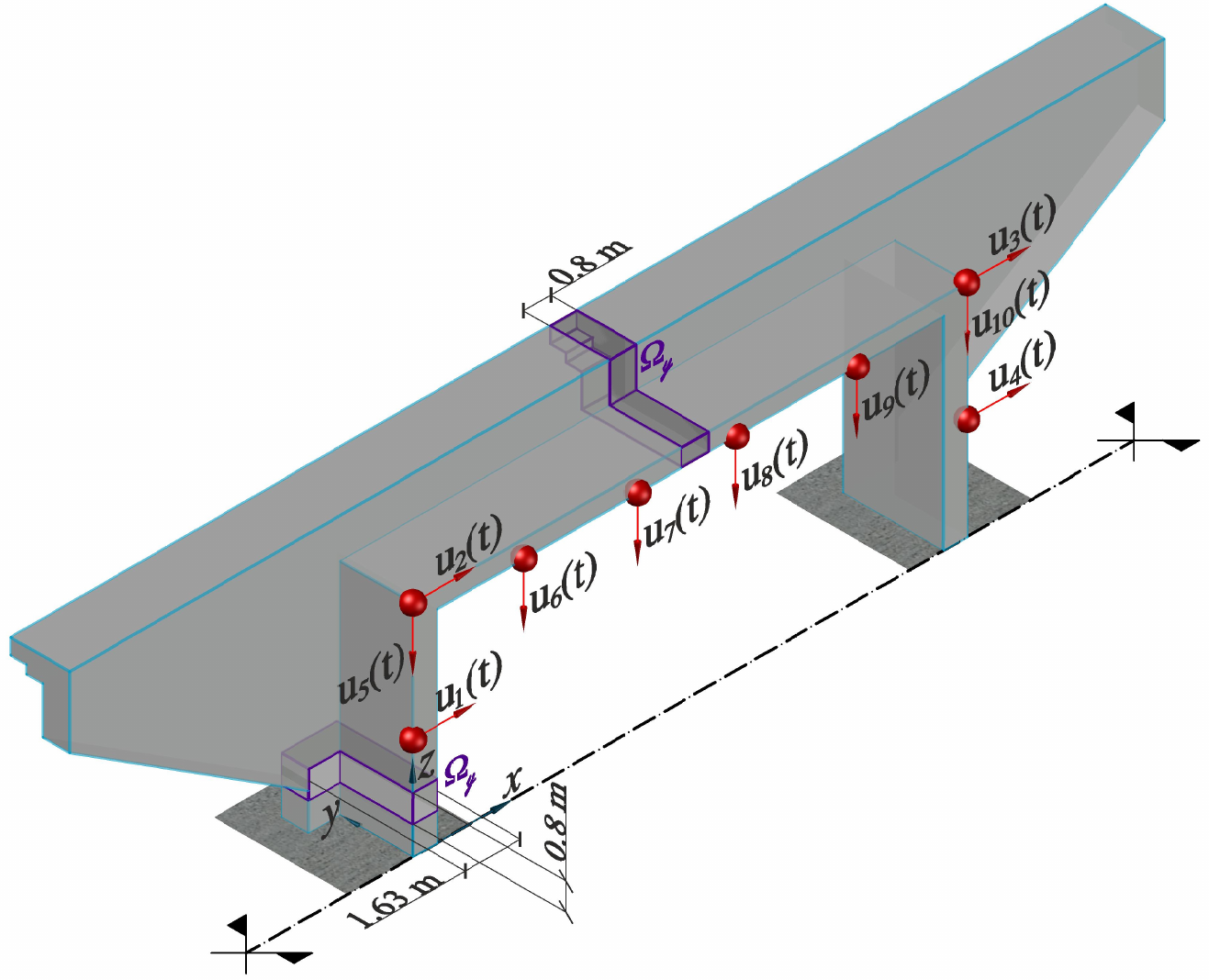}
\caption{{Railway bridge: details of the synthetic recordings of displacements $u_1(t),\dots,u_{10}(t)$ and the damageable region $\Omega_\y$.}\label{fig:bridge_sensors_damage}}
\end{center}
\end{figure}

The monitoring system features $N_u=10$ sensors and is deployed as depicted in \fig\ref{fig:bridge_sensors_damage}. Displacement time histories are provided for a time interval $(0,T)$, with $T=1.5~\textup{s}$, and an acquisition frequency $f_\text{s} = 400~\textup{Hz}$.

The HF numerical model features $N_\text{FE}=17,292$ dofs, resulting from a finite element discretization with an element size of $0.15~\textup{m}$ for the deck, to enable a smooth propagation of the traveling load, and $0.80~\textup{m}$ elsewhere. The ballast layer is accounted for through an increased density for the deck and the edge beams. The embankments are modeled with distributed springs over the surfaces facing the ground, implemented as a Robin mixed boundary condition (elastic coefficient $k_{\textup{Robin}}=10^{8}~\textup{N/m}^3$). The Rayleigh damping matrix accounts for a $5\%$ damping ratio on the first two structural modes. In this case, damage is simulated by means of a localized stiffness reduction that can take place anywhere over the lateral frame walls and the deck, within subdomains $\Omega_{\y}$ featuring a different layout in the two cases (see \fig\ref{fig:bridge_sensors_damage}). The position of $\Omega_{\y}$ is parametrized through $\boldsymbol{\y}= ( x_\Omega, z_\Omega )^\top$, with $x_\Omega$ and $z_\Omega$ varying in the ranges $[-0.115, 16.515]~\textup{m}$ and $[0.4, 6.25]~\textup{m}$, respectively. The stiffness reduction may have a magnitude $\delta\in[40\%,80\%]$, which is kept fixed while a train travels across the bridge. To summarize, the vector of HF input parameters is $\mathbf{x}^\text{HF} = (\upsilon,\phi,x_\Omega,z_\Omega,\delta)^\top$.

The basis matrix $\mathbf{W}$ is obtained from a snapshot matrix $\mathbf{S}$, assembled through $200$ evaluations of the LF-FOM for different values of parameters $\mathbf{x}^\text{LF} = (\upsilon,\phi)^\top$. By setting the error tolerance to $\epsilon=10^{-3}$, $N_\text{RB}=312$ POD modes are retained in $\mathbf{W}$.

The MF-DNN surrogate model is trained using $I_\text{LF}=5000$ LF data instances for $\text{N\hspace{-1px}N}_\text{LF}$, and only $I_\text{HF}=500$ HF data instances for $\text{N\hspace{-1px}N}_\text{HF}$. The MF-DNN surrogate is then employed to populate $\mathbf{D}_\text{train}$ with $I_\text{train}=30,000$ instances for varying values of the $\mathbf{x}^\text{HF}$ input parameters.

The mosaics dataset $\mathbf{D}^\I_\text{train}$ is obtained by encoding each training instance in $\mathbf{D}_\text{train}$ into a $2\times5$ MTF mosaic, with each MTF tessera being a $64\times64$ pixel image. Before undergoing the MTF encoding, the initial $4\%$ of each time history in $\mathbf{D}_\text{train}$ is removed to eliminate potential inaccuracies induced by $\text{N\hspace{-1px}N}_\text{HF}$. Moreover, since in this case the vibration recordings exhibit data distributions predominantly concentrated in the tails, each time history in $\mathbf{D}_\text{train}$ is normalized to take values between $0$ and $1$, and quantized through a uniform bin assignment rather than a Gaussian one. The width of the blurring kernel is set to $9$ in this case.

\begin{figure}[t]
\center
\captionsetup[subfigure]{justification=centering}\subfloat{\includegraphics[width=.33\textwidth]{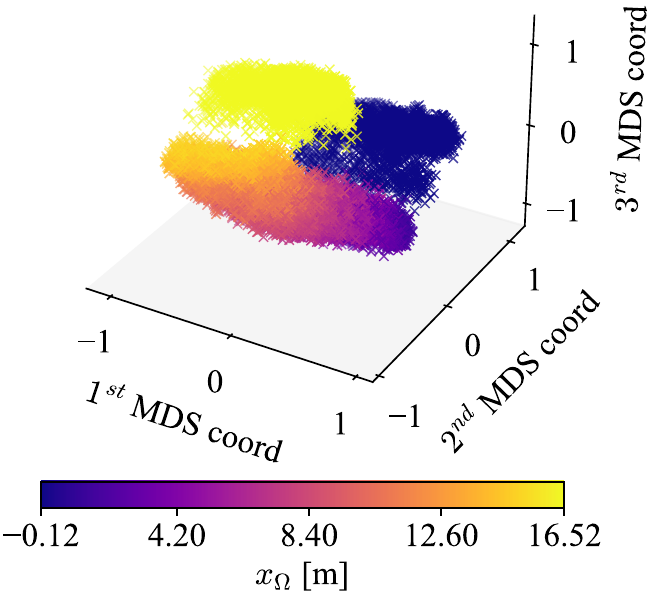}}\subfloat{\includegraphics[width=.33\textwidth]{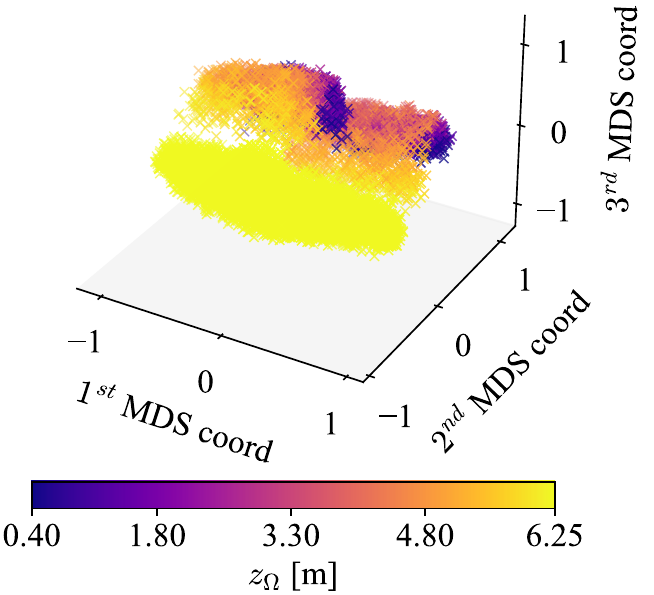}}
\subfloat{\includegraphics[width=.33\textwidth]{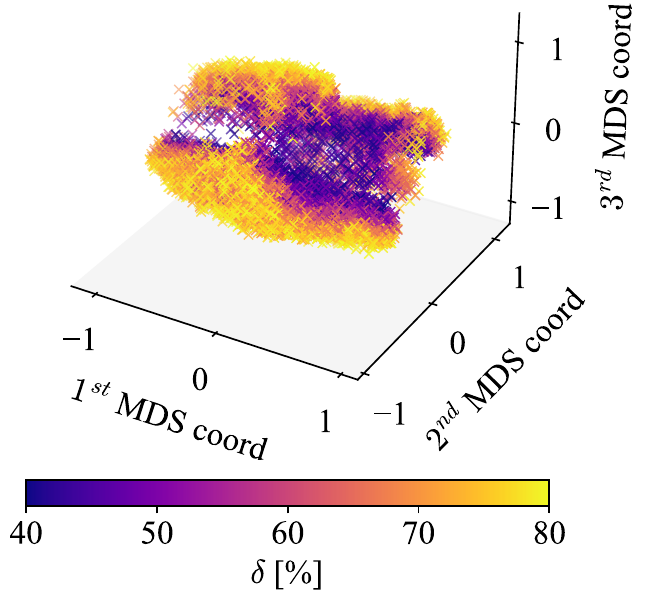}}\\
\subfloat{\includegraphics[width=.33\textwidth]{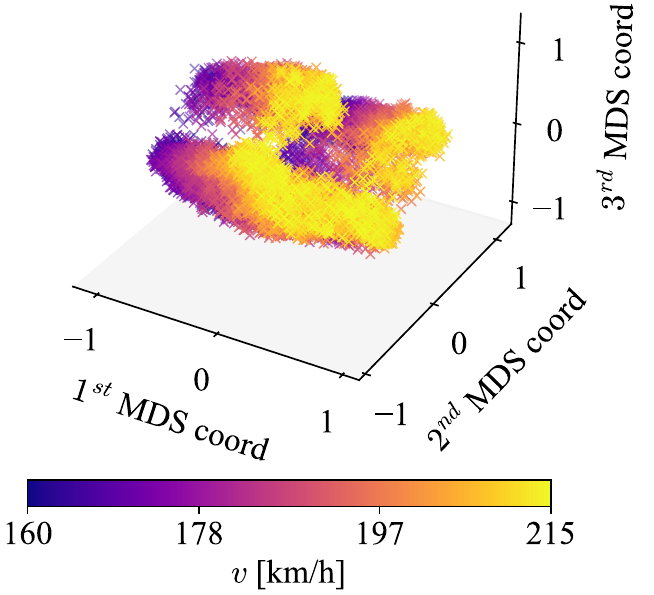}}\subfloat{\includegraphics[width=.33\textwidth]{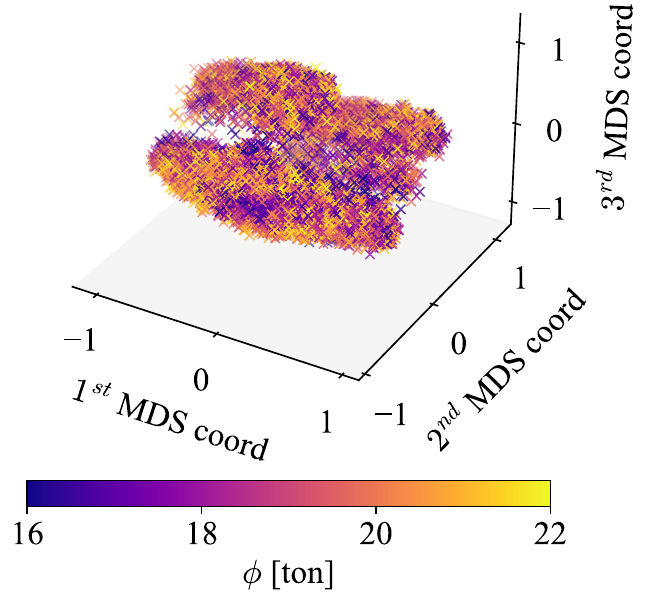}}
\caption{Railway bridge - 3D MDS representations of the low-dimensional features obtained for the validation data, against the target values of (top-left) damage position along the $x$-direction, (top-center) damage position along the $z$-direction, (top-right) damage magnitude, (bottom-left) train velocity, and (bottom-right) axle mass.\label{fig:Bridge_MDS}}
\end{figure}

The visual check on the MDS representation of the features extracted from the validation data is reported in \fig\ref{fig:Bridge_MDS}. In this case, the color channels refer to each entry of $\mathbf{x}^\text{HF}$. It is interesting to note that the overall shape outlined by the scatter plot resembles the structural layout of the bridge (rotated and extruded), which is automatically retrieved from $x_\Omega$ and $z_\Omega$. These plots qualitatively demonstrate a clear sensitivity of the low-dimensional feature space to the damage location, the damage magnitude, and the train velocity. However, the axle mass is characterized by a fuzzier representation, lacking a manifold topology capable of adequately capturing its influence on the processed measurements. 

The MCMC algorithm is fed with batches of $N_\text{obs}=8$ HF observations, all related to the same damage location $\theta_\Omega\in[0.4,26]~\textup{m}$ and damage magnitude $\delta$. However, each observation in a batch is obtained for random values of train velocity $\upsilon$ and axle-carried mass $\phi$, which are presumed to be accurately measured by the train on-board system. Therefore, the relative posterior distribution is deterministically set to the measured values. Table~\ref{tab:bridge_outocomes} reports the results of six MCMC simulations concerning the sampling of the posterior pdf for the damage location and magnitude. The damage location is consistently identified with high accuracy, except in Case 2. Nevertheless, the relative discrepancy between the target and the posterior mean values is only $1.58~\textup{m}$ over an admissible support of $25.6~\textup{m}$. Similarly, for the damage magnitude, the most significant deviation occurs in Case 2, with a discrepancy between the target and the posterior mean values of about $7.5\%$. Figure~\ref{fig:parameter_identification_frame} presents an exemplary MCMC outcome for Case 4.

\begin{table}[t]
\caption{Railway bridge - Damage localization and quantification results for different operational and damage conditions, in terms of target value, posterior mean, posterior mode, standard deviation, and chain length.}\label{tab:bridge_outocomes}
  \centering
   \scriptsize
\begin{tabular}{p{0.5cm} p{2cm} p{2cm} p{2cm} p{2cm} p{0.6cm}}
    \toprule
    \mbox{Case} &  \mbox{Target$(\theta_\Omega;\delta)$} & $\text{Mean}(\theta_\Omega;\delta)$ & \mbox{$\text{Mode}(\theta_\Omega;\delta)$} & \mbox{$\text{Stdv}(\theta_\Omega;\delta)$} & $L_\text{chain}$\\
    \toprule
     \mbox{1} &\mbox{$2.31~\textup{m};73.42\%$} &\mbox{$2.15~\textup{m};70.96\%$} &\mbox{$2.00~\textup{m};75.00\%$} &\mbox{$0.81~\textup{m};7.03\%$} &\mbox{$2650$}\\
     \mbox{2} &\mbox{$3.96~\textup{m};63.75\%$} &\mbox{$2.38~\textup{m};56.28\%$} &\mbox{$2.30~\textup{m};56.50\%$} &\mbox{$0.38~\textup{m};8.43\%$} &\mbox{$2000$}\\
     \mbox{3} &\mbox{$6.07~\textup{m};47.53\%$} &\mbox{$5.72~\textup{m};50.91\%$} &\mbox{$5.75~\textup{m};41.5\%$} &\mbox{$0.18~\textup{m};8.37\%$} &\mbox{$2000$}\\
     \mbox{4} &\mbox{$9.44~\textup{m};51.41\%$} &\mbox{$9.19~\textup{m};49.33\%$} &\mbox{$9.15~\textup{m};50.5\%$} &\mbox{$0.72~\textup{m};5.35\%$} &\mbox{$2000$}\\
     \mbox{5} &\mbox{$13.37~\textup{m};41.25\%$} &\mbox{$13.06~\textup{m};43.97\%$} &\mbox{$13.40~\textup{m};41.50\%$} &\mbox{$0.84~\textup{m};3.95\%$} &\mbox{$2000$}\\
     \mbox{6} &\mbox{$17.13~\textup{m};52.07\%$} &\mbox{$16.27~\textup{m};48.02\%$} &\mbox{$16.20~\textup{m};41.50\%$} &\mbox{$1.45~\textup{m};8.31\%$} &\mbox{$2150$}\\ 
    \bottomrule
\end{tabular}
\end{table}

\begin{figure}[t]
\center
\captionsetup[subfigure]{justification=centering}\subfloat[]{\includegraphics[width=.49\textwidth]{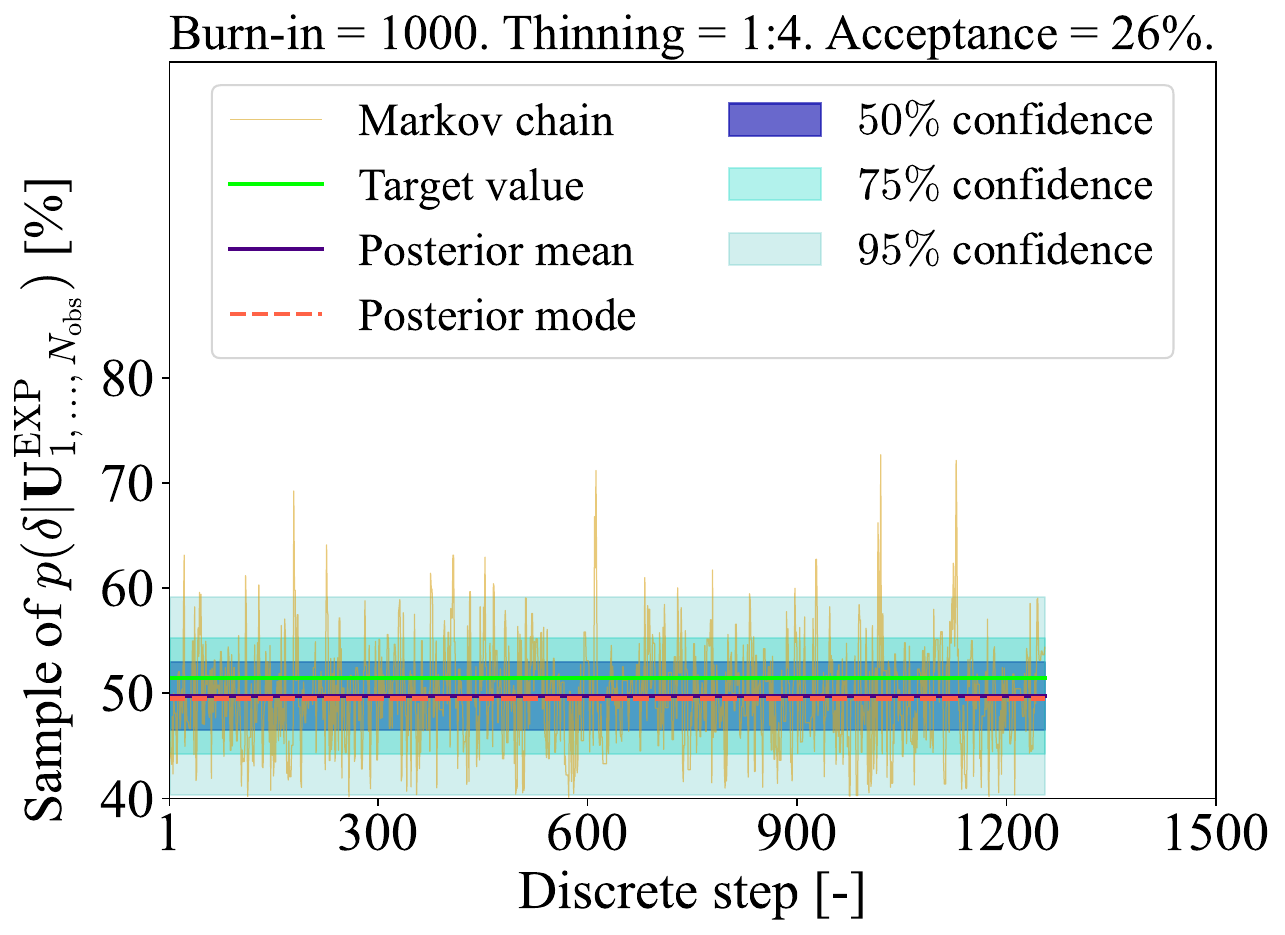}}\hspace{0.25cm}\subfloat[]{\includegraphics[width=.49\textwidth]{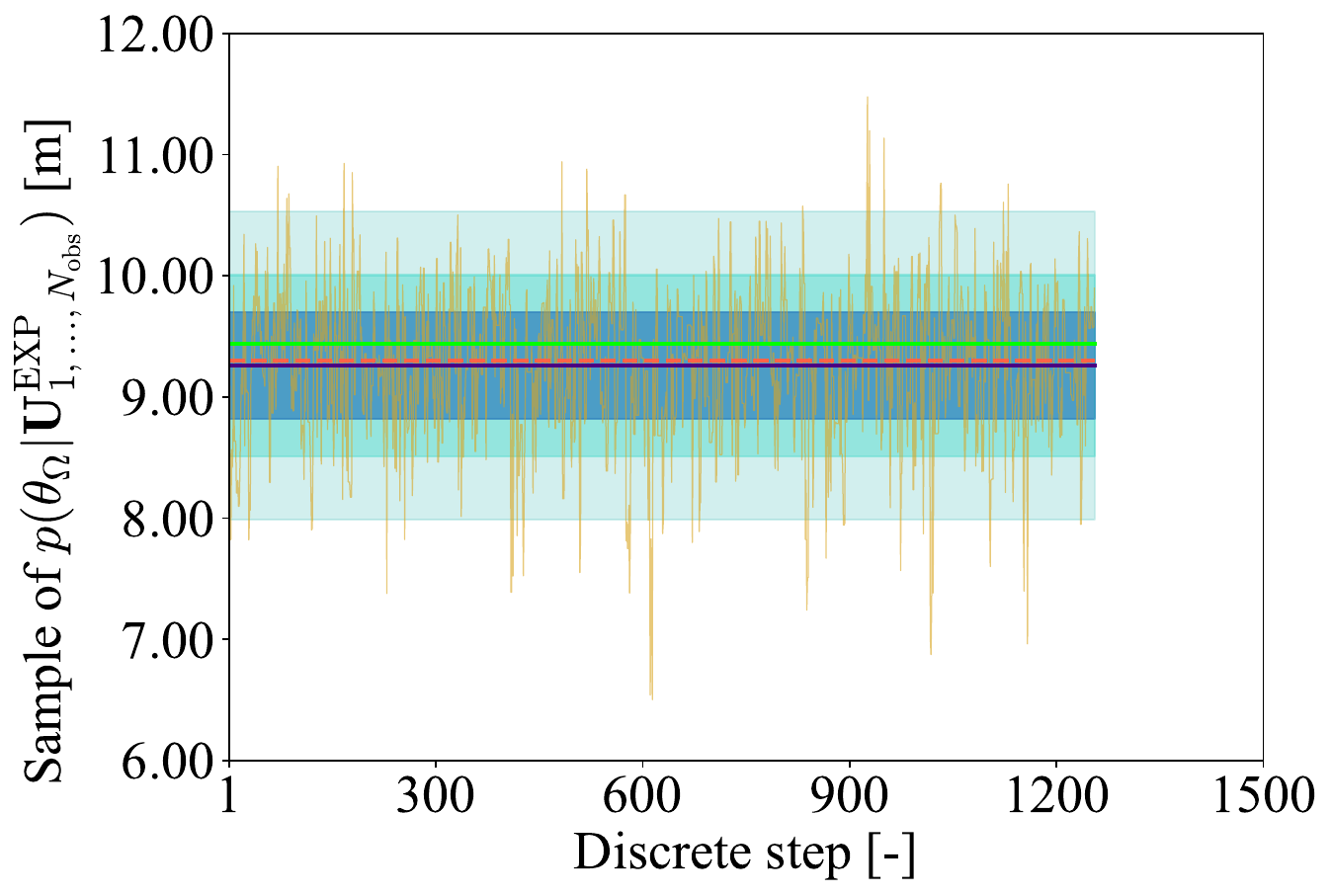}}
\caption{Railway bridge - Exemplary MCMC result (Case 4): Markov chain, target value, posterior mean, posterior mode, and credibility intervals related to the estimation of (a) damage magnitude $\delta$ and (b) damage position $\theta_\Omega$.\label{fig:parameter_identification_bridge}}
\end{figure}

\section{Conclusions}
\label{sec:conclusions}

In this work, we have proposed a deep learning-based strategy to enhance stochastic approaches to structural health monitoring. The presented strategy relies upon a learnable feature extractor and a feature-oriented surrogate model. The two data-driven models are synergistically exploited to improve the accuracy and efficiency of a model updating framework. The feature extractor automates the selection and extraction of informative features from raw sensor recordings. The extracted features encodes the sensitivity of the observational data to the sought parameters onto a low-dimensional metric space. The surrogate model approximates the functional link between the parameters to be inferred and the low-dimensional feature space. The methodology can be easily adapted to solve inverse problems in application domains other than structural health monitoring, such as, e.g., scattering problems, medical diagnoses, and inverse kinematics.

The computational procedure leverages a preliminary offline phase that: (i) uses physics-based numerical models and reduced-order modeling to overcome the lack of experimental data under varying damage and operational conditions; (ii) implements a multi-fidelity surrogate modeling strategy for the generation of a large labeled dataset; (iii) trains both the feature extractor and the feature-oriented surrogate model.

The proposed strategy has been assessed through the simulated monitoring of an L-shaped cantilever beam, a portal frame, and a railway bridge. In the absence of experimental data under the effect of varying operational and damage conditions, the tests have been performed leveraging high-fidelity simulation data corrupted with additive Gaussian noise. The results have demonstrated that the adoption of learnable features, in place of raw vibration recordings,  enables a significant improvement in the accuracy of parameter identification. Moreover, the proposed strategy exhibits high computational efficiency, attributable to the low dimensionality of the extracted features.

The proposed framework is well suited for extension to experimental applications. Provided that the synthetic training data sufficiently capture the operational and damage scenarios encountered in practice, a reasonable degree of generalization to real-world conditions can be expected. Nevertheless, challenges arising from modeling assumptions are likely to emerge. From a practical perspective, all models are intrinsically biased, and model-form uncertainties may compromise the accurate representation of physical systems. Although, in principle, these uncertainties are reducible, they frequently prove irreducible in practice. Despite efforts to refine and calibrate models to reflect the behavior of monitored structures, the inherent complexity of engineering systems imposes limitations on modeling fidelity. The associated concern is that valid regions of the solution space may remain inaccessible, potentially resulting in overconfident yet inaccurate estimates. To address these limitations, latent variable models offer a promising direction. By introducing a probabilistic structure that quantifies uncertainty within the latent space, these models are capable of representing deviations between the actual data-generating process and the training data distribution. This probabilistic approach can help mitigate limitations associated with deterministic feature extraction and model-reality mismatches. In this regard, simulation-based inference~\cite{inference} provides an appealing framework for implementing Bayesian inference engines based on latent variable models. By enabling joint sampling over both the latent distribution and the parameter space, this setup allows for the estimation of observation likelihoods while simultaneously enhancing the overall credibility of the monitoring process.

Future work will be devoted to integrating the proposed strategy within a digital twin framework; for instance, see~\cite{art:Torzoni_DT,art:Kapteyn_nature}. In this context, the assimilation of observational data to deliver real-time structural health estimates would facilitate the optimal planning of maintenance and management actions within a dynamic decision-making environment.


\vspace{6pt} 
\noindent{\bf Acknowledgments:} This work is supported in part by the interdisciplinary Ph.D. Grant ``Physics-Informed Deep Learning for Structural Health Monitoring'' at Politecnico di Milano. Andrea Manzoni acknowledges the project ‘‘Dipartimento di Eccellenza” 2023-2027, funded by MUR, and the project FAIR (Future Artificial Intelligence Research), funded by the NextGenerationEU program within the PNRR-PE-AI scheme (M4C2, Investment 1.3, Line on Artificial Intelligence).

\vspace{6pt} 
\noindent{\bf The authors declare no conflict of interest.}


\medskip

\bibliographystyle{elsarticle-num}

\begin{thebibliography}{10}
\expandafter\ifx\csname url\endcsname\relax
  \def\url#1{\texttt{#1}}\fi
\expandafter\ifx\csname urlprefix\endcsname\relax\def\urlprefix{URL }\fi
\expandafter\ifx\csname href\endcsname\relax
  \def\href#1#2{#2} \def\path#1{#1}\fi

\bibitem{committee2018climate}
A.~S. of~Civil~Engineers, Adaptive Design and Risk Management, American Society
  of Civil Engineers, Reston, Virginia, 2018, Ch.~7, pp. 173–--226.
\newblock \href {https://doi.org/10.1061/9780784415191.ch07}
  {\path{doi:10.1061/9780784415191.ch07}}.

\bibitem{art:Glaser}
S.~D. Glaser, A.~Tolman, {Sense of Sensing: From Data to Informed Decisions for
  the Built Environment}, J Infrastruct Syst 14~(1) (2008) 4--14.
\newblock \href {https://doi.org/10.1061/(ASCE)1076-0342(2008)14:1(4)}
  {\path{doi:10.1061/(ASCE)1076-0342(2008)14:1(4)}}.

\bibitem{art:Achenbach}
J.~D. Achenbach, {Structural health monitoring -- What is the prescription?},
  Mech Res Commun 36~(2) (2009) 137--142.
\newblock \href {https://doi.org/10.1016/j.mechrescom.2008.08.011}
  {\path{doi:10.1016/j.mechrescom.2008.08.011}}.

\bibitem{proc:Rosafalco2022}
L.~Rosafalco, M.~Torzoni, A.~Manzoni, S.~Mariani, A.~Corigliano, {A
  Self-adaptive Hybrid Model/data-Driven Approach to SHM Based on Model Order
  Reduction and Deep Learning}, in: {Structural Health Monitoring Based on Data
  Science Techniques}, Springer International Publishing, 2022, pp. 165--184.
\newblock \href {https://doi.org/10.1007/978-3-030-81716\_9}
  {\path{doi:10.1007/978-3-030-81716\_9}}.

\bibitem{proc:Springer_Ubertini}
E.~Garc{\'\i}a-Mac{\'\i}as, F.~Ubertini, {Integrated SHM Systems: Damage
  Detection Through Unsupervised Learning and Data Fusion}, in: {Structural
  Health Monitoring Based on Data Science Techniques}, Springer International
  Publishing, 2022, pp. 247--268.
\newblock \href {https://doi.org/10.1007/978-3-030-81716-9\_12}
  {\path{doi:10.1007/978-3-030-81716-9\_12}}.

\bibitem{art:Torzoni_DML}
M.~Torzoni, A.~Manzoni, S.~Mariani, {Structural health monitoring of civil
  structures: A diagnostic framework powered by deep metric learning}, Comput
  Struct 271 (2022) 106858.
\newblock \href {https://doi.org/10.1016/j.compstruc.2022.106858}
  {\path{doi:10.1016/j.compstruc.2022.106858}}.

\bibitem{art:Worden_detection}
K.~Worden, Structural fault detection using a novelty measure, J Sound Vib
  201~(1) (1997) 85--101.
\newblock \href {https://doi.org/10.1006/jsvi.1996.0747}
  {\path{doi:10.1006/jsvi.1996.0747}}.

\bibitem{art:Farrar01}
C.~R. Farrar, S.~W. Doebling, D.~A. Nix, {Vibration--Based Structural Damage
  Identification}, Phil Trans R Soc A 359~(1778) (2001) 131--149.
\newblock \href {https://doi.org/10.1098/rsta.2000.0717}
  {\path{doi:10.1098/rsta.2000.0717}}.

\bibitem{book:Bishop}
C.~M. Bishop, {Pattern Recognition and Machine Learning}, {Information Science
  and Statistics}, Springer-Verlag, New York, NY, 2006.

\bibitem{art:Avci_review}
O.~Avci, O.~Abdeljaber, S.~Kiranyaz, M.~Hussein, M.~Gabbouj, D.~Inman, {A
  review of vibration-based damage detection in civil structures: From
  traditional methods to Machine Learning and Deep Learning applications}, Mech
  Syst Signal Process 147 (2021) 107077.
\newblock \href {https://doi.org/10.1016/j.ymssp.2020.107077}
  {\path{doi:10.1016/j.ymssp.2020.107077}}.

\bibitem{art:Ierimonti}
L.~Ierimonti, N.~Cavalagli, E.~Garc{\'\i}a-Mac{\'\i}as, I.~Venanzi,
  F.~Ubertini, {Bayesian-Based Damage Assessment of Historical Structures Using
  Vibration Monitoring Data}, in: {International Workshop on Civil Structural
  Health Monitoring}, Springer, 2021, pp. 415--429.
\newblock \href {https://doi.org/10.1007/978-3-030-74258-4\_28}
  {\path{doi:10.1007/978-3-030-74258-4\_28}}.

\bibitem{art:Demetrio_2}
D.~Cristiani, C.~Sbarufatti, M.~Giglio, Damage diagnosis and prognosis in
  composite double cantilever beam coupons by particle filtering and surrogate
  modelling, Struct Health Monit 20~(3) (2021) 1030--1050.
\newblock \href {https://doi.org/10.1177/1475921720960067}
  {\path{doi:10.1177/1475921720960067}}.

\bibitem{art:kamariotis_voi}
A.~Kamariotis, E.~Chatzi, D.~Straub, {Value of information from vibration-based
  structural health monitoring extracted via Bayesian model updating}, Mech
  Syst Signal Process 166 (2022) 108465.
\newblock \href {https://doi.org/10.1016/j.ymssp.2021.108465}
  {\path{doi:10.1016/j.ymssp.2021.108465}}.

\bibitem{art:Azam_Mariani}
S.~{Eftekhar Azam}, S.~Mariani, {Online damage detection in structural systems
  via dynamic inverse analysis: A recursive Bayesian approach}, Eng Struct 159
  (2018) 28--45.
\newblock \href {https://doi.org/10.1016/j.engstruct.2017.12.031}
  {\path{doi:10.1016/j.engstruct.2017.12.031}}.

\bibitem{art:Siamese}
J.~Bromley, I.~Guyon, Y.~Lecun, E.~S\"ackinger, R.~Shah, {Signature
  Verification using a "Siamese" Time Delay Neural Network}, Int J Pattern
  Recognit Artif Intell 7 (1993) 25.
\newblock \href {https://doi.org/10.5555/2987189.2987282}
  {\path{doi:10.5555/2987189.2987282}}.

\bibitem{proc:LeCun_Contrastive_2005}
S.~Chopra, R.~Hadsell, Y.~LeCun, Learning a similarity metric discriminatively,
  with application to face verification, in: {Proc IEEE Comput Soc Conf Comput
  Vis Pattern Recognit}, 2005, pp. 539--546.
\newblock \href {https://doi.org/10.1109/CVPR.2005.202}
  {\path{doi:10.1109/CVPR.2005.202}}.

\bibitem{proc:LeCun_Contrastive}
R.~Hadsell, S.~Chopra, Y.~Lecun, {Dimensionality Reduction by Learning an
  Invariant Mapping}, in: {Proc IEEE Comput Soc Conf Comput Vis Pattern
  Recognit}, 2006, pp. 1735--1742.
\newblock \href {https://doi.org/10.1109/CVPR.2006.100}
  {\path{doi:10.1109/CVPR.2006.100}}.

\bibitem{art:Deep_metric_survey}
M.~Kaya, H.~Bilge, {Deep Metric Learning: A Survey}, Symmetry 11 (2019) 1066.
\newblock \href {https://doi.org/10.3390/sym11091066}
  {\path{doi:10.3390/sym11091066}}.

\bibitem{art:Metric_survey}
A.~Bellet, A.~Habrard, M.~Sebban, {A Survey on Metric Learning for Feature
  Vectors and Structured Data}, arXiv preprint arXiv:1306.6709 (2013).
\newblock \href {https://doi.org/10.48550/arXiv.1306.6709}
  {\path{doi:10.48550/arXiv.1306.6709}}.

\bibitem{proc:Deep_Metric_Learning_Rank}
F.~Cakir, K.~He, X.~Xia, B.~Kulis, S.~Sclaroff, {Deep Metric Learning to Rank},
  in: {Proc IEEE Comput Soc Conf Comput Vis Pattern Recognit}, 2019, pp.
  1861--1870.
\newblock \href {https://doi.org/10.1109/CVPR.2019.00196}
  {\path{doi:10.1109/CVPR.2019.00196}}.

\bibitem{art:VBI}
R.~Hou, X.~Wang, Y.~Xia, {Vibration-Based Structural Damage Detection Using
  Sparse Bayesian Learning Techniques}, in: {Structural Health Monitoring Based
  on Data Science Techniques}, Springer International Publishing, 2022, pp.
  1--25.
\newblock \href {https://doi.org/10.1007/978-3-030-81716-9\_1}
  {\path{doi:10.1007/978-3-030-81716-9\_1}}.

\bibitem{art:AM_Green}
P.~L. Green, K.~Worden, {Bayesian and Markov chain Monte Carlo methods for
  identifying nonlinear systems in the presence of uncertainty}, Phil Trans R
  Soc A 373 (2015).
\newblock \href {https://doi.org/10.1098/rsta.2014.0405}
  {\path{doi:10.1098/rsta.2014.0405}}.

\bibitem{art:Lam2018}
H.~F. Lam, J.~H. Yang, S.~K. Au, {Markov chain Monte Carlo-based Bayesian
  method for structural model updating and damage detection}, Struct Contr
  Health Monit 25~(4) (2018) 1--22.
\newblock \href {https://doi.org/10.1002/stc.2140}
  {\path{doi:10.1002/stc.2140}}.

\bibitem{art:Metodologico}
L.~Rosafalco, M.~Torzoni, A.~Manzoni, S.~Mariani, A.~Corigliano, Online
  structural health monitoring by model order reduction and deep learning
  algorithms, Comput Struct 255 (2021) 106604.
\newblock \href {https://doi.org/10.1016/j.compstruc.2021.106604}
  {\path{doi:10.1016/j.compstruc.2021.106604}}.

\bibitem{art:Torzoni_temperature}
M.~Torzoni, L.~Rosafalco, A.~Manzoni, S.~Mariani, A.~Corigliano, {SHM under
  varying environmental conditions: An approach based on model order reduction
  and deep learning}, Comput Struct 266 (2022) 106790.
\newblock \href {https://doi.org/10.1016/j.compstruc.2022.106790}
  {\path{doi:10.1016/j.compstruc.2022.106790}}.

\bibitem{art:Torzoni_MF}
M.~Torzoni, A.~Manzoni, S.~Mariani, {A multi-fidelity surrogate model for
  structural health monitoring exploiting model order reduction and artificial
  neural networks}, Mech Syst Signal Process 197 (2023) 110376.
\newblock \href {https://doi.org/10.1016/j.ymssp.2023.110376}
  {\path{doi:10.1016/j.ymssp.2023.110376}}.

\bibitem{book:RB}
A.~Quarteroni, A.~Manzoni, F.~Negri, Reduced basis methods for partial
  differential equations: an introduction, Springer, 2015.
\newblock \href {https://doi.org/10.1007/978-3-319-15431-2}
  {\path{doi:10.1007/978-3-319-15431-2}}.

\bibitem{book:ML_perspective}
C.~Farrar, K.~Worden, {Structural Health Monitoring: A Machine Learning
  Perspective}, John Wiley \& Sons, 2013.
\newblock \href {https://doi.org/10.1002/9781118443118}
  {\path{doi:10.1002/9781118443118}}.

\bibitem{art:Kapteyn_nature}
M.~G. Kapteyn, J.~V.~R. Pretorius, K.~E. Willcox, A probabilistic graphical
  model foundation for enabling predictive digital twins at scale, Nat Comput
  Sci 1~(5) (2021) 337--347.
\newblock \href {https://doi.org/10.1038/s43588-021-00069-0}
  {\path{doi:10.1038/s43588-021-00069-0}}.

\bibitem{art:TEUGHELS2002}
A.~Teughels, J.~Maeck, G.~{De Roeck}, {Damage assessment by FE model updating
  using damage functions}, Comput Struct 80~(25) (2002) 1869--1879.
\newblock \href {https://doi.org/10.1016/S0045-7949(02)00217-1}
  {\path{doi:10.1016/S0045-7949(02)00217-1}}.

\bibitem{art:sirovich}
L.~Sirovich, {Turbulence and the dynamics of coherent structures. I. Coherent
  structures}, Q Appl Math 45~(3) (1987) 561--571.
\newblock \href {https://doi.org/10.1090/qam/910462}
  {\path{doi:10.1090/qam/910462}}.

\bibitem{art:Kerschen_1}
G.~Kerschen, J.-C. Golinval, Physical interpretation of the proper orthogonal
  modes using the singular value decomposition, J Sound Vib 249~(5) (2002)
  849--865.
\newblock \href {https://doi.org/10.1006/jsvi.2001.3930}
  {\path{doi:10.1006/jsvi.2001.3930}}.

\bibitem{art:Kerschen_2}
G.~Kerschen, J.-c. Golinval, A.~F. Vakakis, L.~A. Bergman, {The Method of
  Proper Orthogonal Decomposition for Dynamical Characterization and Order
  Reduction of Mechanical Systems: An Overview}, Nonlinear Dyn 41~(1) (2005)
  147--169.
\newblock \href {https://doi.org/10.1007/s11071-005-2803-2}
  {\path{doi:10.1007/s11071-005-2803-2}}.

\bibitem{art:MF_survey}
B.~Peherstorfer, K.~Willcox, M.~Gunzburger, {Survey of Multifidelity Methods in
  Uncertainty Propagation, Inference, and Optimization}, {SIAM Review} 60~(3)
  (2018) 550--591.
\newblock \href {https://doi.org/10.1137/16M1082469}
  {\path{doi:10.1137/16M1082469}}.

\bibitem{art:Meng2020a}
X.~Meng, G.~E. Karniadakis, {A composite neural network that learns from
  multi-fidelity data: Application to function approximation and inverse PDE
  problems}, {Journal of Computational Physics} 401 (2020) 109020.
\newblock \href {https://doi.org/10.1016/j.jcp.2019.109020}
  {\path{doi:10.1016/j.jcp.2019.109020}}.

\bibitem{art:conti2022MF}
P.~Conti, M.~Guo, A.~Manzoni, J.~Hesthaven, Multi-fidelity surrogate modeling
  using long short-term memory networks, Comput Methods Appl Mech Eng 404
  (2023) 115811.
\newblock \href {https://doi.org/10.1016/j.cma.2022.115811}
  {\path{doi:10.1016/j.cma.2022.115811}}.

\bibitem{proc:Giglioni_EWSHM}
V.~Giglioni, I.~Venanzi, A.~E. Baia, V.~Poggioni, A.~Milani, F.~Ubertini, {Deep
  Autoencoders for Unsupervised Damage Detection with Application to the Z24
  Benchmark Bridge}, in: {European Workshop on Structural Health Monitoring},
  Springer International Publishing, 2023, pp. 1048--1057.
\newblock \href {https://doi.org/10.1007/978-3-031-07258-1\_105}
  {\path{doi:10.1007/978-3-031-07258-1\_105}}.

\bibitem{art:MTF_Fink}
G.~R. Garcia, G.~Michau, M.~Ducoffe, J.~S. Gupta, O.~Fink, {Temporal signals to
  images: Monitoring the condition of industrial assets with deep learning
  image processing algorithms}, Proc Inst Mech Eng O J Risk Reliab 236~(4)
  (2022) 617--627.
\newblock \href {https://doi.org/10.1177/1748006X21994446}
  {\path{doi:10.1177/1748006X21994446}}.

\bibitem{art:MTF_rocking}
I.~M. Mantawy, M.~O. Mantawy, Convolutional neural network based structural
  health monitoring for rocking bridge system by encoding time-series into
  images, Struct Control Health Monit 29~(3) (2022) e2897.
\newblock \href {https://doi.org/10.1002/stc.2897}
  {\path{doi:10.1002/stc.2897}}.

\bibitem{art:MTF}
Z.~Wang, T.~Oates, {Imaging Time-Series to Improve Classification and
  Imputation}, in: {Proc of the International Conference on Artificial
  Intelligence}, Vol.~24, 2015, pp. 3939--3945.

\bibitem{art:rec_plots}
N.~Marwan, M.~{Carmen Romano}, M.~Thiel, J.~Kurths, Recurrence plots for the
  analysis of complex systems, Phys Rep 438~(5) (2007) 237--329.
\newblock \href {https://doi.org/10.1016/j.physrep.2006.11.001}
  {\path{doi:10.1016/j.physrep.2006.11.001}}.

\bibitem{art:gray_scale}
G.~Xu, M.~Liu, Z.~Jiang, W.~Shen, C.~Huang, {Online Fault Diagnosis Method
  Based on Transfer Convolutional Neural Networks}, IEEE Trans Instrum Meas
  69~(2) (2020) 509--520.
\newblock \href {https://doi.org/10.1109/TIM.2019.2902003}
  {\path{doi:10.1109/TIM.2019.2902003}}.

\bibitem{book:DL_book}
I.~Goodfellow, Y.~Bengio, A.~Courville, Deep Learning, MIT Press, Cambridge,
  MA, 2016, \url{http://www.deeplearningbook.org}.

\bibitem{art:Fresca2020}
S.~Fresca, L.~Dede, A.~Manzoni, {A comprehensive deep learning-based approach
  to reduced order modeling of nonlinear time-dependent parametrized PDEs}, J
  Sci Comput 87 (2021) 1--36.
\newblock \href {https://doi.org/10.1007/s10915-021-01462-7}
  {\path{doi:10.1007/s10915-021-01462-7}}.

\bibitem{art:MH}
W.~K. Hastings, {Monte Carlo Sampling Methods Using Markov Chains and Their
  Applications}, Biometrika 57~(1) (1970) 97--109.
\newblock \href {https://doi.org/10.2307/2334940} {\path{doi:10.2307/2334940}}.

\bibitem{art:MHadaptive1}
H.~Haario, E.~Saksman, J.~Tamminen, {An adaptive Metropolis algorithm},
  Bernoulli 7~(2) (2001) 223--242.
\newblock \href {https://doi.org/10.2307/3318737} {\path{doi:10.2307/3318737}}.

\bibitem{art:TMCMC_Straub}
W.~Betz, I.~Papaioannou, D.~Straub, {Transitional Markov chain Monte Carlo:
  observations and improvements}, J Eng Mech 142~(5) (2016) 04016016.
\newblock \href {https://doi.org/10.1061/(ASCE)EM.1943-7889.0001066}
  {\path{doi:10.1061/(ASCE)EM.1943-7889.0001066}}.

\bibitem{art:No_U_Turn}
M.~D. Hoffman, A.~Gelman, {{{The No-U-Turn Sampler: Adaptively Setting Path
  Lengths in Hamiltonian Monte Carlo}}}, J Mach Learn Res 15~(1) (2014)
  1593--1623.
\newblock \href {https://doi.org/10.5555/2627435.2638586}
  {\path{doi:10.5555/2627435.2638586}}.

\bibitem{art:Gelman-Rubin}
A.~Gelman, D.~B. Rubin, {Inference from Iterative Simulation Using Multiple
  Sequences}, Stat Sci 7~(4) (1992) 457--472.
\newblock \href {https://doi.org/10.1214/ss/1177011136}
  {\path{doi:10.1214/ss/1177011136}}.

\bibitem{Redbkit}
F.~Negri, {redbKIT, version 2.2}, \url{http://redbkit.github.io/redbKIT}
  (2016).

\bibitem{chollet2015keras}
F.~Chollet, et~al., Keras, \url{https://keras.io} (2015).

\bibitem{scikit-learn}
F.~Pedregosa, G.~Varoquaux, A.~Gramfort, V.~Michel, B.~Thirion, O.~Grisel,
  M.~Blondel, P.~Prettenhofer, R.~Weiss, V.~Dubourg, J.~Vanderplas, A.~Passos,
  D.~Cournapeau, M.~Brucher, M.~Perrot, E.~Duchesnay, {Scikit-learn: Machine
  Learning in {P}ython}, J Mach Learn Res 12 (2011) 2825--2830.

\bibitem{thesis:kth3}
M.~{\"{U}}lker-Kaustell, Some aspects of the dynamic soil-structure interaction
  of a portal frame railway bridge, Ph.D. thesis, KTH Royal Institute of
  Technology (2009).

\bibitem{thesis:kth2}
T.~Arvidsson, J.~Li, {Dynamic analysis of a portal frame railway bridge using
  frequency dependent soil structure interaction}, Master thesis, KTH Royal
  Institute of Technology (2011).

\bibitem{code:EC1}
{European Committee for Standardization}, {Part 2: Traffic loads on bridges},
  in: {EN 1991-2 Eurocode 1: Actions on structures}, 2003, pp. 66--74.

\bibitem{inference}
K.~Cranmer, J.~Brehmer, G.~Louppe, The frontier of simulation-based inference,
  {Proceedings of the National Academy of Sciences} 117~(48) (2020)
  30055--30062.
\newblock \href {https://doi.org/10.1073/pnas.1912789117}
  {\path{doi:10.1073/pnas.1912789117}}.

\bibitem{art:Torzoni_DT}
M.~Torzoni, M.~Tezzele, S.~Mariani, A.~Manzoni, K.~E. Willcox, {A digital twin
  framework for civil engineering structures}, Comput Methods Appl Mech Eng 418
  (2024) 116584.
\newblock \href {https://doi.org/10.1016/j.cma.2023.116584}
  {\path{doi:10.1016/j.cma.2023.116584}}.

\bibitem{art:Donner_2010}
R.~V. Donner, Y.~Zou, J.~F. Donges, N.~Marwan, J.~Kurths, Recurrence
  networks—a novel paradigm for nonlinear time series analysis, New J Phys
  12~(3) (2010) 033025.
\newblock \href {https://doi.org/10.1088/1367-2630/12/3/033025}
  {\path{doi:10.1088/1367-2630/12/3/033025}}.

\bibitem{art:Campanharo}
A.~S. Campanharo, M.~I. Sirer, R.~D. Malmgren, F.~M. Ramos, L.~A.~N. Amaral,
  {Duality between Time Series and Networks}, {PLoS One} 6~(8) (2011) 1--13.
\newblock \href {https://doi.org/10.1371/journal.pone.0023378}
  {\path{doi:10.1371/journal.pone.0023378}}.

\bibitem{art:SAX}
J.~Lin, E.~Keogh, L.~Wei, S.~Lonardi, {Experiencing SAX: a novel symbolic
  representation of time series}, Data Min Knowl Discov 15 (2007) 107--144.
\newblock \href {https://doi.org/10.1007/s10618-007-0064-z}
  {\path{doi:10.1007/s10618-007-0064-z}}.

\bibitem{art:Glorot}
X.~Glorot, Y.~Bengio, Understanding the difficulty of training deep feedforward
  neural networks, J Mach Learn Res 9 (2010) 249--256.

\bibitem{art:Adam}
D.~Kingma, J.~Ba, {Adam: A Method for Stochastic Optimization}, in: {Int Conf
  Learn Represent}, Vol.~3, 2015, pp. 1--13.

\end{thebibliography}
\biboptions{sort&compress}


\appendix
\newpage

\section{Imaging time series via Markov transition field}
\label{sec:MTF}

In this Appendix, we review the MTF encoding~\cite{art:MTF} employed in this work to transform multivariate time series $\mathbf{U}=[\mathbf{u}_1,\ldots,\mathbf{u}_{N_u}]\in\mathbb{R}^{L\times N_u}$ into images. The technique is detailed with reference to univariate time series, and it is applied identically to all the $N_u$ input channels.

The MTF encoding can be traced back to the use of recurrence networks to analyze the structural properties of time series. As proposed in~\cite{art:Donner_2010}, the recurrence matrix of a time series can be interpreted as the adjacency matrix of an associated complex network. In~\cite{art:Campanharo}, the concept of building adjacency matrices was extended by extracting transition dynamics from first order Markov matrices. Given a time series $\mathbf{u}=(u_1,\ldots,u_L)^\top$, this is first discretized into $N_\omega$ quantile bins. Each entry $u_l$, $l=1,\ldots,L$, is assigned to the corresponding bin $\omega_\jj$, $\jj=1,\ldots,N_\omega$. A weighted adjacency matrix $\mathbf{Z}\in\mathbb{R}^{N_\omega\times N_\omega}$ is then built with entries $z_{\jj,\kk}=\overline{z}_{\jj,\kk}/\sum_\kk\overline{z}_{\jj,\kk}$, where $\kk=1,\ldots,N_\omega$ and $\overline{z}_{\jj,\kk}$ is the number of transitions $\omega_\jj\rightarrow \omega_\kk$ between consecutive time steps. $\mathbf{Z}$ is a Markov transition matrix. From a network perspective, each bin represents a node, and each pair of nodes is connected with a weight proportional to the probability that a data point in bin $\omega_\jj$ is followed by a data point in bin $\omega_\kk$.

The MTF encoding~\cite{art:MTF} $\overline{\mathbf{Z}}\in\mathbb{R}^{L\times L}$ expands $\mathbf{Z}$ by measuring the probabilities of observing a change of value between any pair of points in the time series. Similarly to $\mathbf{Z}$, the matrix $\overline{\mathbf{Z}}$ encodes the Markovian dynamics. However, in $\overline{\mathbf{Z}}$, the transition probabilities are represented sequentially to preserve the time dependence of the conditional relationships. The MTF matrix $\overline{\mathbf{Z}}$ reads:
\begin{linenomath*}
\begin{equation}
\begin{small}
\overline{\mathbf{Z}}=\left[\begin{matrix}z_{\jj,\kk}|u_1\in \omega_\jj,u_1\in \omega_\kk & \ldots & z_{\jj,\kk}|u_1\in \omega_\jj,u_L\in \omega_\kk \\ z_{\jj,\kk}|u_2\in \omega_\jj,u_1\in \omega_\kk & \ldots & z_{\jj,\kk}|u_2\in \omega_\jj,u_L\in \omega_\kk \\ \vdots & \ddots & \vdots \\ z_{\jj,\kk}|u_L\in \omega_\jj,u_1\in \omega_\kk & \ldots & z_{\jj,\kk}|u_L\in \omega_\jj,u_L\in \omega_\kk \end{matrix}\right]~,
\end{small}
\end{equation}
\end{linenomath*}
and measures the probability of a transition $\omega_\jj\rightarrow\omega_\kk$ for each pair of time steps, not necessarily consecutive. This is equivalent to unfolding the matrix $\mathbf{Z}$ along the time axis, taking into account the temporal positions of the data points in $\mathbf{u}$. By measuring the quantiles transition probabilities between two arbitrary time steps, matrix $\overline{\mathbf{Z}}$ captures the multi-span transition structure of the time series. 

The MTF requires discretizing the time series into $N_\omega$ quantile bins. Since this discretization is a surjective transformation, it is not reversible and leads to some loss of information. The amount of retained information is primarily governed by the refinement level of the discretization. Using an equally spaced discretization, a large $N_\omega$ may result in a sparse image, which may not effectively highlight structures and patterns in the data. Conversely, a small $N_\omega$ may cause significant information loss. To achieve a good trade-off between image sparsity and information retention, the symbolic aggregate approximation algorithm~\cite{art:SAX} is employed to perform a non-uniform bin assignment. As proposed in~\cite{art:MTF_Fink}, the time series is discretized into bins that approximately follow a Gaussian distribution. This non-uniform binning strategy is also well suited for handling time histories that exhibit long-tailed distributions, making the choice of the number of bins less critical. In the present work, the number of bins is set to $N_\omega=20$, which provides satisfactory results without imposing a significant computational burden. Finally, to control the image size and enhance the computational efficiency of downstream image processing, the MTF matrix $\overline{\mathbf{Z}}$ is downsized by averaging the pixels within each non-overlapping square patch using a blurring kernel.

\section{Implementation details} 
\label{sec:implementation}

In this Appendix, we discuss the implementation details of the DL models employed in \sez\ref{subsec:DL_models}.  The architectures and training options have been chosen through a preliminary study aimed at minimizing $\mathcal{L}_\text{AE}$ and $\mathcal{L}_\text{SUR}$ while retaining the generalization capabilities of $\text{N\hspace{-1px}N}_\text{ENC}$, $\text{N\hspace{-1px}N}_\text{DEC}$, and $\text{N\hspace{-1px}N}_\text{SUR}$.

$\text{N\hspace{-1px}N}_\text{ENC}$ and $\text{N\hspace{-1px}N}_\text{DEC}$ are defined as the encoder and decoder of a convolutional autoencoder, whose architecture is outlined in \tab\ref{tab:NN_AE_arch}. The encoding branch consists of a stack of four two-dimensional (2D) convolutional layers  followed by max pooling layers. The output is then flattened and passed through a fully-connected layer featuring $D_h=20$ neurons, which provides the low-dimensional features. This bottleneck layer is connected to the decoding branch via another fully-connected layer, whose output is reshaped before undergoing a stack of four transposed 2D convolutional layers to reconstruct the input mosaic. All convolutional layers employ $3\times3$ kernels and Softsign activation function, except for the final layer, which is Sigmoid-activated. Both fully-connected layers are Softsign-activated. 

Using Xavier's weight initialization~\cite{art:Glorot}, the loss function $\mathcal{L}_\text{AE}$ is minimized using Adam~\cite{art:Adam} for a maximum of $100$ epochs. The learning rate $\eta_\text{AE}$ is initially set to $0.001$ and is decreased over $4/5$ of the total training steps according to a cosine decay schedule with weight decay coefficient set to $0.05$. The optimization is carried out using an $80:20$ split of the dataset for training and validation, respectively. An early stopping strategy is adopted to interrupt the training process whenever the loss function evaluated on the validation set fails to decrease for a prescribed number of consecutive patience epochs. The relevant hyperparameters and training options are reported in \tab\ref{tab:NN_AE_hyper}.

\begin{table}[h!]
\caption{$\text{N\hspace{-1px}N}_\text{ENC}$ and $\text{N\hspace{-1px}N}_\text{DEC}$ --- (a) employed architecture and (b) selected hyperparameters and training options.}
      \centering
       \subfloat[\label{tab:NN_AE_arch}]{
       \scriptsize
		\begin{tabular}{lllll}
    \toprule
    \mbox{Layer} & \mbox{Type} & \mbox{Output shape} &  \mbox{Activ.} & \mbox{Input layer}\\
    \toprule
    \mbox{0} & \mbox{Input} & \mbox{$(B_\text{AE}, h_\I, w_\I, 1)$} & \mbox{None} & \mbox{None}\\
    \mbox{1} & \mbox{Conv2D} & \mbox{$(B_\text{AE}, h_\I, w_\I, 4)$} & \mbox{Softsign} & \mbox{0}\\
    \mbox{2} & \mbox{MaxPool2D} & \mbox{$(B_\text{AE}, h_\I/2, w_\I/2, 4)$} & \mbox{None} & \mbox{1}\\
    \mbox{3} & \mbox{Conv2D} & \mbox{$(B_\text{AE}, h_\I/2, w_\I/2, 8)$} & \mbox{Softsign} & \mbox{2}\\
    \mbox{4} & \mbox{MaxPool2D} & \mbox{$(B_\text{AE}, h_\I/4, w_\I/4, 8)$} & \mbox{None} & \mbox{3}\\
    \mbox{5} & \mbox{Conv2D} & \mbox{$(B_\text{AE}, h_\I/4, w_\I/4, 16)$} & \mbox{Softsign} & \mbox{4}\\
    \mbox{6} & \mbox{MaxPool2D} & \mbox{$(B_\text{AE}, h_\I/8, w_\I/8, 16)$} & \mbox{None} & \mbox{5}\\
     \mbox{7} & \mbox{Conv2D} & \mbox{$(B_\text{AE}, h_\I/8, w_\I/8, 32)$} & \mbox{Softsign} & \mbox{6}\\
    \mbox{8} & \mbox{MaxPool2D} & \mbox{$(B_\text{AE}, h_\I/16, w_\I/16, 32)$} & \mbox{None} & \mbox{7}\\   
    \mbox{9} & \mbox{Flatten} & \mbox{$(B_\text{AE}, h_\I w_\I/8)$} & \mbox{None} & \mbox{8}\\   
 \mbox{10} & \mbox{Dense} & \mbox{$(B_\text{AE}, D_h=20)$} & \mbox{Softsign} & \mbox{9}\\
    \mbox{11} & \mbox{Dense} & \mbox{$(B_\text{AE}, h_\I w_\I/8)$} & \mbox{Softsign} & \mbox{10}\\
    \mbox{12} & \mbox{Reshape} & \mbox{$(B_\text{AE}, h_\I/16, w_\I/16, 32)$} & \mbox{None} & \mbox{11}\\
     \mbox{13} & \mbox{Conv2D$^\top$} & \mbox{$(B_\text{AE}, h_\I/8, w_\I/8, 16)$} & \mbox{Softsign} & \mbox{12}\\
     \mbox{14} & \mbox{Conv2D$^\top$} & \mbox{$(B_\text{AE}, h_\I/4, w_\I/4, 8)$} & \mbox{Softsign} & \mbox{13}\\
     \mbox{15} & \mbox{Conv2D$^\top$} & \mbox{$(B_\text{AE}, h_\I/2, w_\I/2, 4)$} & \mbox{Softsign} & \mbox{14}\\
     \mbox{16} & \mbox{Conv2D$^\top$} & \mbox{$(B_\text{AE}, h_\I, w_\I, 1)$} & \mbox{Sigmoid} & \mbox{15}\\
        \bottomrule
          \end{tabular}		
          }\\
       \subfloat[\label{tab:NN_AE_hyper}]{
       \scriptsize
  \begin{tabular}{ll}
    \toprule
    
            \mbox{Convolution kernel size:} & \mbox{$3\times3$}\\
            \mbox{$L^2$ regularization rate:} & \mbox{$\lambda_\text{AE}=10^{-4}$}\\
		\mbox{Weight initializer:} & \mbox{Xavier}\\
		\mbox{Optimizer:} & \mbox{Adam}\\
		\mbox{Batch size:} & \mbox{$B_\text{AE}=128$}\\
		\mbox{Initial learning rate:} & \mbox{$\eta_\text{AE}=0.001$}\\
		\mbox{Allowed epochs:} & \mbox{$100$}\\
		\mbox{Learning schedule:} & \mbox{$\frac{4}{5}$ cosine decay}\\
		\mbox{Weight decay:} & \mbox{$0.05$}\\
		\mbox{Early stop patience:} & \mbox{15 epochs}\\
		\mbox{Positive pairings:} & \mbox{$\zeta_{+}=2$}\\
		\mbox{Negative pairings:} & \mbox{$\zeta_{-}=2$}\\
		\mbox{Similarity margin:} & \mbox{$\psi=1$}\\
		\mbox{Train-val split:} & \mbox{$80:20$}\\
    \bottomrule
  \end{tabular}
		}
\end{table}

$\text{N\hspace{-1px}N}_\text{SUR}$ is a four-layer fully-connected DL model, whose architecture is outlined in \tab\ref{tab:NN_SUR_arch}. The three hidden layers are Softsign-activated, while no activation is applied to the output layer. As in the previous case, the optimization is performed using Adam in combination with Xavier's weight initialization. The learning rate $\eta_\text{SUR}$ is progressively decreased during training according to a cosine decay schedule. An early stopping strategy is adopted to prevent overfitting, using an $80:20$ splitting ratio for training and validation. The relevant hyperparameters and the training options are summarized in \tab\ref{tab:NN_SUR_hyper}.
 
 \begin{table}[h!]
\caption{$\text{N\hspace{-1px}N}_\text{SUR}$ --- (a) employed architecture and (b) selected hyperparameters and training options.}
      \centering
       \subfloat[\label{tab:NN_SUR_arch}]{
       \scriptsize
		\begin{tabular}{lllll}
		    \toprule
    \mbox{Layer} & \mbox{Type} & \mbox{Output shape} &  \mbox{Activ.} & \mbox{Input layer}\\
\toprule
    \mbox{0} & \mbox{Input} & \mbox{$(B_\text{SUR},{N_\text{par}})$} & \mbox{None} & \mbox{None}\\
    \mbox{1} & \mbox{Dense} & \mbox{$(B_\text{SUR},10)$} & \mbox{Softsign} & \mbox{$0$}\\
    \mbox{2} & \mbox{Dense} & \mbox{$(B_\text{SUR},10)$} & \mbox{Softsign} & \mbox{$1$}\\
    \mbox{3} & \mbox{Dense} & \mbox{$(B_\text{SUR},40)$} & \mbox{Softsign} & \mbox{$2$}\\
    \mbox{4} & \mbox{Dense} & \mbox{$(B_\text{SUR},D_h=20)$} & \mbox{None} & \mbox{$3$}\\
        \bottomrule
		\end{tabular}	
          }
       \subfloat[\label{tab:NN_SUR_hyper}]{
       \scriptsize
  \begin{tabular}{ll}
    \toprule
    	\mbox{$L^2$ regularization rate:} & \mbox{$\lambda_\text{SUR}=10^{-4}$}\\
		\mbox{Weight initializer:} & \mbox{Xavier}\\
		\mbox{Optimizer:} & \mbox{Adam}\\
		\mbox{Batch size:} & \mbox{$B_\text{SUR}=128$}\\
		\mbox{Initial learning rate:} & \mbox{$\eta_\text{SUR}=0.001$}\\
		\mbox{Allowed epochs:} & \mbox{$5000$}\\
		\mbox{Learning schedule:} & \mbox{$\frac{4}{5}$ cosine decay}\\
		\mbox{Weight decay:} & \mbox{$0.01$}\\
		\mbox{Early stop patience:} & \mbox{$100$ epochs}\\
		\mbox{Train-val split:} & \mbox{$80:20$}\\
    \bottomrule
  \end{tabular}
		}
\end{table}
\end{document}